\pdfoutput=1
\documentclass{article}

\usepackage[eandd, final]{neurips_2026}

\usepackage[utf8]{inputenc} 
\usepackage[T1]{fontenc}    
\usepackage{hyperref}       
\usepackage{url}            
\usepackage{booktabs}       
\usepackage{amsfonts}       
\usepackage{nicefrac}       
\usepackage{microtype}      
\usepackage{xcolor}         
\usepackage[super]{nth}
\usepackage{amsmath}
\usepackage{amsthm}
\usepackage{mathtools}
\usepackage{bm}
\usepackage{dsfont}
\usepackage{graphicx}
\usepackage{amsmath,amssymb}
\usepackage{caption}
\usepackage{subcaption}
\usepackage{float}
\usepackage{enumitem}
\usepackage{array}
\usepackage{makecell}
\usepackage{hyperref}
\usepackage{multirow}
\usepackage{tabularx}
\usepackage{makecell} 
\usepackage{url}
\usepackage{xcolor}
\usepackage[flushleft]{threeparttable}
\usepackage[ruled,vlined,linesnumbered]{algorithm2e}
\usepackage{algpseudocode}
\usepackage[table]{xcolor}
\usepackage{threeparttable}
\usepackage{tabularx}
\usepackage{array}
\usepackage{amssymb}   
\usepackage{textcomp}  
\usepackage{wrapfig}
\usepackage[dvipsnames, table,xcdraw]{xcolor}

\usepackage{colortbl}

\usepackage{pifont}

\usepackage{minitoc}
\usepackage{tocloft}
\usepackage[toc,page,header]{appendix}
\usepackage{etoolbox}
\usepackage{environ}
\usepackage{placeins}

\usepackage{tcolorbox}
\tcbuselibrary{listings, breakable, skins}
\usepackage{listings}
\usepackage{xcolor}

\definecolor{cycle2}{RGB}{106, 191, 0}
\definecolor{cycle3}{RGB}{191, 0, 0}

\definecolor{amber}{rgb}{1.0, 0.75, 0.0}
\definecolor{awesome}{rgb}{1.0, 0.13, 0.32}
\definecolor{ao(english)}{rgb}{0.0, 0.5, 0.0}

\newcommand{\cmark}{\textcolor{cycle2}{\ding{52}}}
\newcommand{\xmark}{\textcolor{cycle3}{\ding{56}}}

\usepackage[normalem]{ulem}
\useunder{\uline}{\ul}{}

\usepackage{booktabs}

\theoremstyle{definition}

\definecolor{googleblue}{HTML}{4285F4}
\definecolor{googlered}{HTML}{DB4437}
\definecolor{googlepurple}{HTML}{A142F4} 
\definecolor{googlegreen}{HTML}{0F9D58}
\definecolor{googleyellow}{HTML}{F4B400} 
\definecolor{googleorange}{HTML}{FBBC05} 
\definecolor{googlecyan}{HTML}{34A853} 
\definecolor{googlegray}{HTML}{9AA0A6} 
\definecolor{googlepink}{HTML}{EA4335} 
\definecolor{googlelightblue}{HTML}{7BAAF7} 

\usepackage{makecell}

\newcounter{todocounter}
\newtoggle{comments}
 \toggletrue{comments} 
\NewEnviron{doweprint}[1]{%
  \iftoggle{#1}{\BODY}{}%
}

\title{Benchmarking Multi-Modal Graph-based \\Social Media Popularity Prediction \vspace{-0.5em}}

\author{
\\[-3em]
  \textbf{Utkarsh Sahu}$^{1}$ \quad
  \textbf{Zhisheng Qi}$^{2}$ \quad
  \textbf{Li Zhu}$^{2}$ \quad
  \textbf{Yizhao Yang}$^{1}$ \quad \\
  \textbf{Jun Li}$^{1}$ \quad 
  \textbf{Ryan Rossi}$^{3}$ \quad
  \textbf{Yu Wang}$^{2}$ \\[0.35em]
  $^{1}$\textbf{University of Oregon} \quad
  $^{2}$\textbf{University of Georgia} \quad
  $^{3}$\textbf{Adobe Research}
  \\[0.25em]
  \texttt{\{utkarsh, yizhao, lijun\}@uoregon.edu} \\
  \texttt{\{zq03788, ryanlizhu, yuwang\}@uga.edu} \\
  \texttt{ryrossi@adobe.com}
}

\begin{document}

\doparttoc
\faketableofcontents

\maketitle
\vspace{-5ex}
\begin{abstract}
\vspace{-2ex}
Social media popularity prediction aims to forecast the future reach or influence of online content from early-stage observations.
Accurate prediction enables key downstream applications, such as advertising optimization and strategic content planning by users, creators, and platforms.
Despite substantial progress, existing popularity prediction works often fail to jointly consider multimodal content and temporal social interaction signals. Moreover, the literature remains highly fragmented across datasets, modalities, observation windows, prediction targets, and evaluation protocols. This fragmentation prevents fair comparison and obscures a systematic understanding of how textual, visual, temporal, and interaction-based signals jointly shape popularity dynamics.
To address these challenges, we introduce \texttt{MMG-Pop}, a  \underline{\textbf{M}}ulti-\underline{\textbf{m}}odal \underline{\textbf{G}}raph-based \underline{\textbf{Pop}}ularity Prediction benchmark, which unifies datasets, modalities, temporal interaction signals, and representative baselines under a standardized evaluation protocol. Furthermore, we propose \texttt{MMG-PopNet}, a unified multi-modal graph-based network that jointly models the aforementioned multi-modal signals and graph-structured social interactions. Extensive experiments on MMG-Pop across Bluesky and Reddit platforms demonstrate the superior performance of MMG-PopNet and yield new insights into cross-platform training generalization, multi-task prediction benefits, multi-modality contributions, and LLM prediction limitations. These findings establish a foundation for future research on social dynamics modeling and intervention under heterogeneous modalities and socially-aware agentic ecosystem paradigms. The MMG-Pop benchmark and MMG-PopNet code are available at this
\color{blue}\href{https://github.com/utkarshxsahu/MMG-Pop}{Link}.
\vspace{-0.75ex}




\end{abstract}

\vspace{-2ex}
\section{Introduction}\label{sec:intro}
\vspace{-1ex}
Social dynamics refers to the patterns of interactions and relationships among individuals within a society~\cite{socialdynamicsdef,farmer2018social,brock2001discrete}, emerging across diverse real-world contexts such as public health behavior change, collective responses during crises, and political mobilization~\cite{centola2010spread,newman2001structure,zhou2021survey,bail2018exposure}.
Accurately modeling social dynamics provides critical insights for analyzing, anticipating, and potentially intervening in collective social behaviors (e.g., early detection of toxic information cascades and timely intervention strategies on online platforms)~\cite{bak2022combining,zhao2015enquiring,shao2018spread,cheng2017anyone,lin2021early}
In this work, we focus on one of the most important social dynamics modeling problems, social media popularity prediction, which aims to leverage early observations of social content to predict its future popularity/influence (e.g., number of likes, reposts) across diverse social contexts and modalities~\cite{lerman2010using,meghawat2018multimodal,szabo2010predicting}.
Effectively predicting popularity on social media has important implications for both platforms and users. For platforms, it supports content recommendation, trend forecasting, advertising, and efficient allocation of moderation resources~\cite{pinto2013using,cobbe2021algorithmic,tang2017popularity,javari2014accurate} by estimating which content is likely to attract substantial future engagement. For users, creators, and organizations, it helps estimate future reach, optimize posting content and promotion strategies~\cite{mazloom2016multimodal,zhang2018become}, and plan social or marketing campaigns more effectively~\cite{yu2011toward,van2017leading}. 

\begin{table}[t!]
\centering
\caption{Prior social media popularity prediction methods, organized by modality signal, prediction metric, modeling approach, and method category: feature engineering, statistical, and deep learning.}
\small
\setlength{\tabcolsep}{1pt}
\resizebox{\textwidth}{!}{
\begin{tabular}{lccccccr}
\toprule
\multirow{2.5}{*}{\textbf{Existing Work}} & \multicolumn{5}{c}{\textbf{Social Modality Signal}} & \multirow{2.5}{*}{\textbf{Popularity Metric}} & \multirow{2.5}{*}{\textbf{Popularity Prediction Method}} \\
\cmidrule(lr){2-6}
& \textbf{Graph} & \textbf{Text} & \textbf{Image} & \textbf{Time} & \textbf{Video} & & \\
\midrule
@Username~\cite{yang2010predicting}
& \cmark & \xmark & \xmark & \cmark & \xmark
& Speed/Width/Depth
& Cox PH + Log-linear Regression\\

Resubmissions~\cite{lakkaraju2013s}
& \xmark & \cmark & \xmark & \cmark & \xmark
& Reddit Karma
& Supervised LDA + Linear Regression \\

Szabo-Huberman~\cite{szabo2010predicting}
& \xmark & \xmark & \xmark & \cmark & \xmark
& View Count/ Digg votes
& Log-linear Regression \\

Flickr-SVR~\cite{khosla2014makes}
& \xmark & \cmark & \cmark & \xmark & \xmark
& View Count
& Support Vector Regression \\

\midrule
SEISMIC~\cite{zhao2015seismic}
& \xmark & \xmark & \xmark & \cmark & \xmark
& Cascade size
& Self-exciting point process \\

Galton--Watson~\cite{medvedev2019modelling}
& \cmark & \xmark & \xmark & \cmark & \xmark
& Cascade size
& Branching Hawkes process \\

HIP~\cite{lakkaraju2013s}
& \xmark & \cmark & \xmark & \cmark & \xmark
& View Count
& Hawkes Intensity Processes \\

\midrule
DeepHawkes~\cite{cao2017deephawkes}
& \cmark & \xmark & \xmark & \cmark & \xmark
& Cascade size
& GRU + Time Decay \\

DeepCas~\cite{li2017deepcas}
& \cmark & \xmark & \xmark & \xmark & \xmark
& Cascade size
& Bi-directional GRU \\

CasSeqGCN~\cite{wang2022casseqgcn}
& \cmark & \xmark & \xmark & \cmark & \xmark
& Cascade size
& GCN + LSTM \\

TSGNN~\cite{liu2021content}
& \cmark & \xmark & \xmark & \xmark & \xmark
& Cascade size
& GAT + GLU \\

GraphLSTM~\cite{zayats2018conversation}
& \cmark & \cmark & \xmark & \cmark & \xmark
& Reddit Karma
& Graph-structured LSTM \\

UHAN~\cite{zhang2018user}
& \xmark & \cmark & \cmark & \xmark & \xmark
& View Count
& Multi-modal Attention\\

HMMVED~\cite{xie2021micro}
& \xmark & \cmark & \cmark & \xmark & \cmark
& Comment/Repost/Likes/View
& Hierarchical Multimodal VAE\\

MMRA~\cite{zhong2024predicting}
& \xmark & \cmark & \cmark & \xmark & \cmark
& View Count
& Multi-modal Attention + Retrieval\\

MASSL~\cite{zhang2022multi}
& \xmark & \cmark & \cmark & \xmark & \cmark
& View Count
& Multimodal VAE\\
\bottomrule
\end{tabular}
}
\vspace{-3ex}
\label{tab:prior_work}
\end{table}

Prior social media popularity prediction can be categorized into three lines. 
The first predicts social media popularity based on the initial media content
without considering its subsequent spread~\cite{szabo2010predicting,bandari2012pulse,tsur2012s,khosla2014makes,gelli2015image,ding2019social}. 
However, they are content-centric and fail to account for how early social interaction dynamics, such as high-impact comments and influencers' reshares~\cite{garcia2017understanding}, impact eventual popularity.
The second leverages observed interaction patterns (e.g., reshares, replies, and user engagements) to model how a post propagates through the social interaction network. These interaction histories are represented as cascades and modeled using point processes or geometric deep learning (e.g., RNNs/GNNs) to capture the temporal/structural dynamics of information diffusion~\cite{zhao2015seismic,medvedev2019modelling,cao2017deephawkes,li2017deepcas,wang2022casseqgcn}. 
However, they under-utilize the semantic content of posts and responses, such as textual meaning and visual signals that shape engagement.
%
A third line of work jointly leverages both post content and interaction structure~\cite{aragon2017generative,zayats2018conversation,zhang2018conversations,zubiaga2016analysing}. However, these studies primarily focus on tasks such as toxicity detection, sentiment analysis, rumor propagation, or conversation modeling, rather than directly addressing popularity prediction.
More recently, emerging generative models and agentic AI have been applied to social dynamics modeling, either through LLM-based multi-agent simulations~\cite{liu2025popsim,yang2024oasis} or by directly repurposing LLMs as autoregressive cascade predictors or reasoning-augmented regressors for popularity forecasting~\cite{zheng2025autocas,xu2025forecasting}. However, none of these approaches has been systematically benchmarked against prior non-LLM baselines.




In addition to the above limitations, existing social media popularity prediction studies are constructed under heterogeneous yet inconsistent experimental settings as in Table~\ref{tab:prior_work}.
These differences span dataset versions
(e.g., X~\cite{cao2017deephawkes,li2017deepcas,wang2022casseqgcn} versus (vs.) Reddit~\cite{medvedev2019modelling,zayats2018conversation}), modality signals (e.g., text, image and video~\cite{zhang2022multi,zhong2024predicting,xie2021micro} vs. graph topology~\cite{li2017deepcas,cao2017deephawkes,liu2021content}), prediction targets (e.g., cascade size~\cite{liu2021content,zhao2015seismic,wang2022casseqgcn} vs. content view count~\cite{szabo2010predicting,zhang2018user}).
This fragmentation prevents meaningful comparison of existing methods, thereby hindering the derivation of key insights, such as which modalities contribute most and whether cross-platform transferability exists.
Furthermore, existing popularity prediction benchmarks~\cite{xu2025smtpd,wu2023smp} mainly focus on initial media content, overlooking the evolving cascade of user interactions over time.
%
%
To address these gaps, we introduce \texttt{MMG-Pop}, a unified benchmark for evaluating social media popularity prediction from evolving multimodal cascades. As existing methods capture only subsets of these heterogeneous signals, we further propose \texttt{MMG-PopNet}, a unified reference model that jointly integrates multimodal content, temporal dynamics, and graph-structured interactions. Our key contributions are summarized as follows:
\vspace{-1ex}

\begin{itemize}[itemsep=0pt, leftmargin=*]
    \item \textbf{Unified Popularity Prediction Benchmark.} We introduce the \texttt{MMG-Pop} benchmark, standardizing datasets, social modality signals, observation windows, prediction horizons, and popularity measures to enable consistent evaluation across in-domain forecasting, future-horizon prediction, and cross-platform transfer, with representative baselines.
    
    \item \textbf{Unified Multi-Modal Model.} We propose \texttt{MMG-PopNet}, the first unified architecture to jointly model multimodal content, graph-structured interaction dynamics, and temporal signals through bidirectional graph message passing, supporting multi-objective popularity prediction.
    
    \item \textbf{Comprehensive Experiments and Novel Insights.} We evaluate MMG-PopNet across regression and classification tasks, highlighting insights enabled by MMG-Pop, including the importance of jointly modeling multimodal content and cascade structure, generalization gains from cross-community training, benefits of multi-task learning, and LLM limitations in popularity prediction.
\end{itemize}

\newpage
\section{Design Space of MMG-Pop Popularity Prediction Benchmark}

\vspace{-1.5ex}
This section outlines the design space of our proposed \texttt{MMG-Pop} benchmark for social media popularity prediction, encompassing the notation and problem formulation, popularity measurement, training and evaluation, dataset curation, and existing baselines.

\begin{figure}[t!]
    \centering
    \includegraphics[width=1\linewidth]{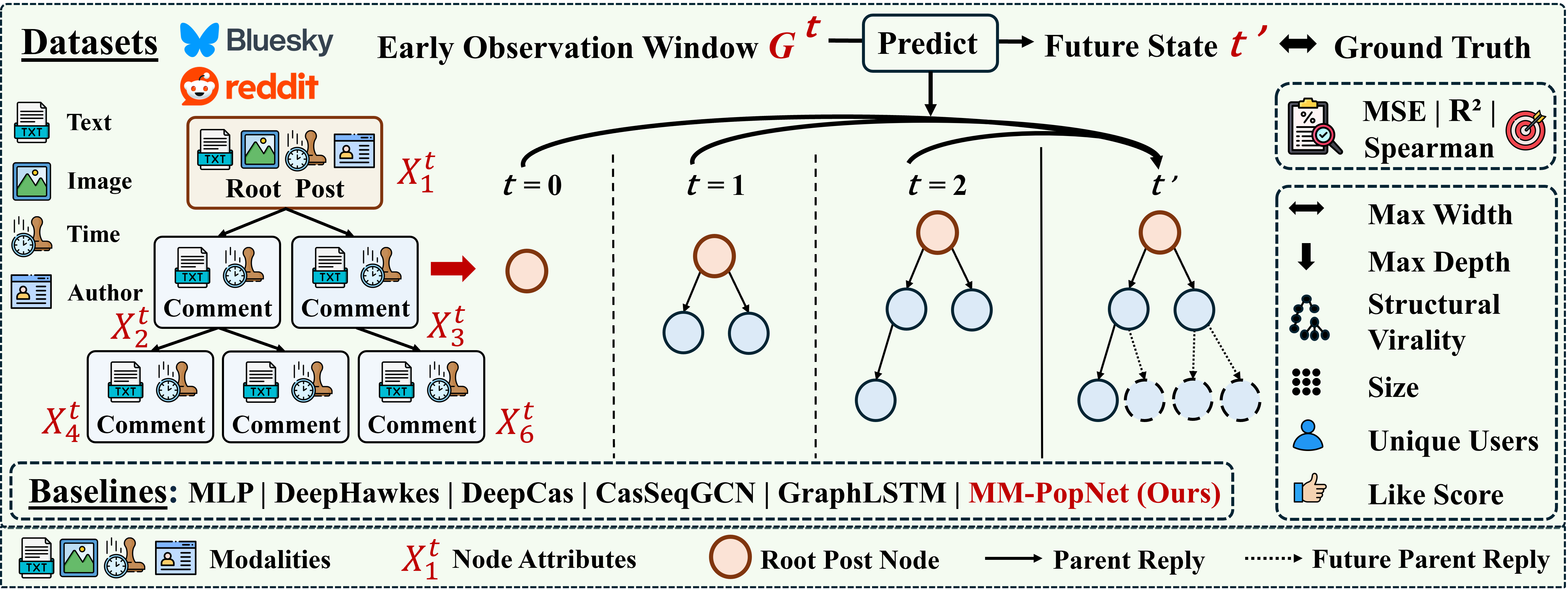}
    \caption{\textbf{Overview of MMG-Pop Benchmark.} Social cascades from Bluesky and Reddit are represented as tree-structured graphs, where each node carries multi-modal attributes. Given only an early observed prefix $G^t$, the task is to predict six complementary popularity dimensions characterizing the future cascade state $G^{t'}$. The benchmark evaluates baselines alongside our proposed \texttt{MMG-PopNet} across multiple observation windows.}
    \label{fig:model}
    \vspace{-3ex}
\end{figure}

\vspace{-1.75ex}
\subsection{Notation and Problem Formulation}\label{sec:prelim}

\vspace{-1.5ex}
\textbf{Notation of Social Media Popularity Prediction.} We represent an online social cascade as a directed tree-structured graph 
\(G = (\mathcal{V}, \mathcal{E})\), 
where \(\mathcal{V}\) denotes the set of nodes and \(\mathcal{E} \subseteq \mathcal{V} \times \mathcal{V}\) denotes the set of directed edges. 
Each node \(v \in \mathcal{V}\) corresponds to a content item (e.g., a post, comment, or reply), and a directed edge \((u, v) \in \mathcal{E}\) indicates that node \(v\) is generated in response to node \(u\), capturing the information propagation (e.g., reply-to, repost/reshare, or quote relationships). 
The graph is rooted at a unique node $v_{\text{root}} \in \mathcal{V}$ representing the initial item of the cascade (e.g., a starter post on Bluesky or a submission on Reddit). Each node \(v \in \mathcal{V}\) is associated with multimodal attributes \(X_v = (X_v^{\text{Text}},\, X_v^{\text{Visual}},\, X_v^{\text{Graph}},\, X_v^{\text{Social}},\, X_v^{\text{Time}})\) where \(X_v^{\text{Text}}\) denotes textual content features (e.g., post/comment text, hashtags, or semantic embeddings~\cite{zhang2018user,lakkaraju2013s}), 
\(X_v^{\text{Visual}}\) denotes visual features when media is present (e.g., images, videos, or visual descriptors~\cite{khosla2014makes, zhang2022multi}), \(X_v^{\text{Graph}}\) denotes local structural or network context (e.g., neighborhood subgraph statistics or position within the cascade~\cite{zayats2018conversation}), 
\(X_v^{\text{Social}}\) denotes author-level social context features (e.g., user profile attributes~\cite{keneshloo2016predicting} and interaction graph-derived influence proxies, including centrality and PageRank scores~\cite{guo2016comparison,brin1998anatomy}), and \(X_v^{\text{Time}}\) denotes temporal features (e.g., global timestamp or relative time to parent nodes within the cascade~\cite{zayats2018conversation}). In addition, each cascade may be associated with thread-level contextual metadata \(X^{\text{Thread}}\), capturing properties of the root post $v_{\text{root}}$ (e.g., topic, presence of visual media, or root-author follower count).

\textbf{Formulation of Social Media Popularity Prediction.} 
The core objective is to predict the future evolution of a social cascade given only its early-stage observations. Let \(t \ge 0\) denote the elapsed time since the root post. 
Given a cascade graph \(G = (\mathcal{V}, \mathcal{E})\), we define the observed prefix at time \(t\) as \(G^{t} = (\mathcal{V}^{t}, \mathcal{E}^{t})\), where \(\mathcal{V}^{t} = \{ v \in \mathcal{V} \mid t_v \le t \}\) and \(\mathcal{E}^{t}\) contains all edges among nodes in \(\mathcal{V}^{t}\). This prefix captures the historical context of the cascade, including time-truncated multimodal node attributes \(\{X_v^{t}\}_{v \in \mathcal{V}^{t}}\) and thread-level context \(X^{\text{Thread}, t}\) observable up to time \(t\).
The prediction target is the future state of the cascade, \(\mathbf{Y}_G \in \mathbb{R}^K\), consisting of \(K\) popularity measures defined in Section~\ref{sec:metric}. We aim to learn a parametric mapping $\mathcal{F}_{\boldsymbol{\Theta}}: \left(G^{t}, \{X_v^{t}\}_{v \in \mathcal{V}^{t}}, X^{\text{Thread}, t} \right) \mapsto \mathbf{Y}_G$.

\begin{table}[t!]
\centering

\caption{
\textbf{Dataset statistics} across observation windows, excluding cascades completed within each window.
\textbf{w/ Images} counts cascades containing a root-post image (``--'' indicates unavailable modality).
\textbf{Final (Target)} describes the complete cascades at the data collection cutoff, while \textbf{Early (Input)} represents the observed cascade prefixes within each observation window.
\textbf{Total}, \textbf{Mean}, and \textbf{Med.} denote total nodes, mean nodes per cascade, and median nodes per cascade.
\textbf{Early/Final (\%)} denotes the percentage of final nodes observed within the window (total early nodes / total final nodes).
}

\label{tab:dataset}
\vspace{1mm}

\fontsize{8.3pt}{9.7pt}\selectfont
\setlength{\tabcolsep}{3.25pt}
\renewcommand{\arraystretch}{0.95}

\begin{tabular}{lllrrrrrrrr}
\toprule
\multirow{2}{*}{\textbf{Dataset}} &
\multirow{2}{*}{\textbf{Window}} &
\multirow{2}{*}{\textbf{Cascades}} &
\multirow{2}{*}{\textbf{w/ Images}} &
\multicolumn{3}{c}{\textbf{Final (Target)}} &
\multicolumn{3}{c}{\textbf{Early (Input)}} &
\multirow{2}{*}{\makecell{\textbf{Early/Final}\\\textbf{(\%)}}} \\

\cmidrule(lr){5-7}
\cmidrule(lr){8-10}

& & & &
\textbf{Total} & \textbf{Mean} & \textbf{Med.} &
\textbf{Total} & \textbf{Mean} & \textbf{Med.} & \\
\midrule

\multirow{3}{*}{Bluesky}
    & 2min  & 63,896 & -- & 1,198,608 & 18.76 & 7.0 & 103,807 & 1.62 & 1.0 & 8.7\% \\
    & 10min & 59,305 & -- & 1,179,377 & 19.89 & 7.0 & 195,275 & 3.29 & 2.0 & 16.6\% \\
    & 20min & 55,486 & -- & 1,159,651 & 20.90 & 7.0 & 262,885 & 4.74 & 2.0 & 22.7\% \\
\midrule

\multirow{3}{*}{r/AMA}
    & 15min & 38,982 & -- & 1,500,181 & 38.5 & 15.0 & 156,139 & 4.0  & 3.0 & 10.4\% \\
    & 30min & 37,682 & -- & 1,491,365 & 39.6 & 16.0 & 259,185 & 6.9  & 5.0 & 17.4\% \\
    & 60min & 35,677 & -- & 1,471,684 & 41.3 & 17.0 & 392,444 & 11.0 & 7.0 & 26.7\% \\
\midrule

\multirow{3}{*}{r/Gaming}
    & 20min & 13,398 & 5,065 & 1,739,720 & 129.8 & 22.0 & 70,939 & 5.3 & 4.0 & 4.1\% \\
    & 50min & 12,882 & 4,875 & 1,735,771 & 134.7 & 24.0 & 143,475 & 11.1 & 7.0 & 8.3\% \\
    & 90min & 12,413 & 4,714 & 1,730,833 & 139.4 & 25.0 & 237,263 & 19.1 & 9.0 & 13.7\% \\
\midrule

\multirow{3}{*}{r/Futurology}
    & 30min  & 9,780 & 6,699 & 1,220,687 & 124.8 & 14.0 & 35,077 & 3.6 & 2.0 & 2.9\% \\
    & 90min  & 8,765 & 6,207 & 1,215,134 & 138.6 & 18.0 & 95,953 & 10.9 & 4.0 & 7.9\% \\
    & 180min & 8,349 & 6,025 & 1,211,576 & 145.1 & 19.0 & 197,398 & 23.6 & 7.0 & 16.3\% \\

\bottomrule
\end{tabular}
\vspace{-0.75em}
\end{table}

\vspace{-1.5ex}
\subsection{Popularity Measurement}\label{sec:metric}

\vspace{-1.15ex}
Social popularity can be quantified through multiple measures. Following prior literature~\cite{vosoughi2018spread,goel2016structural,zhang2021conspiracy,szabo2010predicting}, MMG-Pop considers six distinct dimensions to capture popularity dynamics. We categorize these six metrics into three conceptual dimensions: \textcolor{googleblue}{\bf \scshape structural}, \textcolor{googlegreen}{\bf \scshape participation}, and \textcolor{googlepurple}{\bf \scshape engagement}:


\vspace{-1.15ex}
\begin{itemize}[leftmargin=*, itemsep=0pt]
    \item \textcolor{googleblue}{\textbf{Max Width:}} It measures the largest breadth of the cascade~\cite{vosoughi2018spread,zhang2021conspiracy}. It is quantified as the maximum number of nodes appearing at any depth level in the cascade graph $G$.
    \item \textcolor{googleblue}{\textbf{Max Depth:}\label{task_width}} It measures the length of the longest reply chain in the cascade~\cite{zhang2021conspiracy,szabo2010predicting}. It is quantified as the maximum distance from $v_{\text{root}}$ to any node in the cascade graph $G$.
    \item \textcolor{googleblue}{\textbf{Structural Virality:}} It quantifies whether cascade diffusion is dominated by shallow broadcast spread or deeper multi-hop propagation~\cite{goel2016structural}. It is measured as the average shortest-path distance between all pairs of distinct nodes in the cascade graph.

    \item \textcolor{googleblue}{\textbf{Size:}} It measures how many content items are generated in the cascade~\cite{zhang2021conspiracy,li2017deepcas}. It is quantified as the number of nodes in the cascade graph $G$, i.e., $|\mathcal{V}|$.

    \item \textcolor{googlegreen}{\textbf{Unique Users:}} It measures the number of unique users who participate in the cascade~\cite{zhang2021conspiracy}.

    \item \textcolor{googlepurple}{\textbf{Like Score:}} It captures platform-visible engagement received by the root post content $v_{\text{root}}$ through platform-specific metrics such as likes or up-votes (e.g., Reddit Karma score)~\cite{szabo2010predicting}.
    
\end{itemize}
\vspace{-1ex}
We further examine the relationships among these metrics in Appendix~\ref{sec:popularity_correlation}.
\vspace{-2ex}

\subsection{Training and Evaluation}\label{sec:traineval}
\vspace{-1.15ex}
\textbf{Training.}  
Let \(\mathcal{G}\) denote the set of cascades, partitioned into \(\mathcal{G}^{\text{Train}} \cup \mathcal{G}^{\text{Val}} \cup \mathcal{G}^{\text{Test}}\).  
For each cascade \(G \in \mathcal{G}^{\text{Train}}\), we construct its observed prefix \(G^{t}\) with truncated node/thread-level features \(\{X_v^{t}\}\) and \(X^{\text{Thread},t}\).  
The objective is to predict future outcomes at a later time \(t' > t\). Given heavy-tailed nature of social popularity signals~\cite{cha2009measurement,tatar2014survey}, we define popularity targets in the log-transformed space: $\widetilde{\textbf{Y}}_G^{t'} = \log(\mathbf{I} + \textbf{Y}_G^{t'})$. The model is trained to predict \(\widetilde{\textbf{Y}}_G^{t'}\) by optimizing: $\boldsymbol{\Theta}^{*}
=
\arg\min_{\boldsymbol{\Theta}}
\sum_{G \in \mathcal{G}^{\text{Train}}}
\mathcal{L}(
\mathcal{F}_{\boldsymbol{\Theta}}(G^{t}, \{X_v^{t}\}_{v\in\mathcal{V}^t}, X^{\text{Thread},t}),
\widetilde{\textbf{Y}}_G^{t'}
)$ where $\mathcal{L}$ is the mean squared error (MSE) loss: $
\mathcal{L}(\widehat{\textbf{Y}}, \widetilde{\textbf{Y}})
=
\frac{1}{d}
||
\widehat{\textbf{Y}} - \widetilde{\textbf{Y}}
||_2^2, 
$ with \(d\) denoting the number of prediction targets.

\textbf{Evaluation.}  
During evaluation, the learned model \(\mathcal{F}_{\boldsymbol{\Theta}}^{*}\) is applied to unseen cascades \(G \in \mathcal{G}^{\text{Test}}\), using only their observed prefixes \(G^{t}\) and corresponding features.  
We evaluate performance by comparing the predicted future cascade dynamics at time \(t'\) with the corresponding ground-truth values. 
We report MSE, \(R^2\), and Spearman correlation, all computed in the log-transformed space.

\vspace{-2ex}
\subsection{Dataset Curation}\label{sec:dataset_curation}

\vspace{-1.5ex}
We curate social cascades from Bluesky and Reddit, two platforms with distinct platform dynamics. Each discussion thread is represented as a tree-structured social cascade following the formulation in Section~\ref{sec:prelim}, where the root post is \(v_{\text{root}}\), posts or comments are nodes in \(\mathcal{V}\), and parent--reply relations define directed edges \((u,v)\in\mathcal{E}\). Node attributes  \(X_v\) and thread-level context \(X^{\text{Thread}}\) are instantiated from the available platform metadata.\\
%
\textbf{Bluesky.} We curate our Bluesky subset from \cite{failla2024m}, which originally contains approximately 235 million posts from 4 million users between February 2023 and March 2024. We construct cascades from reply interactions, which provide explicit conversational content for modeling discussion dynamics.\\
%
\textbf{Reddit.} We use Pushshift data dumps \cite{baumgartner2020pushshift} from three communities: r/AMA, r/Gaming, and r/Futurology. These subreddits capture complementary discussion styles: centralized Q\&A, media-rich entertainment discussion, and speculative scientific discourse. We construct cascades from submissions and their comment reply trees. The datasets cover July 2021--December 2024 for r/AMA, January 2023--August 2024 for r/Gaming, and August 2019--December 2024 for r/Futurology.

Raw platform records do not directly provide the temporally structured cascades required for early-stage popularity prediction. We reconstruct reply trees by linking Bluesky posts through thread and parent identifiers, and Reddit comments through their parent references to submissions or other comments. To ensure structural validity, we discard posts with missing parents and cascades with missing roots, while retaining deleted or removed Reddit comments when their interaction structure and metadata remain intact. This preserves observed participation even when textual content is unavailable. We then map platform-specific records into a unified schema of node attributes and thread-level context, from which we construct early observation prefixes and future popularity targets. To account for varying cascade evolution speeds, we adopt dataset-specific observation windows of 2/10/20 minutes for Bluesky, 15/30/60 for r/AMA, 20/50/90 for r/Gaming, and 30/90/180 for r/Futurology, excluding cascades already completed within each window. Table~\ref{tab:dataset} provides the dataset statistics, with additional dataset information in Appendix~\ref{sec:appendix_dataset}.

\vspace{-2ex}
\subsection{Representative Baselines}\label{sec:baselines}

\vspace{-1.5ex}
We evaluate representative baselines spanning structure-agnostic, temporal, sequence-based, graph-based, and content-aware cascade modeling baselines. \textbf{MLP} is a structure-agnostic baseline that represents each cascade using root-post features, aggregated reply features, and global thread metadata. \textbf{DeepHawkes}~\cite{cao2017deephawkes} captures temporal diffusion dynamics through user embeddings, diffusion-path encoding, and time-decay modeling. \textbf{DeepCas}~\cite{li2017deepcas} models cascades as sampled diffusion paths and learns sequence representations with attention. \textbf{CasSeqGCN}~\cite{wang2022casseqgcn} represents temporal graph evolution by encoding graph snapshots with GCN to model time progression. \textbf{GraphLSTM}~\cite{zayats2018conversation} is a content-aware graph sequence baseline that models reply-tree structure with textual, user, temporal, and structural features. We follow the original architectures and modality configurations of the representative baselines to evaluate how the intended modeling approaches would perform on the benchmark. Additional details along with modality encoder analysis are provided in the Appendix~\ref{sec:appendix_baseline}.

\begin{figure}[t!]
    \centering
    \includegraphics[width=1\linewidth]{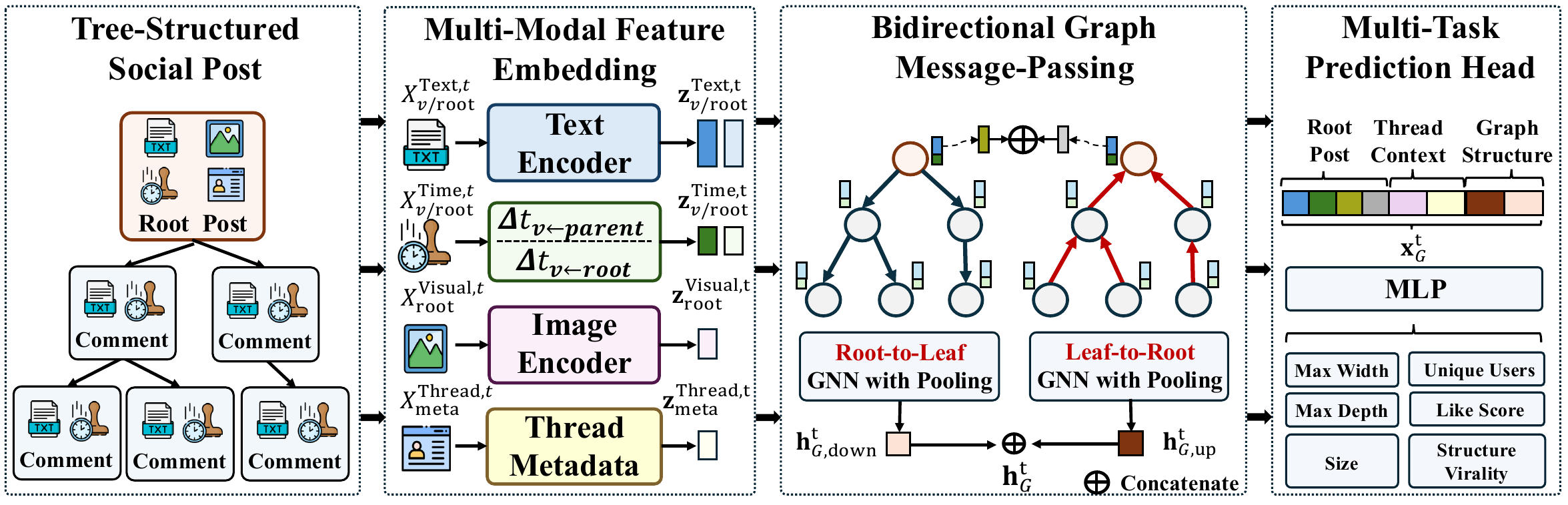}
    \vspace{-3ex}
    \caption{\textbf{Overview of MMG-PopNet Model:} The model embeds node-level text and temporal signals for bidirectional graph message passing over the cascade, and root visual content and thread metadata are encoded as separate contextual features. The learned root, graph, visual, and metadata representations are fused at the prediction stage to support multi-task popularity forecasting.}
    \label{fig:model}
    \vspace{-3ex}
\end{figure}

\vspace{-2.5ex}
\section{Foundational Multi-modal Graph-based Popularity Prediction Network}
\vspace{-2ex}
We introduce MMG-PopNet as a reference framework that jointly models multimodal content and graph-structured social interactions for popularity prediction, as illustrated in Figure~\ref{fig:model}. The model first encodes heterogeneous cascade signals, including textual, temporal, visual, and thread-level context, and then employs bidirectional graph message passing to capture interaction dynamics. The resulting representations are fused to predict multiple dimensions of future popularity. While MMG-Pop defines a model-independent benchmark, MMG-PopNet demonstrates how its complementary signals can be integrated within a unified framework.

\textbf{Multi-Modal Feature Embedding.}
We employ modality-specific encoders to capture the heterogeneous attributes of observed cascades. For each node \(v \in \mathcal{V}^{t}\), textual content is encoded using a Transformer-based encoder to obtain \(\mathbf{z}_v^{\text{Text},t}\). Temporal features capture the elapsed time since the parent reply and the root post, providing information about the pace and stage of cascade evolution. These signals are transformed into temporal representations \(\mathbf{z}_v^{\text{Time},t}\). In addition to node-level features, we incorporate thread-level context consisting of root-post visual content and metadata. When available, root images are encoded using CLIP~\cite{radford2021learning}, while root-author follower count and cyclic posting-time features capture user influence and posting context. Together, these form the thread-level representation \(\mathbf{z}^{\text{Thread},t}\).

\textbf{Bidirectional Graph Message-Passing.}
To model structural dependencies and information propagation within the observed cascade, we initialize node representations by combining textual and temporal embeddings. We then apply bidirectional GraphSAGE message passing over the reply tree, enabling information aggregation in both root-to-leaf and leaf-to-root directions. These complementary directions capture how interactions propagate outward from the root and how responses throughout the cascade contribute to its structural representation. After \(L\) layers, mean pooling aggregates the node embeddings from each direction, and their concatenation yields the graph-level representation \(\mathbf{h}_G^t\). This allows the model to jointly capture content-aware interaction patterns and the evolving topology of social discussions.

\textbf{Multi-Modal Feature Fusion and Multi-Task Prediction.}
We combine four complementary representations: (1) the initial root-node embedding, (2) the final bidirectional root representation, (3) the pooled graph representation, and (4) the thread-level context. These are concatenated into a unified cascade representation:
 $\mathbf{x}_G^t
=
\mathbf{h}_{\text{root}}^{(0)}
\;\Vert\;
\mathbf{h}_{\text{root}}^{(L)}
\;\Vert\;
\mathbf{g}_{G}^{t}
\;\Vert\;
\mathbf{z}^{\text{Thread,t}}.$
A shared multi-layer perceptron (MLP) maps this representation to \(K\) log-space popularity predictions, \(\widehat{\mathbf{Y}}_G^{t'} \in \mathbb{R}^{K}\). The model is trained end-to-end using a multi-task mean squared error objective, jointly learning representations across the popularity targets. Full feature transformations, message-passing formulations, and implementation details are provided in the Appendix~\ref{sec:appendix_mmg_popnet}.
\vspace{-2ex}

\vspace{-1ex}
\section{Experiment}\label{sec-experiments}
\vspace{-2ex}
We conduct extensive experiments with the MMG-Pop benchmark, designed to systematically address the following research questions:

\noindent$\mathbf{Q}_1$: \textit{How do different baselines and MMG-Pop-Net perform in predicting cascade popularity across varying future horizons and early observation windows under our MMG-Pop benchmark?}

\noindent$\mathbf{Q}_2$: \textit{Does unified training across communities and platforms improve social popularity prediction?}

\noindent$\mathbf{Q}_3$: \textit{Can MM-LLMs serve as competitive predictors for multimodal social popularity forecasting?}

\noindent$\mathbf{Q}_4$: \textit{Does multi-objective prediction benefit from jointly predicting popularity targets?}

\noindent$\mathbf{Q}_5$: \textit{How do different modalities contribute to popularity prediction?}

\noindent$\mathbf{Q}_6$: \textit{Can MMG-Pop benchmark support a broader range of prediction tasks beyond regression?}

Detailed experiment settings are described in Appendix~\ref{sec:appendix_experiment}.




\vspace{-2ex}
\subsection{$\mathbf{Q}_1$: Popularity Prediction across Varying Future Horizons/Early Observation Windows.} 
\vspace{-0.6em}
\textbf{Popularity prediction of final cascade state under different early observation.}
Given an early observation window $t$, we aim to predict the final popularity of a social cascade based on the cascade state observed up to the data collection time.
To assess the role of early information, we consider a root-only setting (window 0) in addition to the three dataset-specific observation windows defined in Section~\ref{sec:dataset_curation}. This setting includes only the root post and its thread-level context, allowing us to compare predictions based on initial content and context alone with those incorporating progressively more early interaction information.
Table~\ref{tab:MSE_performance} reports MSE in the log-transformed target space across all datasets, observation windows, and prediction targets. 
MMG-PopNet achieves the best overall average MSE for every target category, with 4.6\% to 17.0\% reductions over the strongest non-MMG-PopNet baselines.
Among the baselines, Graph-LSTM is the strongest competitor on most structural targets, reflecting the value of reply-tree structure, temporal ordering, and textual content. CasSeqGCN also performs competitively, likely by modeling evolving cascade as snapshots, which capture propagation topology. MLP is particularly strong for \textsc{Like Score} prediction because this target measures engagement received by the root post, whose features are directly combined with the mean representation of early observed nodes. However, without message passing, MLP cannot fully model cascade-level dependencies, whereas MMG-PopNet integrates these early signals through bidirectional graph propagation and achieves the lowest average MSE across all targets. 



\begin{table}[t]
\small
\caption{MSE results for final cascade-state prediction under different early observation windows, covering \textcolor{googleblue}{\bf \scshape structural prediction tasks} (\textcolor{googleblue}{max width, max depth, structural virality, and cascade size}), \textcolor{googlegreen}{\bf \scshape unique-user prediction}, and \textcolor{googlepurple}{\bf \scshape like-score prediction}. Lower values indicate better performance. The best results are highlighted in \textbf{bold}, while the second-best results are \underline{underlined}. Statistical significance analyses show that improvements are significant across settings in Table~\ref{tab:significance_final}.}

\centering
\setlength{\tabcolsep}{3pt}
\renewcommand{\arraystretch}{1.25}
\resizebox{\textwidth}{!}{%
\begin{tabular}{c l | ccccc ccccc ccccc ccccc |c}
\toprule
\multirow{2}{*}{\textbf{Task}} & 
\multirow{2}{*}{\textbf{Model}} & 
\multicolumn{5}{c}{\textbf{Bluesky}} & 
\multicolumn{5}{c}{\textbf{r/AMA}} & 
\multicolumn{5}{c}{\textbf{r/Gaming}} & 
\multicolumn{5}{c|}{\textbf{r/Futurology}} & 
\multirow{2}{*}{\textbf{Avg}} \\ 
\cmidrule(r){3-7}
\cmidrule(r){8-12}
\cmidrule(r){13-17}
\cmidrule(lr){18-22}

 &  & 
\textbf{0} & \textbf{2} & \textbf{10} & \textbf{20} & \textbf{Avg} & 
\textbf{0} & \textbf{15} & \textbf{30} & \textbf{60} & \textbf{Avg} & 
\textbf{0} & \textbf{20} & \textbf{50} & \textbf{90} & \textbf{Avg} & 
\textbf{0} & \textbf{30} & \textbf{90} & \textbf{180} & \textbf{Avg} & \\ 
\midrule

\multirow{6}{*}{\textcolor{googleblue}{\bf \scshape Max Width}} & 
MLP & \underline{0.435} & \underline{0.406} & 0.338 & 0.326 & \underline{0.376} & \underline{0.664} & 0.528 & 0.493 & 0.415 & 0.525 & \underline{1.755} & 1.182 & 0.877 & 0.708 & 1.131 & 1.810 & 1.341 & 0.697 & 0.457 & 1.076 & 0.777 \\
 & DeepHawkes & 0.659 & 0.595 & 0.387 & 0.336 & 0.494 & 0.713 & 0.554 & 0.491 & 0.365 & 0.531 & 1.888 & 1.268 & 1.147 & 0.839 & 1.286 & 1.876 & 1.643 & 1.184 & 0.783 & 1.371 & 0.921 \\
 & DeepCas & 0.601 & 0.648 & 0.677 & 0.684 & 0.653 & 0.841 & 0.757 & 0.767 & 0.709 & 0.769 & 2.361 & 1.895 & 1.833 & 1.690 & 1.945 & \underline{1.738} & 1.643 & 1.190 & 1.101 & 1.418 & 1.196 \\
 & CasSeqGCN & 0.649 & 0.518 & 0.332 & 0.290 & 0.447 & 0.713 & 0.534 & 0.446 & 0.313 & 0.502 & 1.864 & \underline{1.140} & 0.841 & 0.587 & 1.108 & 1.820 & 1.258 & 0.679 & \underline{0.366} & 1.031 & 0.772 \\
 & Graph-LSTM & 0.635 & 0.506 & \underline{0.326} & \underline{0.285} & 0.438 & 0.686 & \underline{0.510} & \underline{0.430} & \underline{0.305} & \underline{0.483} & 1.771 & 1.155 & \underline{0.811} & \textbf{0.518} & \underline{1.064} & 1.749 & \underline{1.149} & \textbf{0.631} & \textbf{0.345} & \underline{0.969} & \underline{0.738} \\
 & \textbf{MMG-PopNet} & \textbf{0.430} & \textbf{0.358} & \textbf{0.258} & \textbf{0.239} & \textbf{0.321} & \textbf{0.625} & \textbf{0.491} & \textbf{0.429} & \textbf{0.276} & \textbf{0.455} & \textbf{1.557} & \textbf{1.096} & \textbf{0.714} & \underline{0.562} & \textbf{0.982} & \textbf{1.581} & \textbf{1.132} & \underline{0.665} & 0.372 & \textbf{0.938} & \textbf{0.674} \\ 
\midrule

\multirow{6}{*}{\textcolor{googleblue}{\bf \scshape Max Depth}} & 
MLP & \underline{0.374} & 0.354 & 0.286 & 0.287 & 0.325 & \underline{0.318} & 0.267 & 0.225 & 0.198 & 0.252 & \underline{0.344} & \underline{0.279} & 0.214 & 0.186 & 0.256 & 0.593 & 0.484 & \underline{0.289} & \underline{0.221} & 0.397 & 0.307 \\
 & DeepHawkes & 0.384 & 0.372 & 0.336 & 0.361 & 0.363 & 0.329 & 0.280 & 0.253 & 0.240 & 0.276 & 0.346 & 0.290 & 0.269 & 0.229 & 0.284 & 0.602 & 0.567 & 0.429 & 0.325 & 0.481 & 0.351 \\
 & DeepCas & 0.376 & 0.381 & 0.369 & 0.415 & 0.385 & 0.432 & 0.375 & 0.341 & 0.333 & 0.370 & 0.577 & 0.409 & 0.387 & 0.340 & 0.428 & 0.596 & 0.554 & 0.410 & 0.374 & 0.483 & 0.417 \\
 & CasSeqGCN & 0.381 & 0.367 & 0.292 & 0.311 & 0.338 & 0.328 & 0.281 & 0.227 & 0.193 & 0.257 & 0.347 & 0.296 & 0.247 & 0.210 & 0.275 & 0.586 & 0.477 & 0.317 & 0.230 & 0.403 & 0.318 \\
 & Graph-LSTM & 0.379 & \underline{0.351} & \underline{0.282} & \underline{0.284} & \underline{0.324} & 0.323 & \underline{0.264} & \textbf{0.215} & \underline{0.176} & \underline{0.245} & 0.353 & 0.288 & \underline{0.209} & \underline{0.161} & \underline{0.253} & \underline{0.570} & \underline{0.453} & 0.298 & 0.232 & \underline{0.388} & \underline{0.302} \\
 & \textbf{MMG-PopNet} & \textbf{0.370} & \textbf{0.340} & \textbf{0.274} & \textbf{0.274} & \textbf{0.315} & \textbf{0.314} & \textbf{0.256} & \underline{0.219} & \textbf{0.172} & \textbf{0.240} & \textbf{0.335} & \textbf{0.275} & \textbf{0.200} & \textbf{0.160} & \textbf{0.243} & \textbf{0.517} & \textbf{0.418} & \textbf{0.282} & \textbf{0.205} & \textbf{0.356} & \textbf{0.288} \\ 
\midrule

\multirow{6}{*}{\makecell{\textcolor{googleblue}{\bf \scshape Structural}\\ \textcolor{googleblue}{\bf \scshape Virality} }} & 
MLP & \underline{0.159} & \underline{0.151} & \underline{0.115} & \underline{0.122} & \underline{0.137} & \textbf{0.129} & \underline{0.105} & 0.089 & 0.074 & 0.099 & \underline{0.096} & \textbf{0.078} & \underline{0.059} & 0.049 & \underline{0.071} & 0.209 & 0.164 & \underline{0.102} & \underline{0.077} & 0.138 & \underline{0.111} \\
 & DeepHawkes & 0.170 & 0.165 & 0.137 & 0.152 & 0.156 & 0.134 & 0.111 & 0.101 & 0.094 & 0.110 & \textbf{0.095} & 0.085 & 0.083 & 0.069 & 0.082 & 0.200 & 0.195 & 0.157 & 0.127 & 0.170 & 0.130 \\
 & DeepCas & 0.165 & 0.169 & 0.160 & 0.187 & 0.170 & 0.218 & 0.168 & 0.146 & 0.141 & 0.168 & 0.265 & 0.154 & 0.140 & 0.117 & 0.169 & 0.213 & 0.200 & 0.165 & 0.145 & 0.181 & 0.172 \\
 & CasSeqGCN & 0.169 & 0.159 & 0.120 & 0.131 & 0.145 & 0.134 & 0.111 & 0.089 & 0.070 & 0.102 & \textbf{0.095} & 0.086 & 0.074 & 0.066 & 0.080 & 0.196 & 0.162 & 0.116 & 0.081 & 0.139 & 0.116 \\
 & Graph-LSTM & 0.167 & 0.154 & 0.116 & 0.123 & 0.140 & 0.131 & 0.106 & \textbf{0.085} & \underline{0.065} & \underline{0.097} & 0.103 & 0.083 & 0.060 & \textbf{0.044} & 0.073 & \underline{0.195} & \underline{0.157} & 0.106 & 0.084 & \underline{0.136} & \underline{0.111} \\
 & \textbf{MMG-PopNet} & \textbf{0.158} & \textbf{0.144} & \textbf{0.112} & \textbf{0.118} & \textbf{0.133} & \underline{0.130} & \textbf{0.100} & \underline{0.087} & \textbf{0.064} & \textbf{0.095} & 0.098 & \underline{0.081} & \textbf{0.057} & \underline{0.045} & \textbf{0.070} & \textbf{0.182} & \textbf{0.146} & \textbf{0.101} & \textbf{0.073} & \textbf{0.126} & \textbf{0.106} \\ 
\midrule

\multirow{6}{*}{\textcolor{googleblue}{\bf \scshape \textbf{Size}} } & 
MLP & \underline{0.701} & \underline{0.652} & \underline{0.492} & 0.486 & \underline{0.583} & \underline{0.918} & 0.711 & 0.626 & 0.519 & 0.694 & \underline{2.018} & \underline{1.399} & 0.993 & 0.801 & 1.303 & 2.747 & 2.094 & 1.066 & 0.637 & 1.637 & 1.054 \\
 & DeepHawkes & 0.900 & 0.813 & 0.587 & 0.529 & 0.707 & 0.982 & 0.738 & 0.635 & 0.478 & 0.709 & 2.137 & 1.472 & 1.314 & 0.939 & 1.466 & 2.903 & 2.556 & 1.904 & 1.159 & 2.130 & 1.253 \\
 & DeepCas & 0.824 & 0.880 & 0.883 & 0.934 & 0.880 & 1.262 & 1.093 & 1.053 & 1.002 & 1.103 & 2.952 & 2.228 & 2.152 & 1.918 & 2.313 & 2.673 & 2.517 & 1.767 & 1.605 & 2.141 & 1.609 \\
 & CasSeqGCN & 0.882 & 0.749 & 0.504 & 0.483 & 0.655 & 0.979 & 0.735 & 0.599 & 0.421 & 0.684 & 2.101 & 1.400 & 1.021 & 0.715 & 1.309 & 2.760 & 2.015 & 1.124 & 0.583 & 1.621 & 1.067 \\
 & Graph-LSTM & 0.868 & 0.732 & 0.493 & \underline{0.466} & 0.640 & 0.945 & \underline{0.700} & \textbf{0.572} & \underline{0.405} & \underline{0.656} & 2.031 & 1.418 & \underline{0.988} & \textbf{0.641} & \underline{1.270} & \underline{2.640} & \underline{1.852} & \underline{1.052} & \underline{0.563} & \underline{1.527} & \underline{1.023} \\
 & \textbf{MMG-PopNet} & \textbf{0.696} & \textbf{0.597} & \textbf{0.435} & \textbf{0.419} & \textbf{0.537} & \textbf{0.863} & \textbf{0.661} & \underline{0.573} & \textbf{0.366} & \textbf{0.616} & \textbf{1.829} & \textbf{1.317} & \textbf{0.842} & \underline{0.653} & \textbf{1.160} & \textbf{2.356} & \textbf{1.743} & \textbf{0.997} & \textbf{0.559} & \textbf{1.414} & \textbf{0.932} \\ 
\midrule

\multirow{6}{*}{\makecell{\textcolor{googlegreen}{\bf \scshape Unique}\\ \textcolor{googlegreen}{\bf \scshape Users}
}} & 
MLP & \underline{0.447} & \underline{0.417} & \underline{0.332} & 0.321 & \underline{0.379} & \underline{0.657} & 0.531 & 0.502 & 0.418 & 0.527 & \underline{1.883} & 1.306 & 0.949 & 0.756 & 1.224 & 2.139 & 1.642 & 0.847 & 0.512 & 1.285 & 0.854 \\
 & DeepHawkes & 0.707 & 0.635 & 0.410 & 0.359 & 0.528 & 0.707 & 0.559 & 0.505 & 0.370 & 0.535 & 2.009 & 1.392 & 1.257 & 0.917 & 1.394 & 2.228 & 1.996 & 1.439 & 0.896 & 1.639 & 1.024 \\
 & DeepCas & 0.633 & 0.694 & 0.711 & 0.723 & 0.690 & 0.869 & 0.767 & 0.771 & 0.716 & 0.781 & 2.629 & 2.019 & 2.002 & 1.796 & 2.112 & \underline{2.073} & 1.976 & 1.381 & 1.269 & 1.675 & 1.314 \\
 & CasSeqGCN & 0.695 & 0.562 & 0.351 & 0.308 & 0.479 & 0.706 & 0.543 & 0.468 & 0.328 & 0.511 & 1.979 & \underline{1.294} & 0.966 & 0.671 & 1.228 & 2.157 & 1.583 & 0.872 & \underline{0.449} & 1.265 & 0.871 \\
 & Graph-LSTM & 0.678 & 0.542 & 0.339 & \underline{0.295} & 0.464 & 0.679 & \underline{0.514} & \underline{0.448} & \underline{0.320} & \underline{0.490} & 1.898 & 1.307 & \underline{0.928} & \textbf{0.599} & \underline{1.183} & 2.074 & \underline{1.441} & \underline{0.816} & 0.456 & \underline{1.197} & \underline{0.833} \\
 & \textbf{MMG-PopNet} & \textbf{0.443} & \textbf{0.374} & \textbf{0.270} & \textbf{0.248} & \textbf{0.334} & \textbf{0.622} & \textbf{0.494} & \textbf{0.447} & \textbf{0.290} & \textbf{0.463} & \textbf{1.693} & \textbf{1.218} & \textbf{0.792} & \underline{0.613} & \textbf{1.079} & \textbf{1.859} & \textbf{1.374} & \textbf{0.785} & \textbf{0.439} & \textbf{1.114} & \textbf{0.748} \\ 
\midrule
\multirow{6}{*}{\textcolor{googlepurple}{\bf \scshape Like Score}} & 
MLP & \underline{1.297} & \underline{1.291} & \underline{1.227} & \underline{1.217} & \underline{1.258} & \underline{1.351} & \underline{1.284} & \underline{1.292} & \underline{1.193} & \underline{1.280} & \underline{5.633} & 5.593 & \underline{4.991} & \underline{4.660} & \underline{5.219} & 6.462 & 6.104 & \underline{4.061} & \underline{3.746} & 5.093 & \underline{3.213} \\
 & DeepHawkes & 2.501 & 2.395 & 1.916 & 1.900 & 2.178 & 1.438 & 1.411 & 1.406 & 1.325 & 1.395 & 6.240 & 6.122 & 5.973 & 6.130 & 6.116 & 6.885 & 6.880 & 6.046 & 5.304 & 6.279 & 3.992 \\
 & DeepCas & 2.350 & 2.496 & 2.497 & 2.498 & 2.460 & 1.444 & 1.428 & 1.483 & 1.435 & 1.448 & 6.359 & 6.099 & 6.048 & 6.268 & 6.194 & \underline{5.725} & 5.785 & 4.576 & 4.905 & 5.248 & 3.837 \\
 & CasSeqGCN & 2.501 & 2.256 & 1.844 & 1.728 & 2.082 & 1.436 & 1.406 & 1.400 & 1.307 & 1.387 & 6.206 & 6.127 & 6.043 & 6.004 & 6.095 & 6.778 & 6.604 & 5.428 & 5.000 & 5.953 & 3.879 \\
 & Graph-LSTM & 2.442 & 2.185 & 1.771 & 1.616 & 2.004 & 1.360 & 1.311 & 1.337 & 1.237 & 1.311 & 5.707 & \underline{5.576} & 5.302 & 4.866 & 5.363 & 6.239 & \underline{5.395} & 4.343 & 4.393 & \underline{5.092} & 3.443 \\
 & \textbf{MMG-PopNet} & \textbf{1.256} & \textbf{1.117} & \textbf{1.003} & \textbf{0.949} & \textbf{1.081} & \textbf{1.311} & \textbf{1.212} & \textbf{1.166} & \textbf{1.028} & \textbf{1.179} & \textbf{5.067} & \textbf{4.654} & \textbf{4.307} & \textbf{4.073} & \textbf{4.525} & \textbf{4.999} & \textbf{4.596} & \textbf{3.214} & \textbf{2.750} & \textbf{3.890} & \textbf{2.669} \\ 
\bottomrule
\end{tabular}
}
\label{tab:MSE_performance}
\vspace{-1.5em}
\end{table}

\begin{figure}[t]
    \centering
\includegraphics[width=1\linewidth]{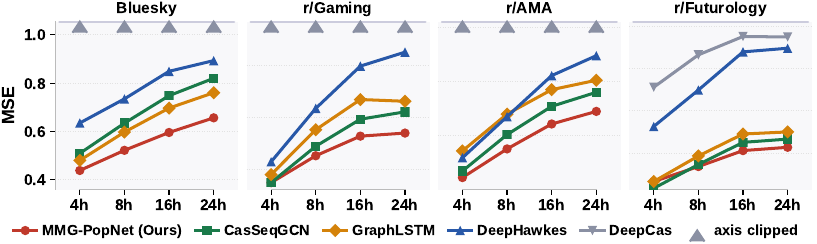}
    \vspace{-1.5em}
    \caption{MSE-Loss trajectories across datasets, comparing different models for target \textsc{Size}. \textbf{Lower is better}. 
    MMG-PopNet achieves the lowest MSE, with the strongest gains at later horizons.}
    \label{fig:future_fig}
    \vspace{-5ex}
\end{figure}

\textbf{Popularity Prediction of Cascade States at Future Horizons.} Beyond prediction at the terminal state, we evaluate social media popularity prediction at intermediate future horizons of cascades. 
Given an early observed prefix $G^t$ (as described in previous section), each method predicts future cascade outcomes at later times $t' > t$, using targets computed from the cascade state at horizons $\{4\text{h}, 8\text{h}, 16\text{h}, 24\text{h}\}$.
This setting tests whether a model can forecast not only terminal popularity, but also the trajectory of cascade growth over time. 
Figure~\ref{fig:future_fig} reports MSE trajectories for the \textsc{Size} target across datasets. Prediction error generally increases with the forecasting horizon, reflecting the greater uncertainty of longer-range cascade growth.
Despite this increased difficulty, MMG-PopNet consistently achieves the lowest error across datasets and horizons. Graph-LSTM and CasSeqGCN are the closest baselines, with comparable performance at 4h and 8h, but they fall behind at 16h and 24h as forecasting uncertainty increases. In contrast, DeepCas has substantially higher error in several cases, with clipped values indicating that its trajectory predictions fall outside the plotted range.
Similar trends are observed for other popularity targets in Appendix~\ref{sec:appendix_future}.

\vspace{-1.25em}
\subsection{$\mathbf{Q}_2$: Unified Training across Communities and Platforms}\label{sec:unified_train}
\begin{figure}[t!]
    \centering
\includegraphics[width=1\linewidth]{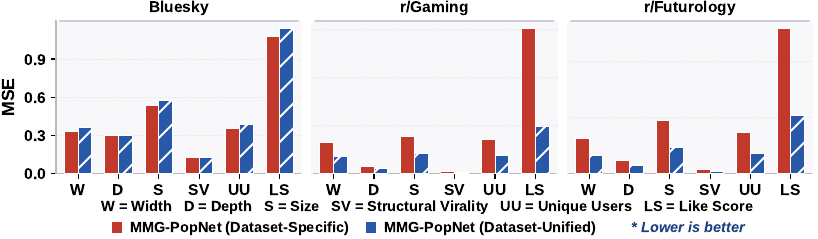}
\vspace{-1.2em}
\caption{\textbf{Dataset-Specific vs. Unified Training.}
Avg MSE of MMG-PopNet under dataset-specific and unified training. Lower is better. Unified training greatly improves performance on Reddit communities and remains competitive on Bluesky.
}
    \label{fig:foundational_compare}
    \vspace{-1em}
\end{figure}

\vspace{-0.75em}
We investigate whether popularity prediction benefits from unified training across datasets from multiple platforms. Instead of training isolated MMG-PopNet models per dataset and observation window, we train a single model on the combined cascades from all datasets.
This evaluates whether joint supervision over heterogeneous cascades improves generalization compared to dataset-specific training.
Figure~\ref{fig:foundational_compare} demonstrates that unified training yields substantial performance gains on Reddit while maintaining comparable accuracy on Bluesky.
On Reddit, the unified model reduces average MSE across all popularity targets, achieving dramatic error reductions for \textsc{Like Score}, \textsc{Size}, and \textsc{Unique Users}.
Conversely, dataset-specific training retains a marginal edge on Bluesky.
This pattern suggests that the unified model benefits most when cross-community training shares a platform-level interaction structure, while Bluesky introduces distinct dynamics less represented in the combined training distribution.
Detailed results and analysis can be found in Appendix~\ref{sec:appendix_foundational}.

\begin{figure}[t!]
    \vspace{0ex}
    \centering
    \includegraphics[width=1\linewidth]{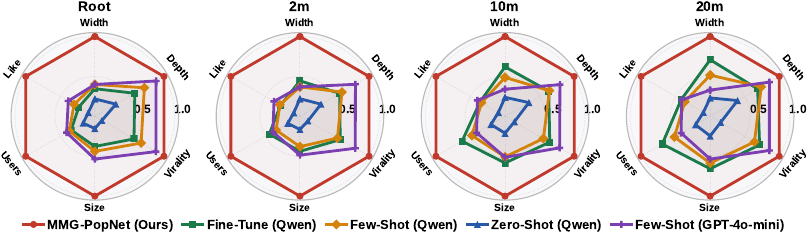}
\caption{\textbf{Normalized LLM Performance on Bluesky.}
Scores are normalized with MMG-PopNet as the reference baseline, fixed at $1.0$ on all axes, where smaller areas indicate worse performance. MMG-PopNet outperforms LLM baselines across all settings. Among LLMs, retrieval-augmented few-shot prompting performs better in sparse early windows, while fine-tuning becomes stronger as longer cascade prefixes provide richer temporal and structural training signals.
}
    \label{fig:llm_fig}
    \vspace{-1.6em}
\end{figure}

\vspace{-1em}
\subsection{$\mathbf{Q}_3$: Comparison with LLM-Based Approaches}
\vspace{-0.75em}
We compare MMG-PopNet with multimodal LLM models under three settings: zero-shot prompting, retrieval-augmented few-shot prompting, and supervised fine-tuning.
Zero-shot setting serialized the observed cascade prefix as a structured JSON input prompt for LLM-based prediction.
The retrieval-augmented few-shot setting includes four training examples with similar root posts.
Fine-tuning setting trains LLM with early-observation cascade inputs.
Figure~\ref{fig:llm_fig} shows the normalized comparison on Bluesky. Scores are computed as $\mathrm{MSE}_{\mathrm{model}} / \mathrm{MSE}_{\mathrm{MMG-PopNet}}$
and smaller polygons indicate worse performance. 
MMG-PopNet uniformly outperforms all LLM baselines across all targets and observation windows. Zero-shot prompting yields the highest error, proving that direct prompting lacks the numerical calibration of LLMs for popularity prediction.
Few-shot prompting rivals or exceeds fine-tuning given root-only or early windows, where historical examples provide crucial context for sparse cascades.
Fine-tuning overtakes prompting as the window expands, likely because it better exploits the richer reply structure, temporal progression, and participation signals in longer observation windows.
Overall, MMG-PopNet performs better likely because it models topology, timing features, and multimodal context for popularity prediction more directly, rather than relying on prompt-driven inference over serialized cascade inputs. Detailed setting in Appendix~\ref{sec-llm-app}.


\vspace{-1.2em}
\subsection{$\mathbf{Q}_4$: Single-Task vs. Multi-Task Training}\label{sec:single_vs_multi}
\begin{figure}[t!]
\centering
\begin{minipage}[t]{0.58\linewidth}
\captionof{table}{Comparison of multi-task versus single-task MSE results across datasets using the following windows: Bluesky @ 10min, r/AMA @ 30min, r/Gaming @ 50min, and r/Futurology @ 90min. \textbf{Best in bold.}}
\label{tab:multitask_compare}
\scriptsize
\raggedright
\setlength{\tabcolsep}{3.25pt}
\renewcommand{\arraystretch}{1}
\begin{tabular}{l cc cc cc cc}
\toprule
\multirow{2}{*}{\textbf{Task}} 
& \multicolumn{2}{c}{\textbf{Bluesky}} 
& \multicolumn{2}{c}{\textbf{r/AMA}}
& \multicolumn{2}{c}{\textbf{r/Gaming}} 
& \multicolumn{2}{c}{\textbf{r/Futurology}} \\
\cmidrule(lr){2-3}
\cmidrule(lr){4-5}
\cmidrule(lr){6-7}
\cmidrule(lr){8-9}
& \textbf{Single} & \textbf{Multi}
& \textbf{Single} & \textbf{Multi}
& \textbf{Single} & \textbf{Multi}
& \textbf{Single} & \textbf{Multi} \\
\midrule
\textcolor{googleblue}{\bf \scshape Max Width} 
& \textbf{0.280} & 0.284 & \textbf{0.411} & 0.429 & \textbf{0.663} & 0.714 & \textbf{0.583} & 0.665 \\
\textcolor{googleblue}{\bf \scshape Max Depth} 
& \textbf{0.272} & 0.273 & \textbf{0.207} & 0.219 & \textbf{0.182} & 0.200 & \textbf{0.272} & 0.282 \\
\makecell[l]{\textcolor{googleblue}{\bf \scshape Structural}\\\textcolor{googleblue}{\bf \scshape Virality}}
& 0.115 & \textbf{0.114} & \textbf{0.082} & 0.087 & \textbf{0.050} & 0.057 & \textbf{0.094} & 0.101 \\
\textcolor{googleblue}{\bf \scshape Size} 
& 0.494 & \textbf{0.470} & 0.578 & \textbf{0.573} & 0.843 & \textbf{0.842} & \textbf{0.987} & 0.997 \\
\textcolor{googlegreen}{\bf \scshape Unique Users} 
& \textbf{0.295} & 0.302 & \textbf{0.431} & 0.447 & \textbf{0.766} & 0.792 & \textbf{0.774} & 0.785 \\
\textcolor{googlepurple}{\bf \scshape Like Score} 
& 1.575 & \textbf{1.026} & 1.284 & \textbf{1.166} & 4.482 & \textbf{4.307} & 3.708 & \textbf{3.214} \\
\bottomrule
\end{tabular}
\end{minipage}
\hfill
\begin{minipage}[t]{0.35\linewidth}
\captionof{figure}{\textbf{Modality Ablation}: Avg. MSE \textbf{increase} per excluded modality relative to full MMG-PopNet.}
\label{fig:modality_ablation}
\vspace{-2pt}
\centering
\includegraphics[width=\linewidth]{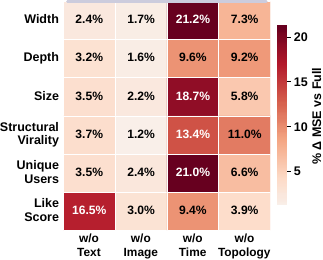}
\end{minipage}
\vspace{-5ex}
\end{figure}

\vspace{-0.6em}
We examine whether multi-objective prediction benefits from jointly modeling complementary popularity targets. Here, we compare the multi-task MMG-PopNet with task-specific variants trained independently for each target. Table~\ref{tab:multitask_compare} shows that the benefit of joint training depends on the target.
Single-task training is stronger for topology-driven objectives. It achieves lower MSE for \textsc{Max Width} and \textsc{Max Depth} across all datasets, indicating that these structural properties benefit from dedicated training. Similar trends appear for \textsc{Structural Virality} and \textsc{Unique Users}, although the gaps are smaller. Multi-task training remains competitive for \textsc{Size}, matching or slightly improving over single-task models on three datasets.
Joint modeling is most beneficial for engagement. Multi-task MMG-PopNet lowers \textsc{Like Score} MSE on every dataset, with large gains on Bluesky and r/Futurology. This suggests that engagement prediction can benefit from shared signals of cascade structure and user participation.
Overall, joint modeling does not improve every target. However, it offers a useful deployment trade-off by remaining competitive on most outcomes while consistently improving \textsc{Like Score} prediction.

\vspace{-1.2em}
\subsection{$\mathbf{Q}_5$: Modality Ablation Analysis}\label{sec:modality_analysis}
\vspace{-0.6em}
We evaluate how each modality contributes by removing one input source from MMG-PopNet at a time and measuring the relative MSE increase over the full model. The ablations remove textual semantics $X^{\text{Text}}$, root visual content $X^{\text{Visual}}$, node temporal features $X^{\text{Time}}$, or reply-tree topology $X^{\text{Graph}}$.Figure~\ref{fig:modality_ablation} reports results on r/Gaming and r/Futurology, where all four modalities are available.
All removals increase error, showing that each modality adds useful information.
Temporal features have the largest effect on cascade growth and participation. They encode each node's time since its parent reply and since the root post.
Removing these features sharply hurts \textsc{Max Width} (21.2\%), \textsc{Unique Users} (21.0\%), and \textsc{Size} (18.7\%). These features capture the pace of early discussion. Fast replies signal bursty growth relevant to width and size, while slower temporal medians can indicate longer-lived discussions with more distinct users.
Text is most important for engagement. Removing textual semantics increases \textsc{Like Score} error by 16.5\%, while only mildly affecting structural targets. This suggests that audience approval depends strongly on what is said, not only how the cascade grows. Reply-tree topology mainly supports structural prediction, especially \textsc{Structural Virality} and \textsc{Max Depth}. Root visual content has the smallest effect, but its consistent gains indicate a modest complementary role. Detailed results in Appendix~\ref{sec:appendix_modality}.

\vspace{-1.2em}
\subsection{$\mathbf{Q}_6$: Graph Classification Tasks}
\vspace{-0.6em}
To broaden MMG-Pop beyond regression, we introduce two graph-classification tasks that predict binary outcomes for observed cascade graphs, enabling threshold-based decisions about future diffusion and engagement rather than exact-value prediction. \textbf{(1) Future-Shape Classification:} Following~\cite{10.1145/2566486.2567997}, models predict whether a cascade's final Structural Virality exceeds the training-set median, assessing whether early interaction patterns can predict its eventual diffusion structure. \textbf{(2) Future-Popularity Classification:} Following~\cite{info11090453}, models predict whether the root post's final Like Score exceeds the 75th percentile of the training distribution, evaluating the ability to identify high-engagement posts from early observations. We evaluate both tasks across all four datasets using a representative observation window per dataset, with two additional windows for r/Futurology to examine the effect of observation duration. Table~\ref{tab:rebuttal_f1_auc} reports Macro-F1 and ROC-AUC, where MMG-PopNet achieves the highest average performance on both tasks and consistently outperforms the baselines in ROC-AUC across all settings. Performance also generally improves with longer observation windows, highlighting the value of additional early interaction signals. Detailed experimental details and analysis are provided in Appendix~\ref{sec:classification}. 

\begin{table*}[t]
\centering
\scriptsize
\caption{Shape and popularity classification performance across observation windows. Higher is better. \textbf{Bold} = best and \underline{underlined} = second-best. \textbf{Ours}: MMG-PopNet, \textbf{G-LSTM}: Graph-LSTM}
\vspace{-1em}
\label{tab:rebuttal_f1_auc}
\setlength{\tabcolsep}{3.5pt}
\renewcommand{\arraystretch}{1.22}
\resizebox{\textwidth}{!}{%
\begin{tabular}{ll | ccc ccc | ccc ccc}
\toprule
\multirow{3}{*}{\textbf{Dataset}} &
\multirow{3}{*}{\textbf{Window}} &
\multicolumn{6}{c|}{\textbf{Shape}} &
\multicolumn{6}{c}{\textbf{Popularity}} \\
\cmidrule(lr){3-8}\cmidrule(l){9-14}
& & \multicolumn{3}{c}{\textbf{F1}} & \multicolumn{3}{c|}{\textbf{AUC}}
  & \multicolumn{3}{c}{\textbf{F1}} & \multicolumn{3}{c}{\textbf{AUC}} \\
\cmidrule(lr){3-5}\cmidrule(lr){6-8}\cmidrule(lr){9-11}\cmidrule(l){12-14}
& & \textbf{MLP} & \textbf{G-LSTM} & \textbf{Ours}
  & \textbf{MLP} & \textbf{G-LSTM} & \textbf{Ours}
  & \textbf{MLP} & \textbf{G-LSTM} & \textbf{Ours}
  & \textbf{MLP} & \textbf{G-LSTM} & \textbf{Ours} \\
\midrule
\textbf{Bluesky} & \textbf{10} & 0.347 & \underline{0.616} & \textbf{0.635} & 0.704 & \underline{0.738} & \textbf{0.739} & 0.603 & \underline{0.701} & \textbf{0.711} & 0.663 & \underline{0.804} & \textbf{0.808} \\
\textbf{r/AMA} & \textbf{30} & 0.708 & \underline{0.730} & \textbf{0.744} & 0.798 & \underline{0.817} & \textbf{0.820} & 0.600 & \underline{0.639} & \textbf{0.643} & 0.694 & \underline{0.739} & \textbf{0.752} \\
\textbf{r/Gaming} & \textbf{50} & 0.712 & \textbf{0.736} & \underline{0.735} & 0.810 & \underline{0.831} & \textbf{0.836} & 0.615 & \underline{0.677} & \textbf{0.719} & 0.743 & \underline{0.823} & \textbf{0.829} \\
\textbf{r/Futurology} & \textbf{30} & 0.552 & \textbf{0.688} & \underline{0.682} & 0.755 & \underline{0.789} & \textbf{0.805} & 0.584 & \underline{0.691} & \textbf{0.703} & 0.696 & \underline{0.776} & \textbf{0.804} \\
 & \textbf{90} & \underline{0.765} & 0.748 & \textbf{0.784} & 0.849 & \underline{0.867} & \textbf{0.868} & \underline{0.743} & 0.738 & \textbf{0.763} & 0.831 & \underline{0.842} & \textbf{0.868} \\
 & \textbf{180} & \underline{0.802} & \textbf{0.816} & \textbf{0.816} & 0.881 & \underline{0.885} & \textbf{0.889} & 0.774 & \underline{0.792} & \textbf{0.821} & 0.876 & \underline{0.889} & \textbf{0.908} \\
\midrule
\textbf{Avg} &  & 0.648 & \underline{0.722} & \textbf{0.733} & 0.800 & \underline{0.821} & \textbf{0.826} & 0.653 & \underline{0.706} & \textbf{0.727} & 0.751 & \underline{0.812} & \textbf{0.828} \\
\bottomrule
\end{tabular}%
}
\vspace{-2em}
\end{table*}

\vspace{-3ex}
\section{Related Work}\label{sec-relatedwork}
\vspace{-2ex}
\textbf{Social Dynamics Modeling.}
Social dynamics modeling studies how local interactions among individuals give rise to collective outcomes such as consensus, segregation, polarization, and information diffusion~\cite{castellano2009statistical,flache2017models}. Classical models explain these phenomena through simple but expressive mechanisms, where~\cite{schelling1971dynamic} showed how individual preferences can produce macro-level segregation, while threshold and cascade models describe how behaviors spread once social reinforcement exceeds adoption barriers~\cite{granovetter1978threshold,watts2002simple}. Opinion dynamics and social-influence models further categorize how network structure, homophily, and repeated exposure shape agreement, diversity, and polarization~\cite{degroot1974reaching,friedkin1990social,flache2017models}. With online platforms, this perspective has expanded to large-scale information diffusion, misinformation spread, and intervention analysis, where temporal interactions and network topology jointly determine collective trajectories~\cite{guille2013information,bak2022combining,bail2018exposure}. Recent agent-based and LLM-driven simulations enrich this line by modeling adaptive, language-mediated agents to simulate broad social dynamics~\cite{yang2024oasis, guo2024large}.

\vspace{-1ex}
\textbf{Social Media Popularity Prediction.} Social media popularity prediction models and forecasts social dynamics by predicting the future influence, engagement, or diffusion of online content~\cite{szabo2010predicting,zhou2021survey}. It has broad applications in trend forecasting, advertisement targeting, and public opinion analysis. Existing studies can be categorized based on social signals and modeling paradigms. From the signal perspective, prior work includes content-based methods leveraging textual or visual information~\cite{szabo2010predicting,bandari2012pulse,tsur2012s,khosla2014makes,gelli2015image,ding2019social}, structure-based methods modeling diffusion topology and user interactions~\cite{zhao2015seismic,medvedev2019modelling,cao2017deephawkes,li2017deepcas,wang2022casseqgcn}, and hybrid approaches combining both~\cite{aragon2017generative,zayats2018conversation,zhang2018conversations,zubiaga2016analysing}. From the modeling perspective, earlier studies mainly relied on handcrafted features with classical machine learning~\cite{zhao2015seismic, medvedev2019modelling, lakkaraju2013s}, followed by sequential and geometric deep learning methods, including LSTMs and GNNs, to capture temporal and structural dynamics~\cite{cao2017deephawkes, li2017deepcas, wang2022casseqgcn, liu2021content, zayats2018conversation, zhang2018user, xie2021micro, zhong2024predicting, zhang2022multi}. More recently, agentic social simulation approaches~\cite{liu2025popsim,yang2024oasis, zheng2025autocas,xu2025forecasting} have been explored to model contextual user behaviors and social interactions. However, they remain fragmented across modalities, platforms, prediction targets, and evaluation protocols, motivating the need for a unified benchmark.

\vspace{-3ex}
\section{Conclusion and Future Work}\label{sec-conclusion}
\vspace{-3ex}
In this paper, we introduced MMG-Pop, a unified benchmark for multi-modal social media popularity prediction, and MMG-PopNet, a unified model that captures content, temporal dynamics, and reply structure to forecast multiple forms of popularity. Our experiments show that MMG-Pop enables systematic evaluation across datasets, communities, observation windows, prediction horizons, and engagement targets. Results show that MMG-PopNet improves prediction by jointly modeling multimodal content and cascade structure, with different modalities offering complementary signals. In addition, cross-community training improves generalization, while multi-task training captures shared engagement patterns. In contrast, LLMs remain limited in predicting social popularity, suggesting that language understanding alone is insufficient for modeling social dynamics. Graph-classification tasks further demonstrate MMG-Pop's broader applicability beyond regression, extending its evaluation scope to different prediction tasks. Together, these findings establish MMG-Pop as a useful benchmark and MMG-PopNet as an effective model for integrating the signals that shape social media popularity. Furthermore, we have conducted real-world case studies with MMG-PopNet in Appendix~\ref{sec-casestudy}. Future work can explore agentic social simulation to predict popularity. Limitations of this work and additional discussion are in Appendix~\ref{sec:limit}.

\section{Acknowledgment}
\vspace{-1ex}
This research is supported by the National Science Foundation (NSF) under grant numbers IIS 2524379, NAIRR 260016, and Adobe Research. We thank Amelie Wei for helpful feedback on the presentation and interpretation of social media popularity metrics.

\bibliographystyle{unsrt}
\bibliography{reference}

\clearpage
\appendix

\setcounter{tocdepth}{3}
\setcounter{parttocdepth}{3}

\renewcommand \thepart{}
\renewcommand \partname{}

\part{Appendix}
\parttoc
\clearpage

\section{Dataset Details}\label{sec:appendix_dataset}
This section provides additional details on the construction and curation of the MMG-Pop datasets, complementing the overview and dataset statistics presented in Section~\ref{sec:dataset_curation}.

\subsection{Cascade Construction.}
For Bluesky, the metadata supports multiple interaction networks, including reply, repost, and quote networks. We use the reply network because replies capture explicit conversational interactions and provide richer content signals than diffusion-oriented actions such as reposts. Posts are first grouped by their discussion thread identifier, and parent references are then used to add reply edges within each thread. For Reddit, which organizes discussions into topic-specific communities called subreddits (denoted by the ``r/'' prefix, e.g., r/Gaming), each submission defines a thread, and each comment provides a parent identifier indicating whether it replies to the root submission or another comment.

\subsection{Node attributes and thread-level context.}
For both Bluesky and Reddit, each post is associated with textual content and timestamp information, which instantiate \(X_v^{\text{Text}}\) and \(X_v^{\text{Time}}\), respectively. Platform-specific engagement signals are also available at the post level. Bluesky provides a like count for each post, whereas Reddit provides a karma score, computed as upvotes minus downvotes. These engagement signals are available for individual posts, but our Like Score popularity target is defined only on the root post \(v_{\text{root}}\) and is evaluated using its final engagement count. For Bluesky, this translates to the number of likes that the social cascade initiating post receives, while on Reddit, it means the Karma score, which is defined as upvotes minus downvotes for the initial submission post. Thread-level context \(X^{\text{Thread}}\) is derived from metadata associated with the root post, including its timestamp. For Bluesky, this additionally includes the follower count of the root post's author, which provides a proxy for the initiating user's social influence.

\subsection{Modalities.}
The r/Gaming and r/Futurology subsets contain root-post images, whereas Bluesky and r/AMA are text-based in our benchmark. Visual content is not uniformly available across cascades, as summarized in Table~\ref{tab:dataset}. Since images are available only at the root-post level in the multimodal Reddit subsets, their visual representations are incorporated into thread-level context rather than individual reply-node attributes.
 
\subsection{Missing and removed content.}
We discard posts whose parent post is missing. If the root post of a cascade is missing but associated replies are present, we discard the entire cascade, since \(v_{\text{root}}\) is required to define the discussion tree. For Reddit, some posts or comments may no longer contain their original textual content because they were deleted by users or removed by moderators at the time of collection and are instead represented by markers such as \texttt{[deleted]} or \texttt{[removed]}. We retain such cascades when the remaining thread structure and metadata are intact, since these posts still correspond to observed participation in the cascade. Although the original text is no longer accessible, the presence of deletion or removal markers may still provide information to the model about moderation or deletion patterns in the dataset communities. Dataset sizes are computed after these filtering steps.

\subsection{Dataset Sampling and Imbalance Mitigation.}
In the raw Bluesky dataset, cascade sizes are heavily skewed, with small conversation trees containing 3 to 10 posts comprising approximately 87\% of the filtered data. Training directly on this distribution would bias the model toward shallow dynamics and obscure structural patterns present in more complex cascades. To address this imbalance, we employ square-root sampling \cite{mahajan2018exploring}. Specifically, each size bin is sampled proportionally to the square root of its empirical frequency, yielding a more balanced subset of roughly 64,000 Bluesky discussion trees for robust model training and evaluation.

\FloatBarrier
\section{Popularity Metrics Relationship Analysis}\label{sec:popularity_correlation}

The six popularity metrics are grouped into the three conceptual dimensions of \textcolor{googleblue}{\bf \scshape structural}, \textcolor{googlegreen}{\bf \scshape participation}, and \textcolor{googlepurple}{\bf \scshape engagement} tasks rather than statistically independent categories.

\textbf{Structural} popularity captures the scale and shape of diffusion. \textit{Size} measures the total volume of activity; \textit{Max Width} captures broad, broadcast-like expansion; \textit{Max Depth} records the longest reply chain; and \textit{Structural Virality} summarizes the overall extent of multi-hop diffusion through average pairwise distances~\cite{goel2016structural}. These metrics are related but not interchangeable. Cascades of the same size may have very different widths and depths, while cascades with the same maximum depth may differ in structural virality depending on whether deep propagation is concentrated in a single branch or distributed across the cascade.

\textbf{Participation} is measured by \textit{Unique Users}, which captures how widely activity is distributed across individuals. This differs from Size because the same number of posts may be generated by many distinct participants or through repeated contributions from a smaller group.

\textbf{Engagement} is measured by the root post’s \textit{Like Score}. It captures direct audience reactions to the original content, including reactions from users who do not participate in the visible reply cascade. It therefore represents a form of popularity that is not fully reflected by cascade size.

Thus, these metrics characterize different properties of popularity, even when their values are correlated. To quantify their relationships, we compute pairwise Spearman correlations among the ground-truth targets for each dataset, as reported in Table~\ref{tab:popularity_correlation}.

The pairwise Spearman correlations show that some targets are related through shared cascade-growth processes, but these relationships are neither consistently strong nor stable across platforms. For example, the Max Width–Max Depth correlation ranges from 0.299 on Bluesky to 0.749 on r/Futurology. Like Score is less correlated with structural metrics, ranging from 0.106 to 0.420 with Max Depth and from 0.277 to 0.501 with Size, which suggests that it captures a distinct engagement signal.
Overall, the results show that the six metrics are related but not interchangeable. Their correlations vary substantially across datasets, and each captures a distinct or complementary aspect of popularity. We therefore retain all six as a standardized core metric suite.


\newcommand{\structmetric}[1]{%
  \textcolor{googleblue}{\bfseries\scshape #1}}
\newcommand{\partmetric}[1]{%
  \textcolor{googlegreen}{\bfseries\scshape #1}}
\newcommand{\engmetric}[1]{%
  \textcolor{googlepurple}{\bfseries\scshape #1}}

\begin{table}[t!]
\centering
\caption{Pairwise Spearman correlations among the six
popularity targets across the four benchmark datasets.
Metric colors indicate \textcolor{googleblue}{\bf \scshape structural}, \textcolor{googlegreen}{\bf \scshape participation}, and \textcolor{googlepurple}{\bf \scshape engagement} metrics.}
\label{tab:popularity_correlation}
\small
\setlength{\tabcolsep}{9pt}
\begin{tabular}{lcccc}
\toprule
\textbf{Target Pair} &
\textbf{r/Gaming} &
\textbf{r/AMA} &
\textbf{Bluesky} &
\textbf{r/Futurology} \\
\midrule
\structmetric{Max Width}--\structmetric{Max Depth}
& .686 & .515 & .299 & .749 \\

\structmetric{Max Width}--\structmetric{Structural Virality}
& .637 & .604 & .618 & .781 \\

\structmetric{Max Width}--\structmetric{Size}
& .966 & .896 & .790 & .946 \\

\structmetric{Max Width}--\partmetric{Unique Users}
& .986 & .970 & .932 & .959 \\

\structmetric{Max Width}--\engmetric{Like Score}
& .374 & .283 & .660 & .506 \\

\structmetric{Max Depth}--\structmetric{Structural Virality}
& .944 & .955 & .897 & .974 \\

\structmetric{Max Depth}--\structmetric{Size}
& .822 & .797 & .776 & .903 \\

\structmetric{Max Depth}--\partmetric{Unique Users}
& .740 & .534 & .335 & .833 \\

\structmetric{Max Depth}--\engmetric{Like Score}
& .305 & .167 & .106 & .420 \\

\structmetric{Structural Virality}--\structmetric{Size}
& .795 & .873 & .946 & .930 \\

\structmetric{Structural Virality}--\partmetric{Unique Users}
& .701 & .620 & .617 & .865 \\

\structmetric{Structural Virality}--\engmetric{Like Score}
& .284 & .178 & .311 & .446 \\

\structmetric{Size}--\partmetric{Unique Users}
& .981 & .893 & .773 & .974 \\

\structmetric{Size}--\engmetric{Like Score}
& .382 & .277 & .464 & .501 \\

\partmetric{Unique Users}--\engmetric{Like Score}
& .401 & .278 & .693 & .529 \\
\bottomrule
\end{tabular}
\end{table}

\FloatBarrier
\section{MMG-PopNet Additional Details}\label{sec:appendix_mmg_popnet}

This section provides the detailed architectural formulations of MMG-PopNet, complementing the overview presented in Section~3 and Figure~\ref{fig:model}. We elaborate on the three key components: multimodal feature embedding, bidirectional graph message passing, and multimodal feature fusion with multi-task prediction.

\textbf{Multi-Modal Feature Embedding.}
To encode heterogeneous cascade attributes, we design modality-specific encoders tailored to the semantics of each signal. 
For each node \(v \in \mathcal{V}^{t}\), we encode the available node-level textual and temporal attributes. 
For textual content \(X_v^{\text{Text}}\), we use Transformer to obtain \(\mathbf{z}_v^{\text{Text}, t} = F_{\boldsymbol{\Theta}_{\text{Enc}}^{\text{Text}}}^{\text{Text}}(X_v^{\text{Text},t})\).
For temporal information, we encode relative timing signals via \(\Delta_v^{\text{Time}} = [\Delta t_{v \leftarrow \text{parent}}, \Delta t_{v \leftarrow \text{root}}]\), where \(\Delta t_{v \leftarrow \text{parent}}\) and \(\Delta t_{v \leftarrow \text{root}}\) denote the elapsed time since the parent node and the root post, reflecting the immediacy of user engagement and overall temporal stage. These timing signals are log-transformed and z-score normalized to obtain fixed temporal features \(\mathbf{z}_v^{\text{Time,t}}\).
We also encode thread-level context as \(X^{\text{Thread},t} = (X_{\text{root}}^{\text{Visual},t}, X_{\text{meta}}^{\text{Thread},t})\), denoting root post visual content and non-visual thread context related to the root post. 
The root image is encoded with a CLIP~\cite{radford2021learning}, \(\mathbf{z}_{\text{root}}^{\text{Visual,t}} = F_{\boldsymbol{\Theta}_{\text{Enc}}^{\text{Visual}}}^{\text{Visual}}(X_{\text{root}}^{\text{Visual},t})\). Finally, non-visual thread context is modeled with initial user's influence and posting time, i.e., \(X_{\text{meta}}^{\text{Thread},t} = [\mathrm{Followers}(v_{\text{root}}), \phi(t_{\text{post}})]\), where the follower count is log-transformed and standardized, and \(\phi(\cdot)\) is a cyclic encoding of time-of-day as content posting time during a day also influences interactions. The resulting root visual \(\mathbf{z}_{\text{root}}^{\text{Visual,t}}\) and non-visual \(X_{\text{meta}}^{\text{Thread},t}\) representations together concatenate into the thread-level contextual representation \(\mathbf{z}^{\text{Thread,t}}\).

\textbf{Bidirectional Graph Message-Passing.}
While node-level textual and temporal encodings capture rich local signals at each node, modeling the structural context of the cascade is essential to understand how information propagates and evolves over time. 
In particular, early-stage cascade dynamics, such as branching patterns and response depth, provide strong indicators of future popularity. 
To encode such structural dependencies, we perform graph message-passing over the cascade. We first initialize each node \(v \in \mathcal{V}^{t}\) representation by concatenating semantic and temporal features: 
\(\mathbf{h}_v^{(0)} = \mathrm{Concat}\bigl(\mathbf{z}_v^{\text{Text}}, \mathbf{z}_v^{\text{Time}}\bigr)\), capturing both content and temporal context. 
We then apply bidirectional message-passing to simultaneously model root-to-leaf and leaf-to-root propagation. The node embeddings are initialized in both directions \(\mathbf{h}_{v,\text{down}}^{(0)} = \mathbf{h}_{v,\text{up}}^{(0)} = \mathbf{h}_v^{(0)}\) are updated at layer \(\ell\):
\vspace{-1ex}
\begin{equation}
\scriptsize
\mathbf{h}_{v,\text{down}}^{(\ell)}
=
\mathrm{SAGE}_{\text{down}}^{(\ell)}
(
\mathbf{h}_{v,\text{down}}^{(\ell-1)},
\{
\mathbf{h}_{u,\text{down}}^{(\ell-1)} : u \in \mathcal{N}_{\text{down}}(v)
\}
), 
\mathbf{h}_{v,\text{up}}^{(\ell)}
=
\mathrm{SAGE}_{\text{up}}^{(\ell)}
(
\mathbf{h}_{v,\text{up}}^{(\ell-1)},
\{
\mathbf{h}_{u,\text{up}}^{(\ell-1)} : u \in \mathcal{N}_{\text{up}}(v)
\}),
\end{equation}
where \(\mathcal{N}_{\text{down}}(v)\) and \(\mathcal{N}_{\text{up}}(v)\) denote the parent/child-side neighbors of node \(v\). 
After \(L\) layers, we aggregate node representations via mean pooling to obtain direction-specific summaries $\mathbf{h}_{G, \text{down}}^t, \mathbf{h}_{G, \text{up}}^t$, which are then concatenated to form the final graph representation:
\(\mathbf{h}_{G}^{t} = \mathrm{Concat}(\mathbf{h}_{G, \text{down}}^t, \mathbf{h}_{G, \text{up}}^t)\).

\textbf{Multi-Modal Feature Fusion and Multi-Task Prediction.}
After encoding node/graph/thread-level signals, we fuse them into a unified cascade representation for prediction.
Specifically, we aggregate (1) the raw root node representation $\mathbf{h}^0_{\text{root}}$ before message passing, (2) the final bidirectional root representation, \(\mathbf{h}_{\text{root}}^{(L)} = \mathrm{Concat}(\mathbf{h}_{\text{root},\text{down}}^{(L)}, \mathbf{h}_{\text{root},\text{up}}^{(L)})\), (3) the graph structural representation $\textbf{h}_{G}^t$, and (4) the thread-level contextual representation, into a single vector: $\mathbf{x}_G^t
=
\mathbf{h}_{\text{root}}^{(0)}
\;\Vert\;
\mathbf{h}_{\text{root}}^{(L)}
\;\Vert\;
\mathbf{g}_{G}^{t}
\;\Vert\;
\mathbf{z}^{\text{Thread,t}}.$ We then apply a shared multi-layer perceptron (MLP) to map the fused representation into a multi-task output space:
\(\widehat{\mathbf{Y}}_G^{t'} = \mathrm{MLP}_{\boldsymbol{\Theta}_{\text{Pred}}}(\mathbf{x}_G^t)\), where \(\widehat{\mathbf{Y}}_G \in \mathbb{R}^{K}\) contains log-space predictions for \(K\) targets (e.g., final cascade size, unique users, or structural properties). 
The model is trained using a multi-task objective
\(\mathcal{L} = \frac{1}{K}\sum_{k=1}^{K} \mathcal{L}_k(\widehat{Y}_G^{(k)}, Y_G^{(k)})\), where all parameters (including the text encoder ${\boldsymbol{\Theta}_{\text{Enc}}^{\text{Text}}}$, image encoder ${\boldsymbol{\Theta}_{\text{Enc}}^{\text{Image}}}$, two GNNs, and prediction heads ${\boldsymbol{\Theta}_{\text{Pred}}}$) are jointly optimized end-to-end.

\FloatBarrier
\section{Other Baselines Implementation Details}\label{sec:appendix_baseline}

\subsection{MLP.}
MLP is a structure-agnostic baseline that represents each social cascade as a fixed vector rather than a reply tree. For each observed cascade prefix, the model builds a cascade-level representation by concatenating the initial root post's feature vector, the mean feature vector over all observed posts, and the thread-level context vector. Each post feature vector contains a projected text embedding and two temporal features: time since the root post and time since the parent post. The thread-level context contains thread-level metadata, including posting-time and follower count. 
The model passes the resulting cascade representation through a shared multilayer perceptron where it predicts all 6 popularity targets. All targets are predicted in log-transformed space. 

\textbf{Hyperparameter Settings:} The precomputed text embedding dimension is 384, and the projected text embedding dimension is 32. The temporal features are log-transformed and standardized using training-set statistics. The MLP has 2 layers of hidden dimensions 128, dropout is 0.3, and the output dimension is 6, corresponding to the six popularity targets. Training uses Adam with learning rate $10^{-3}$, batch size 256, a maximum of 200 epochs and with early stopping patience 10.

\subsection{DeepHawkes.}
We implemented a DeepHawkes-style baseline~\cite{cao2017deephawkes} for cascade prediction. DeepHawkes represents an information cascade as a set of diffusion paths and learns path-level representations that capture user influence, self-excitation, and temporal decay in an end-to-end neural architecture. In our setting, each MMG-Pop training instance corresponds to one social cascade, represented by the observed root-to-node diffusion paths within the observation window.

Following the DeepHawkes formulation, each path is encoded as a sequence of user embeddings and summarized with a recurrent encoder. To incorporate temporal effects, we assign each path to one of 10 recency bins according to the time of its final post relative to the end of the observation window. The model learns a positive scalar weight for each recency bin, and the cascade representation is obtained by a weighted aggregation of path representations. This preserves the main DeepHawkes design while adapting it to the MMG-Pop cascade format.

\textbf{Hyperparameter Settings:} We use 50-dimensional user embeddings and a GRU hidden size of 32. The prediction MLP has hidden dimensions 32 and 16 with ReLU activations and dropout rate 0.5. The output dimension is 6, corresponding to the six prediction targets. Models are trained with Adam using batch size 32, up to 200 epochs, early stopping patience 10, and weight decay $10^{-4}$. User embeddings use learning rate $5\times10^{-4}$, while the remaining parameters use learning rate $5\times10^{-3}$.

\subsection{DeepCas.}
We implement DeepCas~\cite{li2017deepcas} as a path-based neural cascade encoder. Each cascade is represented by random walks sampled from the observed early-window conversation tree, using only nodes and edges available within the observation window. Each walk starts at the initial post of the social cascade. During sampling, the walker either moves to a child node or jumps to another observed node and where, both child transitions and jump targets are sampled using degree-based weights. If the current node is a leaf, the walker performs a jump using the same weighting rule.
The sampled user sequences are mapped to a pretrained user-embedding vocabulary. The embeddings are trained separately on a train-split global interaction graph constructed from reply links across all cascades in the training set. In this graph, each directed weighted edge connects a replying user to the author being replied to, and repeated reply interactions are aggregated as edge weights. This graph is used to capture user-user interactions across the dataset, rather than within a single cascade. The pretrained embeddings are then loaded into DeepCas and kept fixed during supervised training. Here, the model is adjusted to predict all 6 popularity targets as a 6 dimension vector.

\textbf{Hyperparameter Settings:} We use $K$ = 200 sampled walks, $T$ = 10 steps per walk, pretrained interaction-graph embeddings with dimension 128, GRU hidden dimension 128 per direction, random-walk group size 5, two MLP hidden layers of size 128 with ReLU and dropout 0.4, and a 6-dimensional output layer. Training uses Adam over trainable parameters only, learning rate $10^{-3}$, weight decay $ 5\times10^{-4}$ batch size 256, maximum 200 epochs, early-stopping patience 10, and gradient clipping with maximum norm 1.0.

\subsection{CasSeqGCN.}
CasSeqGCN~\cite{wang2022casseqgcn} represents each cascade as a temporal sequence of graph snapshots constructed from the observed early-window conversation tree. The original model combines structural and temporal cascade information by first encoding each snapshot with graph convolution, then aggregating node embeddings into a snapshot representation with dynamic routing, and finally processing the snapshot sequence with an LSTM before prediction.
Given this ordered sequence, snapshots are formed by adding posts in fixed increments of \(Q=5\). We keep at most \(K_{\max}=15\) snapshots and force the final snapshot to include the last observed post if it is not already included by the fixed stride. Thus, each cascade is represented as a sequence of up to 15 partial graph snapshots. Here, the model is adjusted to predict all 6 popularity targets as a 6 dimension vector.

\textbf{Hyperparameter Settings:} Node embedding dimension is 32, the snapshot embedding dimension is 32, the GCN hidden dimension is 32, the number of GCN layers is 2, the LSTM hidden dimension is 32, the number of LSTM layers is 2, the number of dynamic routing iterations is 3, and dropout is 0.5. For model selection, we search over learning rates $\{0.005, 0.01,0.03,0.05\}$ separately for each dataset/window. Each candidate is trained for at most 20 epochs with patience 5, and the candidate with the lowest validation loss is selected for full training. Full training uses Adam with the selected learning rate,  weight decay $ 5\times10^{-5}$ batch size 256, at most 200 epochs, early-stopping patience 10, gradient clipping with maximum norm 1.0.

\subsection{GraphLSTM.}
GraphLSTM~\cite{zayats2018conversation} represents each cascade as a conversation tree. Each post is a node, and each reply forms an edge from the parent post to the reply post. For every node, the model builds an input vector by concatenating structural features with a mean-pooled text embedding. The structural features describe the node’s timing and position in the observed tree. The text embedding summarizes the post text using learned token embeddings.
The model applies two graph-LSTM passes over the tree. The forward pass uses information from the parent node and the previous sibling node. This gives each node a representation of the conversation context that came before it. The backward pass uses information from the first child node and the next sibling node. This gives each node a representation of the response context that follows it. The forward and backward states are then concatenated to obtain a context-aware representation for each node. To obtain a cascade-level representation, we mean-pool the concatenated node states over all nodes in the tree. Here, the model is adjusted to predict all 6 popularity targets as a 6 dimension vector.

\textbf{Hyperparameter Settings:} The token embedding dimension is 100, the graph-LSTM hidden dimension is 128, the final MLP hidden dimension is 128, dropout is 0.3, the maximum post length is 100 tokens, the vocabulary minimum frequency is 10. Training uses AdamW with learning rate $10^{-3}$, embedding learning rate $10^{-4}$, weight decay $10^{-5}$, batch size 256, a maximum of 200 epochs, early stopping patience 10, and gradient clipping with maximum norm 1.0.

\subsection{Modality Encoder Analysis}
To ensure meaningful comparisons, we preserved each baseline's original architecture and modality configuration while using consistent datasets, observation windows, data splits, and evaluation metrics. MLP uses the same all-MiniLM-L6-v2 text encoder and CLIP ViT-B/32 image encoder as MMG-PopNet, whereas the original Graph-LSTM uses averaged Bag-of-Words text representations. To assess the effect of encoder differences, we additionally evaluate Graph-LSTM with the same all-MiniLM-L6-v2 text encoder as MMG-PopNet. DeepHawkes, DeepCas, and CasSeqGCN retain their original configurations, as introducing content encoders would substantially alter their architectures.

As shown in Table~\ref{tab:encoder_fairness}, standardizing the text encoder improves Graph-LSTM's average MSE across all six popularity targets. MMG-PopNet nevertheless achieves the lowest overall MSE for every target, although standardized Graph-LSTM performs slightly better on Max Width for r/Futurology. These results indicate that the performance gains of MMG-PopNet cannot be attributed solely to differences in text encoders.

\begin{table*}[t]
\centering
\caption{
Comparison of Graph-LSTM with its original and standardized text encoders against MMG-PopNet. MSE is averaged across observation windows for each dataset, with overall averages reported in the last column. \textbf{Lower is better}. Best values are \textbf{bolded} and second-best values are \underline{underlined}. Graph-LSTM (Standardized) uses the same all-MiniLM-L6-v2 text encoder as MMG-PopNet.
}
\label{tab:encoder_fairness}

\vspace{1ex}
\small
\setlength{\tabcolsep}{5pt}
\renewcommand{\arraystretch}{1.18}

\resizebox{\textwidth}{!}{%
\begin{tabular}{ll|ccccc}
\toprule
\textbf{Task} &
\textbf{Model} &
\textbf{Bluesky} &
\textbf{r/AMA} &
\textbf{r/Gaming} &
\textbf{r/Futurology} &
\textbf{Overall Avg.} \\
\midrule

\multirow{3}{*}{\textcolor{googleblue}{\bf \scshape Max Width}}
& Graph-LSTM (Original)
& 0.452 & 0.483 & 1.064 & 0.969 & 0.742 \\
& Graph-LSTM (Standardized)
& \underline{0.422} & \underline{0.458} & \underline{0.998} & \textbf{0.916} & \underline{0.699} \\
& \textbf{MMG-PopNet}
& \textbf{0.336} & \textbf{0.455} & \textbf{0.982} & \underline{0.938} & \textbf{0.678} \\
\midrule

\multirow{3}{*}{\textcolor{googleblue}{\bf \scshape Max Depth}}
& Graph-LSTM (Original)
& 0.308 & \underline{0.245} & 0.253 & 0.388 & 0.298 \\
& Graph-LSTM (Standardized)
& \underline{0.301} & \textbf{0.240} & \underline{0.247} & \underline{0.366} & \underline{0.288} \\
& \textbf{MMG-PopNet}
& \textbf{0.300} & \textbf{0.240} & \textbf{0.243} & \textbf{0.356} & \textbf{0.285} \\
\midrule

\multirow{3}{*}{\textcolor{googleblue}{\bf \scshape Structural Virality}}
& Graph-LSTM (Original)
& 0.132 & \underline{0.097} & 0.073 & 0.136 & 0.109 \\
& Graph-LSTM (Standardized)
& \underline{0.128} & \textbf{0.095} & \underline{0.072} & \underline{0.129} & \underline{0.106} \\
& \textbf{MMG-PopNet}
& \textbf{0.126} & \textbf{0.095} & \textbf{0.070} & \textbf{0.126} & \textbf{0.104} \\
\midrule

\multirow{3}{*}{\textcolor{googleblue}{\bf \scshape Size}}
& Graph-LSTM (Original)
& 0.642 & 0.656 & 1.270 & 1.527 & 1.024 \\
& Graph-LSTM (Standardized)
& \underline{0.604} & \underline{0.619} & \underline{1.187} & \underline{1.429} & \underline{0.960} \\
& \textbf{MMG-PopNet}
& \textbf{0.540} & \textbf{0.616} & \textbf{1.160} & \textbf{1.414} & \textbf{0.932} \\
\midrule

\multirow{3}{*}{\textcolor{googlegreen}{\bf \scshape Unique Users}}
& Graph-LSTM (Original)
& 0.481 & 0.490 & 1.183 & 1.197 & 0.838 \\
& Graph-LSTM (Standardized)
& \underline{0.447} & \underline{0.466} & \underline{1.080} & \underline{1.118} & \underline{0.778} \\
& \textbf{MMG-PopNet}
& \textbf{0.352} & \textbf{0.463} & \textbf{1.079} & \textbf{1.114} & \textbf{0.752} \\
\midrule

\multirow{3}{*}{\textcolor{googlepurple}{\bf \scshape Like Score}}
& Graph-LSTM (Original)
& 2.013 & 1.311 & 5.363 & 5.092 & 3.445 \\
& Graph-LSTM (Standardized)
& \underline{1.803} & \underline{1.200} & \underline{4.763} & \underline{4.411} & \underline{3.044} \\
& \textbf{MMG-PopNet}
& \textbf{1.081} & \textbf{1.179} & \textbf{4.525} & \textbf{3.890} & \textbf{2.669} \\

\bottomrule
\end{tabular}%
}
\vspace{-0.75em}
\end{table*}

\FloatBarrier
\section{Experimental Setup}\label{sec:appendix_experiment}

\subsection{MMG-PopNet Settings.}
MMG-PopNet consists of three modality-specific components and a prediction head. We use \texttt{all-MiniLM-L6-v2} as the text encoder, \texttt{CLIP ViT-B/32} as the vision encoder, and GraphSAGE~\cite{hamilton2017inductive} as the graph message-passing backbone.

\textbf{Hyperparameter Settings:} 
The text encoder \texttt{all-MiniLM-L6-v2} produces a 384-dimensional representation. This representation is mapped to a 32-dimensional text embedding using a two-layer projection network with hidden dimension 64 and dropout rate 0.15. All layers of \texttt{all-MiniLM-L6-v2} are updated during training.
The vision encoder \texttt{CLIP ViT-B/32} produces a 512-dimensional representation. This representation is mapped to a 64-dimensional image embedding using a linear projection layer with dropout rate 0.3. During training, only the final layer of the CLIP vision encoder is updated.
The graph component uses a 3-layer GraphSAGE network with mean aggregation, hidden dimension 128, and dropout rate 0.3. The fused representation is passed to a two-layer MLP prediction head with hidden dimension 128. The prediction head outputs a 6-dimensional vector, with one output corresponding to each popularity target.
We train MMG-PopNet using Adam with separate learning rates for different parameter groups. The learning rate is $5 \times 10^{-6}$ for the text encoder, $10^{-6}$ for the trainable CLIP vision parameters, $10^{-3}$ for the image projection parameters, and $10^{-3}$ for the GNN parameters. The image projection parameters use weight decay $10^{-2}$.
Training uses variable batch sizes with gradient accumulation to obtain an effective batch size of 256. The maximum number of training epochs is 100, and early stopping is applied with patience 10. We use mixed-precision training and clip gradients to a maximum norm of 0.5. The learning-rate schedule consists of linear warmup followed by cosine decay, with warmup ratio 0.05 and minimum learning-rate factor 0.0.

\subsection{Completed Cascade Exclusion.}
Cascades that complete before the observation window are excluded from that setting to avoid trivial prediction cases. For example, a cascade that ends after 8 minutes is excluded from the 10-minute window because its observed prefix would already equal the final cascade.

\subsection{Training Split and Compute Resources.}\label{appendix:split_compute} Dataset was divided into 80/10/10 splits of train, validation and test respectively. For computation, 4 $\times$ Nvidia L40s GPUs were used to train the MMG-PopNet to allow for handling the Out-of-Memory issue due to large text content associated with long social cascades. To ensure fair comparison, all models had an effective batch size of 256. Other representative baselines were trained on 1 L40s GPU. Certain Bluesky experiments were conducted on NVIDIA A100 GPUs, with all models evaluated on the same GPU type for those experiments to ensure a fair comparison.

\FloatBarrier

\section{Popularity Prediction Across Future Horizons}\label{sec:appendix_future}
\subsection{Popularity prediction of final cascade state under different early observation.}\label{sec:appendix_final_state}
In addition to the MSE results reported in Table~\ref{tab:MSE_performance}, we report $R^2$ and Spearman rank correlation results for the same final cascade state prediction setting. 
These two metrics offer complementary perspectives on model performance, where $R^2$ evaluates the exact predictive fit of the model's estimates against true values, while Spearman correlation assesses the rank-order agreement between predicted and actual outcomes.

The $R^2$ metric (coefficient of determination) measures the proportion of variance in the final cascade states that can be explained by the early observation signals. A higher $R^2$ score indicates that a model's numerical predictions tightly fit the actual popularity distributions. As shown in Table~\ref{tab:R2_performance}, MMG-PopNet achieves the strongest overall performance, obtaining the highest average score for every target metric. The gains are consistent across the Bluesky and Reddit datasets, indicating that the model successfully explains social cascade variance across platforms with diverse dynamics. Furthermore, the advantage is maintained across all early observation windows. In contrast, models relying on narrower temporal or structural signals, such as DeepHawkes and DeepCas, show unstable results with $R^2$ values frequently approaching zero or dropping negative, meaning they fail to capture the variance better than simply predicting the mean.

Conversely, the Spearman rank correlation evaluates how well a model preserves the relative ordering of cascades, independent of absolute numerical errors. This is particularly important for downstream applications where the goal is to identify which threads will become the most viral or structurally complex. Table~\ref{tab:Spearman_performance} demonstrates that MMG-PopNet consistently achieves the highest overall average for every prediction task. While baselines like Graph-LSTM and MLP perform reasonably well at ranking tasks compared to their $R^2$ fit, they still fall short of MMG-PopNet. The strong Spearman results confirm that MMG-PopNet's multimodal design not only minimizes numerical error but reliably orders final cascade outcomes, making it highly effective for trend identification across heterogeneous platforms.

\newpage

\begin{table}[t]
\small
\caption{$R^2$ results for final cascade state prediction under different early observation windows, including \textcolor{googleblue}{\bf \scshape structural tasks} (\textcolor{googleblue}{max width, max depth, structural virality, size}), \textcolor{googlegreen}{\bf \scshape unique users}, and \textcolor{googlepurple}{\bf \scshape like score}. Higher is better. Best values are \textbf{bolded}, and second-best values are \underline{underlined}.}
\vspace{0.5em}
\centering
\setlength{\tabcolsep}{3pt}
\renewcommand{\arraystretch}{1.25}
\resizebox{\textwidth}{!}{%
\begin{tabular}{c l| ccccc ccccc ccccc ccccc |c}
\toprule
\multirow{2}{*}{\textbf{Task}} & 
\multirow{2}{*}{\textbf{Model}} & 
\multicolumn{5}{c}{\textbf{Bluesky}} & 
\multicolumn{5}{c}{\textbf{r/AMA}} & 
\multicolumn{5}{c}{\textbf{r/Gaming}} & 
\multicolumn{5}{c|}{\textbf{r/Futurology}} & 
\multirow{2}{*}{\textbf{Avg}} \\ 
\cmidrule(r){3-7}
\cmidrule(r){8-12}
\cmidrule(r){13-17}
\cmidrule(lr){18-22}

 &  & \textbf{0} & \textbf{2} & \textbf{10} & \textbf{20} & \textbf{Avg} & \textbf{0} & \textbf{15} & \textbf{30} & \textbf{60} & \textbf{Avg} & \textbf{0} & \textbf{20} & \textbf{50} & \textbf{90} & \textbf{Avg} & \textbf{0} & \textbf{30} & \textbf{90} & \textbf{180} & \textbf{Avg} & \\ 
\midrule

\multirow{6}{*}{\textcolor{googleblue}{\bf \scshape Max Width}} & MLP & \underline{0.330} & \underline{0.374} & 0.501 & 0.524 & \underline{0.432} & \underline{0.069} & 0.259 & 0.341 & 0.398 & 0.267 & \underline{0.058} & 0.366 & 0.530 & 0.617 & 0.393 & -0.005 & 0.255 & 0.570 & 0.694 & 0.378 & 0.368 \\
 & DeepHawkes & -0.015 & 0.083 & 0.428 & 0.508 & 0.251 & -0.001 & 0.223 & 0.344 & 0.470 & 0.259 & -0.013 & 0.320 & 0.385 & 0.547 & 0.310 & -0.042 & 0.087 & 0.269 & 0.475 & 0.197 & 0.254 \\
 & DeepCas & 0.074 & 0.001 & -0.001 & 0.001 & 0.019 & -0.180 & -0.062 & -0.025 & -0.030 & -0.074 & -0.267 & -0.017 & 0.018 & 0.087 & -0.045 & \underline{0.035} & 0.087 & 0.265 & 0.262 & 0.162 & 0.016 \\
 & CasSeqGCN & 0.000 & 0.201 & 0.509 & 0.576 & 0.322 & 0.000 & 0.250 & 0.404 & 0.545 & 0.300 & 0.000 & \underline{0.388} & 0.549 & 0.683 & 0.405 & -0.011 & 0.301 & 0.581 & \underline{0.755} & 0.406 & 0.358 \\
 & Graph-LSTM & 0.021 & 0.220 & \underline{0.518} & \underline{0.583} & 0.336 & 0.038 & \underline{0.285} & \underline{0.425} & \underline{0.557} & \underline{0.326} & 0.050 & 0.380 & \underline{0.565} & \textbf{0.720} & \underline{0.429} & 0.028 & \underline{0.361} & \textbf{0.611} & \textbf{0.769} & \underline{0.442} & \underline{0.383} \\
 & \textbf{MMG-PopNet (Ours)} & \textbf{0.336} & \textbf{0.448} & \textbf{0.618} & \textbf{0.651} & \textbf{0.513} & \textbf{0.124} & \textbf{0.312} & \textbf{0.427} & \textbf{0.598} & \textbf{0.365} & \textbf{0.164} & \textbf{0.412} & \textbf{0.618} & \underline{0.696} & \textbf{0.472} & \textbf{0.121} & \textbf{0.371} & \underline{0.590} & 0.751 & \textbf{0.458} & \textbf{0.452} \\
\midrule

\multirow{6}{*}{\textcolor{googleblue}{\bf \scshape Max Depth}} & MLP & \underline{0.018} & 0.070 & 0.224 & 0.307 & 0.155 & \underline{0.031} & 0.186 & 0.299 & 0.386 & 0.226 & \underline{0.008} & \underline{0.194} & 0.387 & 0.444 & 0.258 & -0.017 & 0.170 & \underline{0.434} & \underline{0.537} & 0.281 & 0.230 \\
 & DeepHawkes & -0.009 & 0.023 & 0.087 & 0.130 & 0.058 & -0.003 & 0.146 & 0.214 & 0.256 & 0.153 & 0.002 & 0.163 & 0.228 & 0.315 & 0.177 & -0.034 & 0.027 & 0.159 & 0.319 & 0.118 & 0.126 \\
 & DeepCas & 0.014 & -0.001 & -0.000 & -0.000 & 0.003 & -0.316 & -0.143 & -0.061 & -0.032 & -0.138 & -0.664 & -0.180 & -0.109 & -0.017 & -0.243 & -0.023 & 0.050 & 0.197 & 0.215 & 0.110 & -0.067 \\
 & CasSeqGCN & 0.000 & 0.036 & 0.207 & 0.250 & 0.123 & 0.000 & 0.143 & 0.295 & 0.402 & 0.210 & -0.001 & 0.146 & 0.294 & 0.373 & 0.203 & -0.006 & 0.182 & 0.380 & 0.517 & 0.268 & 0.201 \\
 & Graph-LSTM & 0.006 & \underline{0.079} & \underline{0.234} & \underline{0.316} & \underline{0.159} & 0.015 & \underline{0.194} & \textbf{0.330} & \underline{0.455} & \underline{0.248} & -0.020 & 0.168 & \underline{0.401} & \underline{0.518} & \underline{0.267} & \underline{0.021} & \underline{0.223} & 0.417 & 0.514 & \underline{0.294} & \underline{0.242} \\
 & \textbf{MMG-PopNet (Ours)} & \textbf{0.030} & \textbf{0.108} & \textbf{0.256} & \textbf{0.340} & \textbf{0.184} & \textbf{0.042} & \textbf{0.219} & \underline{0.318} & \textbf{0.466} & \textbf{0.261} & \textbf{0.033} & \textbf{0.205} & \textbf{0.427} & \textbf{0.523} & \textbf{0.297} & \textbf{0.112} & \textbf{0.282} & \textbf{0.447} & \textbf{0.571} & \textbf{0.353} & \textbf{0.274} \\
\midrule

\multirow{6}{*}{\makecell{\textcolor{googleblue}{\bf \scshape Structural}\\ \textcolor{googleblue}{\bf \scshape Virality}}} & MLP & \underline{0.058} & \underline{0.108} & \underline{0.278} & \underline{0.347} & \underline{0.198} & \textbf{0.037} & \underline{0.211} & 0.328 & 0.430 & 0.252 & -0.007 & \textbf{0.184} & \underline{0.370} & 0.468 & \underline{0.254} & -0.070 & 0.160 & \underline{0.430} & \underline{0.539} & 0.265 & \underline{0.242} \\
 & DeepHawkes & -0.008 & 0.026 & 0.141 & 0.187 & 0.087 & -0.007 & 0.170 & 0.236 & 0.283 & 0.170 & \underline{-0.001} & 0.111 & 0.112 & 0.252 & 0.118 & -0.027 & 0.002 & 0.126 & 0.244 & 0.086 & 0.115 \\
 & DeepCas & 0.021 & -0.001 & -0.000 & -0.001 & 0.005 & -0.632 & -0.257 & -0.099 & -0.078 & -0.267 & -1.780 & -0.613 & -0.490 & -0.260 & -0.786 & -0.091 & -0.025 & 0.079 & 0.139 & 0.026 & -0.256 \\
 & CasSeqGCN & 0.000 & 0.056 & 0.252 & 0.297 & 0.151 & -0.001 & 0.170 & 0.329 & 0.461 & 0.240 & \textbf{0.000} & 0.098 & 0.211 & 0.288 & 0.149 & -0.004 & 0.171 & 0.350 & 0.515 & 0.258 & 0.200 \\
 & Graph-LSTM & 0.008 & 0.086 & 0.274 & 0.343 & 0.178 & 0.016 & 0.209 & \textbf{0.359} & \underline{0.505} & \underline{0.272} & -0.079 & 0.127 & 0.361 & \textbf{0.521} & 0.232 & \underline{0.001} & \underline{0.195} & 0.407 & 0.498 & \underline{0.275} & 0.239 \\
 & \textbf{MMG-PopNet (Ours)} & \textbf{0.065} & \textbf{0.145} & \textbf{0.298} & \textbf{0.368} & \textbf{0.219} & \underline{0.023} & \textbf{0.251} & \underline{0.343} & \textbf{0.508} & \textbf{0.281} & -0.025 & \underline{0.151} & \textbf{0.389} & \underline{0.513} & \textbf{0.257} & \textbf{0.067} & \textbf{0.251} & \textbf{0.437} & \textbf{0.564} & \textbf{0.330} & \textbf{0.272} \\
\midrule

\multirow{6}{*}{\textcolor{googleblue}{\bf \scshape \textbf{Size}}} & MLP & \underline{0.204} & \underline{0.261} & \underline{0.443} & 0.480 & \underline{0.347} & \underline{0.063} & 0.274 & 0.380 & 0.455 & 0.293 & \underline{0.039} & \underline{0.334} & 0.529 & 0.616 & 0.380 & -0.009 & 0.231 & 0.560 & 0.717 & 0.375 & 0.349 \\
 & DeepHawkes & -0.021 & 0.078 & 0.335 & 0.434 & 0.207 & -0.003 & 0.247 & 0.371 & 0.498 & 0.278 & -0.018 & 0.299 & 0.377 & 0.550 & 0.302 & -0.066 & 0.061 & 0.214 & 0.486 & 0.174 & 0.240 \\
 & DeepCas & 0.065 & 0.002 & -0.000 & 0.001 & 0.017 & -0.288 & -0.116 & -0.044 & -0.053 & -0.125 & -0.406 & -0.061 & -0.021 & 0.081 & -0.102 & 0.018 & 0.076 & 0.270 & 0.288 & 0.163 & -0.012 \\
 & CasSeqGCN & 0.000 & 0.151 & 0.429 & 0.483 & 0.266 & 0.000 & 0.250 & \underline{0.407} & 0.558 & 0.304 & -0.001 & 0.333 & 0.516 & 0.657 & 0.376 & -0.013 & 0.260 & 0.536 & 0.742 & 0.381 & 0.332 \\
 & Graph-LSTM & 0.016 & 0.169 & 0.442 & \underline{0.502} & 0.282 & 0.035 & \underline{0.285} & \textbf{0.433} & \underline{0.575} & \underline{0.332} & 0.033 & 0.325 & \underline{0.531} & \textbf{0.693} & \underline{0.395} & \underline{0.031} & \underline{0.320} & \underline{0.566} & \underline{0.750} & \underline{0.417} & \underline{0.357} \\
 & \textbf{MMG-PopNet (Ours)} & \textbf{0.210} & \textbf{0.323} & \textbf{0.507} & \textbf{0.552} & \textbf{0.398} & \textbf{0.118} & \textbf{0.325} & \textbf{0.433} & \textbf{0.615} & \textbf{0.373} & \textbf{0.129} & \textbf{0.373} & \textbf{0.601} & \underline{0.687} & \textbf{0.448} & \textbf{0.135} & \textbf{0.360} & \textbf{0.588} & \textbf{0.752} & \textbf{0.459} & \textbf{0.419} \\
\midrule

\multirow{6}{*}{\makecell{\textcolor{googlegreen}{\bf \scshape Unique}\\ \textcolor{googlegreen}{\bf \scshape Users}}} & MLP & \underline{0.357} & \underline{0.399} & \underline{0.532} & 0.557 & \underline{0.461} & \underline{0.070} & 0.249 & 0.327 & 0.389 & 0.259 & \underline{0.048} & 0.340 & 0.523 & 0.617 & 0.382 & -0.003 & 0.230 & 0.551 & 0.710 & 0.372 & \underline{0.369} \\
 & DeepHawkes & -0.018 & 0.086 & 0.423 & 0.503 & 0.249 & -0.001 & 0.209 & 0.322 & 0.459 & 0.247 & -0.015 & 0.296 & 0.368 & 0.536 & 0.296 & -0.044 & 0.064 & 0.238 & 0.492 & 0.188 & 0.245 \\
 & DeepCas & 0.089 & 0.002 & -0.000 & 0.001 & 0.023 & -0.231 & -0.086 & -0.034 & -0.048 & -0.100 & -0.329 & -0.021 & -0.006 & 0.090 & -0.067 & \underline{0.028} & 0.074 & 0.268 & 0.280 & 0.163 & 0.005 \\
 & CasSeqGCN & 0.000 & 0.191 & 0.506 & 0.575 & 0.318 & 0.000 & 0.231 & 0.373 & 0.521 & 0.281 & 0.000 & \underline{0.346} & 0.515 & 0.660 & 0.380 & -0.011 & 0.258 & 0.538 & \underline{0.745} & 0.383 & 0.341 \\
 & Graph-LSTM & 0.024 & 0.220 & 0.523 & \underline{0.593} & 0.340 & 0.039 & \underline{0.272} & \underline{0.399} & \underline{0.532} & \underline{0.310} & 0.040 & 0.339 & \underline{0.534} & \textbf{0.697} & \underline{0.402} & \underline{0.028} & \underline{0.325} & \underline{0.568} & 0.741 & \underline{0.415} & 0.367 \\
 & \textbf{MMG-PopNet (Ours)} & \textbf{0.363} & \textbf{0.462} & \textbf{0.620} & \textbf{0.657} & \textbf{0.526} & \textbf{0.119} & \textbf{0.300} & \textbf{0.400} & \textbf{0.576} & \textbf{0.349} & \textbf{0.144} & \textbf{0.384} & \textbf{0.602} & \underline{0.690} & \textbf{0.455} & \textbf{0.129} & \textbf{0.356} & \textbf{0.584} & \textbf{0.751} & \textbf{0.455} & \textbf{0.446} \\
\midrule

\multirow{6}{*}{\textcolor{googlepurple}{\bf \scshape Like Score}} & MLP & \underline{0.481} & \underline{0.484} & \underline{0.509} & \underline{0.513} & \underline{0.497} & \underline{0.059} & \underline{0.105} & \underline{0.134} & \underline{0.165} & \underline{0.116} & \underline{0.092} & 0.099 & \underline{0.204} & \underline{0.306} & \underline{0.175} & 0.041 & 0.094 & \underline{0.332} & \underline{0.406} & \underline{0.218} & \underline{0.252} \\
 & DeepHawkes & -0.001 & 0.042 & 0.232 & 0.240 & 0.128 & -0.002 & 0.017 & 0.057 & 0.073 & 0.036 & -0.006 & 0.013 & 0.048 & 0.087 & 0.035 & -0.022 & -0.021 & 0.006 & 0.158 & 0.030 & 0.058 \\
 & DeepCas & 0.060 & 0.001 & -0.000 & 0.001 & 0.016 & -0.006 & 0.005 & 0.006 & -0.005 & 0.000 & -0.025 & 0.017 & 0.036 & 0.066 & 0.024 & \underline{0.150} & 0.141 & 0.247 & 0.222 & 0.190 & 0.057 \\
 & CasSeqGCN & -0.001 & 0.097 & 0.261 & 0.309 & 0.167 & -0.001 & 0.020 & 0.061 & 0.085 & 0.041 & 0.000 & 0.013 & 0.036 & 0.106 & 0.039 & -0.006 & 0.020 & 0.107 & 0.206 & 0.082 & 0.082 \\
 & Graph-LSTM & 0.023 & 0.126 & 0.291 & 0.354 & 0.199 & 0.052 & 0.087 & 0.104 & 0.134 & 0.094 & 0.080 & \underline{0.101} & 0.155 & 0.275 & 0.153 & 0.074 & \underline{0.199} & 0.286 & 0.303 & 0.215 & 0.165 \\
 & \textbf{MMG-PopNet (Ours)} & \textbf{0.498} & \textbf{0.553} & \textbf{0.598} & \textbf{0.620} & \textbf{0.567} & \textbf{0.087} & \textbf{0.156} & \textbf{0.218} & \textbf{0.281} & \textbf{0.185} & \textbf{0.183} & \textbf{0.250} & \textbf{0.313} & \textbf{0.393} & \textbf{0.285} & \textbf{0.258} & \textbf{0.318} & \textbf{0.471} & \textbf{0.564} & \textbf{0.403} & \textbf{0.360} \\
\bottomrule
\end{tabular}
}
\label{tab:R2_performance}

\vspace{1ex}
\begin{flushleft} 

\normalsize

\textbf{Description: } The $R^2$ metric indicates the proportion of variance in the final popularity outcomes that the models successfully explain. MMG-PopNet consistently outperforms all baselines, achieving the highest average scores across structural, user, and engagement prediction targets. Notably, traditional baselines like DeepCas and DeepHawkes struggle significantly, often yielding near-zero or negative values, while MMG-PopNet maintains a robust predictive fit regardless of the observation horizon.

\end{flushleft}
\end{table}

\newpage
\begin{table}[t]
\small
\caption{Spearman results for final cascade state prediction under different early observation windows, including \textcolor{googleblue}{\bf \scshape structural tasks} (\textcolor{googleblue}{max width, max depth, structural virality, size}), \textcolor{googlegreen}{\bf \scshape unique users}, and \textcolor{googlepurple}{\bf \scshape like score}. Higher is better. Best values are \textbf{bolded}, and second-best values are \underline{underlined}.}
\vspace{0.5em}
\centering
\setlength{\tabcolsep}{3pt}
\renewcommand{\arraystretch}{1.25}
\resizebox{\textwidth}{!}{%
\begin{tabular}{c l| ccccc ccccc ccccc ccccc |c}
\toprule
\multirow{2}{*}{\textbf{Task}} & 
\multirow{2}{*}{\textbf{Model}} & 
\multicolumn{5}{c}{\textbf{Bluesky}} & 
\multicolumn{5}{c}{\textbf{r/AMA}} & 
\multicolumn{5}{c}{\textbf{r/Gaming}} & 
\multicolumn{5}{c|}{\textbf{r/Futurology}} & 
\multirow{2}{*}{\textbf{Avg}} \\ 
\cmidrule(r){3-7}
\cmidrule(r){8-12}
\cmidrule(r){13-17}
\cmidrule(lr){18-22}

 &  & \textbf{0} & \textbf{2} & \textbf{10} & \textbf{20} & \textbf{Avg} & \textbf{0} & \textbf{15} & \textbf{30} & \textbf{60} & \textbf{Avg} & \textbf{0} & \textbf{20} & \textbf{50} & \textbf{90} & \textbf{Avg} & \textbf{0} & \textbf{30} & \textbf{90} & \textbf{180} & \textbf{Avg} & \\ 
\midrule

\multirow{6}{*}{\textcolor{googleblue}{\bf \scshape Max Width}} & MLP & \underline{0.525} & \underline{0.544} & \underline{0.644} & \underline{0.677} & \underline{0.598} & \underline{0.274} & 0.517 & 0.611 & 0.651 & 0.513 & \underline{0.244} & 0.597 & 0.710 & 0.770 & 0.580 & 0.153 & 0.519 & 0.729 & 0.798 & 0.550 & \underline{0.560} \\
 & DeepHawkes & 0.082 & 0.260 & 0.530 & 0.638 & 0.378 & -0.006 & 0.507 & 0.636 & 0.728 & 0.466 & 0.004 & 0.552 & 0.690 & 0.773 & 0.505 & -0.035 & 0.409 & 0.615 & 0.731 & 0.430 & 0.445 \\
 & DeepCas & 0.017 & 0.098 & 0.083 & 0.069 & 0.067 & 0.023 & 0.126 & 0.140 & 0.140 & 0.107 & 0.158 & 0.244 & 0.246 & 0.306 & 0.238 & \underline{0.220} & 0.359 & 0.523 & 0.569 & 0.418 & 0.208 \\
 & CasSeqGCN & 0.000 & 0.319 & 0.577 & 0.661 & 0.389 & 0.000 & 0.515 & 0.670 & 0.760 & 0.486 & 0.000 & \underline{0.609} & 0.749 & 0.819 & 0.544 & 0.000 & 0.573 & 0.755 & \textbf{0.865} & 0.548 & 0.492 \\
 & Graph-LSTM & 0.140 & 0.322 & 0.581 & 0.658 & 0.425 & 0.197 & \underline{0.536} & \underline{0.677} & \underline{0.771} & \underline{0.545} & 0.221 & 0.605 & \underline{0.781} & \textbf{0.846} & \underline{0.613} & 0.192 & \underline{0.598} & \textbf{0.774} & 0.854 & \underline{0.605} & 0.547 \\
 & \textbf{MMG-PopNet (Ours)} & \textbf{0.535} & \textbf{0.586} & \textbf{0.691} & \textbf{0.736} & \textbf{0.637} & \textbf{0.366} & \textbf{0.578} & \textbf{0.681} & \textbf{0.779} & \textbf{0.601} & \textbf{0.430} & \textbf{0.652} & \textbf{0.783} & \underline{0.838} & \textbf{0.676} & \textbf{0.410} & \textbf{0.634} & \underline{0.771} & \underline{0.855} & \textbf{0.667} & \textbf{0.645} \\
\midrule

\multirow{6}{*}{\textcolor{googleblue}{\bf \scshape Max Depth}} & MLP & \underline{0.134} & 0.209 & 0.395 & 0.498 & 0.309 & \underline{0.172} & \underline{0.428} & 0.553 & 0.622 & 0.444 & \underline{0.121} & \underline{0.449} & 0.623 & 0.662 & 0.464 & 0.122 & 0.413 & 0.648 & \underline{0.718} & 0.475 & 0.423 \\
 & DeepHawkes & 0.013 & 0.160 & 0.342 & 0.385 & 0.225 & -0.025 & 0.398 & 0.503 & 0.556 & 0.358 & 0.012 & 0.418 & 0.582 & 0.637 & 0.412 & -0.057 & 0.330 & 0.530 & 0.688 & 0.373 & 0.342 \\
 & DeepCas & -0.012 & 0.035 & 0.034 & 0.056 & 0.028 & 0.057 & 0.126 & 0.135 & 0.139 & 0.114 & 0.081 & 0.161 & 0.164 & 0.230 & 0.159 & \underline{0.196} & 0.304 & 0.464 & 0.491 & 0.364 & 0.166 \\
 & CasSeqGCN & 0.000 & 0.194 & 0.380 & 0.457 & 0.258 & 0.000 & 0.388 & 0.542 & 0.626 & 0.389 & 0.000 & 0.386 & 0.556 & 0.611 & 0.388 & 0.000 & 0.402 & 0.613 & 0.705 & 0.430 & 0.366 \\
 & Graph-LSTM & 0.100 & \underline{0.236} & \underline{0.421} & \underline{0.525} & \underline{0.321} & 0.134 & 0.425 & \underline{0.571} & \underline{0.670} & \underline{0.450} & 0.115 & 0.417 & \underline{0.637} & \underline{0.719} & \underline{0.472} & 0.164 & \underline{0.469} & \underline{0.656} & 0.706 & \underline{0.499} & \underline{0.436} \\
 & \textbf{MMG-PopNet (Ours)} & \textbf{0.187} & \textbf{0.293} & \textbf{0.447} & \textbf{0.552} & \textbf{0.370} & \textbf{0.255} & \textbf{0.469} & \textbf{0.583} & \textbf{0.681} & \textbf{0.497} & \textbf{0.257} & \textbf{0.473} & \textbf{0.667} & \textbf{0.724} & \textbf{0.530} & \textbf{0.395} & \textbf{0.543} & \textbf{0.663} & \textbf{0.745} & \textbf{0.587} & \textbf{0.496} \\
\midrule

\multirow{6}{*}{\makecell{\textcolor{googleblue}{\bf \scshape Structural}\\ \textcolor{googleblue}{\bf \scshape Virality}}} & MLP & \underline{0.286} & \underline{0.351} & \underline{0.508} & \underline{0.597} & \underline{0.436} & \underline{0.190} & \underline{0.465} & 0.592 & 0.666 & 0.478 & \underline{0.131} & \underline{0.440} & 0.616 & 0.675 & \underline{0.465} & 0.128 & 0.422 & 0.664 & \underline{0.737} & 0.488 & \underline{0.467} \\
 & DeepHawkes & 0.034 & 0.219 & 0.470 & 0.530 & 0.313 & -0.026 & 0.439 & 0.556 & 0.621 & 0.398 & 0.001 & 0.395 & 0.550 & 0.590 & 0.384 & -0.051 & 0.346 & 0.535 & 0.683 & 0.378 & 0.368 \\
 & DeepCas & -0.012 & 0.058 & 0.051 & 0.064 & 0.040 & 0.052 & 0.128 & 0.147 & 0.137 & 0.116 & 0.081 & 0.154 & 0.151 & 0.215 & 0.150 & \underline{0.207} & 0.308 & 0.459 & 0.486 & 0.365 & 0.168 \\
 & CasSeqGCN & 0.000 & 0.267 & 0.489 & 0.574 & 0.333 & 0.000 & 0.431 & 0.590 & 0.682 & 0.426 & 0.000 & 0.359 & 0.520 & 0.569 & 0.362 & 0.000 & 0.424 & 0.629 & 0.729 & 0.446 & 0.392 \\
 & Graph-LSTM & 0.109 & 0.272 & 0.491 & 0.589 & 0.365 & 0.149 & 0.456 & \underline{0.609} & \underline{0.712} & \underline{0.481} & 0.088 & 0.389 & \underline{0.626} & \underline{0.726} & 0.457 & 0.178 & \underline{0.482} & \underline{0.674} & 0.715 & \underline{0.512} & 0.454 \\
 & \textbf{MMG-PopNet (Ours)} & \textbf{0.304} & \textbf{0.388} & \textbf{0.519} & \textbf{0.616} & \textbf{0.457} & \textbf{0.276} & \textbf{0.508} & \textbf{0.618} & \textbf{0.721} & \textbf{0.531} & \textbf{0.239} & \textbf{0.476} & \textbf{0.668} & \textbf{0.733} & \textbf{0.529} & \textbf{0.405} & \textbf{0.538} & \textbf{0.684} & \textbf{0.761} & \textbf{0.597} & \textbf{0.529} \\
\midrule

\multirow{6}{*}{\textcolor{googleblue}{\bf \scshape \textbf{Size}}} & MLP & \underline{0.400} & \underline{0.449} & \underline{0.604} & \underline{0.666} & \underline{0.530} & \underline{0.253} & 0.527 & 0.641 & 0.698 & 0.530 & \underline{0.207} & \underline{0.584} & 0.721 & 0.775 & 0.572 & 0.137 & 0.495 & 0.732 & 0.814 & 0.544 & \underline{0.544} \\
 & DeepHawkes & 0.056 & 0.261 & 0.567 & 0.635 & 0.380 & -0.013 & 0.522 & 0.658 & 0.758 & 0.481 & 0.009 & 0.550 & 0.712 & 0.788 & 0.515 & -0.042 & 0.394 & 0.619 & 0.760 & 0.433 & 0.452 \\
 & DeepCas & 0.003 & 0.078 & 0.068 & 0.075 & 0.056 & 0.041 & 0.141 & 0.157 & 0.151 & 0.122 & 0.151 & 0.237 & 0.232 & 0.307 & 0.232 & \underline{0.222} & 0.344 & 0.524 & 0.573 & 0.416 & 0.207 \\
 & CasSeqGCN & 0.000 & 0.318 & 0.570 & 0.650 & 0.385 & 0.000 & 0.512 & 0.663 & 0.763 & 0.485 & 0.000 & 0.575 & 0.741 & \underline{0.805} & 0.530 & 0.000 & 0.525 & 0.731 & \textbf{0.847} & 0.526 & 0.482 \\
 & Graph-LSTM & 0.124 & 0.309 & 0.567 & 0.652 & 0.413 & 0.197 & \underline{0.534} & \underline{0.676} & \underline{0.778} & \underline{0.546} & 0.197 & 0.567 & \underline{0.768} & \textbf{0.833} & \underline{0.591} & 0.190 & \underline{0.566} & \underline{0.752} & 0.832 & \underline{0.585} & 0.534 \\
 & \textbf{MMG-PopNet (Ours)} & \textbf{0.415} & \textbf{0.483} & \textbf{0.622} & \textbf{0.694} & \textbf{0.554} & \textbf{0.359} & \textbf{0.581} & \textbf{0.681} & \textbf{0.788} & \textbf{0.602} & \textbf{0.401} & \textbf{0.624} & \textbf{0.779} & \textbf{0.833} & \textbf{0.659} & \textbf{0.418} & \textbf{0.618} & \textbf{0.762} & \underline{0.845} & \textbf{0.661} & \textbf{0.619} \\
\midrule

\multirow{6}{*}{\makecell{\textcolor{googlegreen}{\bf \scshape Unique}\\ \textcolor{googlegreen}{\bf \scshape Users}}} & MLP & \underline{0.553} & \underline{0.566} & \underline{0.659} & \underline{0.690} & \underline{0.617} & \underline{0.278} & 0.515 & 0.611 & 0.652 & 0.514 & \underline{0.229} & 0.586 & 0.712 & 0.774 & 0.575 & 0.156 & 0.495 & 0.718 & 0.805 & 0.543 & \underline{0.562} \\
 & DeepHawkes & 0.081 & 0.237 & 0.514 & 0.628 & 0.365 & -0.009 & 0.510 & 0.635 & 0.728 & 0.466 & 0.006 & 0.546 & 0.692 & 0.773 & 0.504 & -0.033 & 0.390 & 0.602 & 0.751 & 0.427 & 0.441 \\
 & DeepCas & 0.018 & 0.099 & 0.089 & 0.064 & 0.068 & 0.028 & 0.132 & 0.145 & 0.141 & 0.111 & 0.169 & 0.258 & 0.253 & 0.326 & 0.252 & \underline{0.218} & 0.342 & 0.524 & 0.576 & 0.415 & 0.212 \\
 & CasSeqGCN & 0.000 & 0.277 & 0.551 & 0.643 & 0.368 & 0.000 & 0.512 & 0.659 & 0.750 & 0.480 & 0.000 & \underline{0.589} & 0.741 & 0.809 & 0.535 & 0.000 & 0.524 & 0.722 & \textbf{0.850} & 0.524 & 0.477 \\
 & Graph-LSTM & 0.141 & 0.330 & 0.581 & 0.667 & 0.430 & 0.199 & \underline{0.530} & \underline{0.668} & \underline{0.762} & \underline{0.540} & 0.209 & 0.584 & \underline{0.774} & \textbf{0.840} & \underline{0.602} & 0.189 & \underline{0.566} & \underline{0.751} & 0.824 & \underline{0.582} & 0.539 \\
 & \textbf{MMG-PopNet (Ours)} & \textbf{0.560} & \textbf{0.603} & \textbf{0.694} & \textbf{0.737} & \textbf{0.649} & \textbf{0.370} & \textbf{0.574} & \textbf{0.669} & \textbf{0.767} & \textbf{0.595} & \textbf{0.422} & \textbf{0.641} & \textbf{0.780} & \underline{0.836} & \textbf{0.670} & \textbf{0.422} & \textbf{0.619} & \textbf{0.758} & \underline{0.843} & \textbf{0.660} & \textbf{0.644} \\
\midrule

\multirow{6}{*}{\textcolor{googlepurple}{\bf \scshape Like Score}} & MLP & \underline{0.680} & \underline{0.681} & \underline{0.703} & \underline{0.700} & \underline{0.691} & \underline{0.232} & \underline{0.266} & \underline{0.262} & \underline{0.270} & \underline{0.258} & 0.249 & 0.258 & \underline{0.366} & \underline{0.444} & 0.329 & 0.219 & 0.298 & 0.524 & \underline{0.548} & 0.397 & \underline{0.419} \\
 & DeepHawkes & 0.103 & 0.184 & 0.394 & 0.421 & 0.276 & -0.014 & 0.050 & 0.095 & 0.110 & 0.060 & -0.020 & 0.010 & 0.112 & 0.184 & 0.072 & -0.002 & 0.151 & 0.213 & 0.317 & 0.170 & 0.145 \\
 & DeepCas & 0.041 & 0.110 & 0.070 & 0.078 & 0.075 & 0.039 & 0.068 & 0.133 & 0.101 & 0.085 & 0.086 & 0.156 & 0.186 & 0.255 & 0.171 & \underline{0.411} & 0.418 & 0.459 & 0.456 & \underline{0.436} & 0.192 \\
 & CasSeqGCN & 0.000 & 0.215 & 0.405 & 0.467 & 0.272 & 0.000 & 0.038 & 0.083 & 0.103 & 0.056 & 0.000 & 0.015 & 0.088 & 0.177 & 0.070 & 0.000 & 0.140 & 0.258 & 0.369 & 0.192 & 0.148 \\
 & Graph-LSTM & 0.160 & 0.283 & 0.456 & 0.516 & 0.354 & 0.189 & 0.207 & 0.208 & 0.246 & 0.212 & \underline{0.298} & \underline{0.292} & 0.343 & 0.433 & \underline{0.342} & 0.280 & \underline{0.462} & \underline{0.543} & 0.445 & 0.433 & 0.335 \\
 & \textbf{MMG-PopNet (Ours)} & \textbf{0.698} & \textbf{0.739} & \textbf{0.769} & \textbf{0.781} & \textbf{0.747} & \textbf{0.251} & \textbf{0.308} & \textbf{0.347} & \textbf{0.356} & \textbf{0.316} & \textbf{0.397} & \textbf{0.432} & \textbf{0.489} & \textbf{0.525} & \textbf{0.461} & \textbf{0.532} & \textbf{0.575} & \textbf{0.689} & \textbf{0.726} & \textbf{0.630} & \textbf{0.539} \\
\bottomrule
\end{tabular}
}
\label{tab:Spearman_performance}
\vspace{1ex}
\begin{flushleft} 

\normalsize

\textbf{Description: } The Spearman rank correlation evaluates the models' ability to accurately preserve the relative ordinal ranking of cascades based on their final states. MMG-PopNet achieves the highest average correlation across all tasks and platforms, demonstrating superior capability in ranking future popularity trends. While sequence-based models like Graph-LSTM provide competitive baseline rankings, MMG-PopNet's multimodal approach captures complementary signals that result in more reliable and consistent cascade orderings.

\end{flushleft}
\end{table}

\newpage
\subsection{Popularity Prediction of Cascade States at Future Horizon}
We provide the full future-horizon prediction results in Tables~\ref{tab:bluesky_future_horizon}, \ref{tab:ama_future_horizon}, \ref{tab:gaming_future_horizon}, and \ref{tab:futurology_future_horizon}. 
These tables extend the main results by reporting all target variables across all datasets, observation windows, and prediction horizons. 
In addition to the intermediate horizons $\{4\text{h}, 8\text{h}, 16\text{h}, 24\text{h}\}$, we also report prediction error for the final cascade state. 
Thus, in this setting, each observed prefix $G^t$ can produce multiple supervised instances, where each instance pairs the same prefix with a different target state $\textbf{Y}_G^{t'}$. 
To specify which future state is being predicted, we append a one-hot encoding of the target horizon to the input representation. 
This training setup differs from terminal-state-only prediction because each model receives supervision from both intermediate cascade states and the final state. 
The learned representation is therefore shaped by multiple stages of cascade evolution rather than only by the terminal outcome.

If a cascade has already terminated before a given horizon, the corresponding intermediate target is unavailable and is excluded from that horizon-specific training and evaluation set.
The final-state target differs from the fixed intermediate horizons because it is defined for every completed cascade. 
For this reason, the final-state MSE can be lower than the error at longer horizons such as 16h or 24h. 
Some cascades terminate before those later horizons, so their final state occurs earlier than the fixed horizon and is easier to infer from the observed prefix. 
This effect should be considered when comparing the final-state column with the 16h and 24h columns. 

For engagement target \textsc{Like Score}, ground truth is available only at the final cascade state.  
Consequently, those targets are evaluated only in the Final column and are marked as unavailable at intermediate horizons.

Across the four datasets, MMG-PopNet achieves the lowest MSE in most settings across the popularity targets. It consistently outperforms the structure-only baselines such as DeepCas and DeepHawkes, and this underscores the necessity of multimodal representations for capturing both the semantic and structural dynamics that influence a social cascade.
A notable pattern is that MLP is often competitive for intermediate horizons and sometimes obtains the best result. 
This is particularly visible under root-only inputs and shorter forecasting horizons, where a fixed cascade-level representation can capture much of the available predictive signal. 

The MLP baseline uses the root-post features, the mean representation of observed posts, and thread-level metadata, so its strong performance indicates that early content, aggregate response features, posting context, and author-level context are highly informative for near-term cascade growth. 
This result also shows that complex structural encoders are not always necessary when the target horizon is close to the observation window or when little cascade topology is available.

Graph-LSTM and CasSeqGCN form the strongest structure-aware baselines in many settings. 
They are especially competitive when longer observation windows expose more of the reply tree, and their strongest results occur for \textsc{Max Depth} and \textsc{Structural Virality}. 

Overall, the appendix results support the trends reported in the main section while adding a more detailed view across targets. 

\begin{table}[H]
\centering
\caption{Bluesky future-horizon prediction MSE across observation windows and popularity targets. \textbf{Lower is better}. Best values are \textbf{bolded} and second-best values are \underline{underlined}. Column groups correspond to Root Only, 2 min, 10 min, and 20 min observation windows, and columns within each group report prediction error at $\{4\text{h}, 8\text{h}, 16\text{h}, 24\text{h}\}$ and at the final cascade state. \textsc{Like Score} is reported only for the final state because its intermediate-horizon labels are unavailable.}
\label{tab:bluesky_future_horizon}

\vspace{1ex}
\small
\setlength{\tabcolsep}{3.2pt}
\renewcommand{\arraystretch}{1.18}

\resizebox{\textwidth}{!}{%
\begin{tabular}{l l| ccccc ccccc ccccc ccccc}
\toprule
\multirow{2}{*}{\textbf{Task}} &
\multirow{2}{*}{\textbf{Model}} &
\multicolumn{5}{c}{\textbf{Root Only}} &
\multicolumn{5}{c}{\textbf{2 min}} &
\multicolumn{5}{c}{\textbf{10 min}} &
\multicolumn{5}{c}{\textbf{20 min}} \\
\cmidrule(lr){3-7}
\cmidrule(lr){8-12}
\cmidrule(lr){13-17}
\cmidrule(lr){18-22}
& &
\textbf{4h} & \textbf{8h} & \textbf{16h} & \textbf{24h} & \textbf{Final} &
\textbf{4h} & \textbf{8h} & \textbf{16h} & \textbf{24h} & \textbf{Final} &
\textbf{4h} & \textbf{8h} & \textbf{16h} & \textbf{24h} & \textbf{Final} &
\textbf{4h} & \textbf{8h} & \textbf{16h} & \textbf{24h} & \textbf{Final} \\
\midrule

\multirow{6}{*}{\textcolor{googleblue}{\bf \scshape Max Width}}
& MLP & \textbf{0.472} & \underline{0.570} & \underline{0.694} & \underline{0.800} & \underline{0.458} & \underline{0.410} & \underline{0.499} & \underline{0.620} & \underline{0.721} & \underline{0.429} & 0.318 & 0.398 & \underline{0.497} & \underline{0.584} & 0.373 & 0.291 & 0.364 & 0.454 & 0.532 & 0.348 \\
& DeepHawkes & 0.809 & 0.999 & 1.262 & 1.444 & 0.728 & 0.683 & 0.856 & 1.087 & 1.270 & 0.634 & 0.369 & 0.494 & 0.661 & 0.786 & 0.436 & 0.261 & 0.347 & 0.462 & 0.549 & 0.330 \\
& DeepCas & \underline{0.707} & 0.864 & 1.051 & 1.198 & 0.628 & 0.787 & 0.968 & 1.175 & 1.320 & 0.675 & 0.790 & 0.967 & 1.174 & 1.323 & 0.676 & 0.790 & 0.968 & 1.173 & 1.320 & 0.677 \\
& CasSeqGCN & 0.789 & 0.969 & 1.173 & 1.317 & 0.676 & 0.538 & 0.666 & 0.831 & 0.974 & 0.542 & 0.289 & 0.390 & 0.508 & 0.619 & 0.360 & \underline{0.211} & 0.296 & 0.398 & 0.493 & 0.292 \\
& Graph-LSTM & 0.751 & 0.905 & 1.083 & 1.217 & 0.660 & 0.517 & 0.632 & 0.778 & 0.886 & 0.523 & \underline{0.284} & \underline{0.382} & 0.498 & 0.599 & \underline{0.354} & 0.212 & \underline{0.293} & \underline{0.386} & \underline{0.474} & \underline{0.288} \\
& \textbf{MMG-PopNet} & \textbf{0.472} & \textbf{0.559} & \textbf{0.667} & \textbf{0.762} & \textbf{0.448} & \textbf{0.383} & \textbf{0.447} & \textbf{0.532} & \textbf{0.612} & \textbf{0.386} & \textbf{0.257} & \textbf{0.320} & \textbf{0.404} & \textbf{0.478} & \textbf{0.295} & \textbf{0.199} & \textbf{0.256} & \textbf{0.327} & \textbf{0.394} & \textbf{0.243} \\
\addlinespace[1mm]
\midrule

\multirow{6}{*}{\textcolor{googleblue}{\bf \scshape Max Depth}}
& MLP & \underline{0.498} & \underline{0.504} & \underline{0.558} & \underline{0.565} & \textbf{0.356} & \underline{0.450} & \underline{0.459} & \underline{0.510} & \underline{0.527} & 0.347 & 0.348 & 0.358 & 0.410 & 0.428 & \underline{0.295} & 0.301 & 0.323 & 0.380 & 0.400 & 0.268 \\
& DeepHawkes & 0.596 & 0.612 & 0.679 & 0.689 & 0.403 & 0.560 & 0.556 & 0.615 & 0.626 & 0.384 & 0.428 & 0.472 & 0.501 & 0.562 & 0.340 & 0.404 & 0.414 & 0.467 & 0.483 & 0.334 \\
& DeepCas & 0.558 & 0.575 & 0.638 & 0.646 & \underline{0.362} & 0.575 & 0.595 & 0.662 & 0.671 & 0.371 & 0.578 & 0.594 & 0.663 & 0.673 & 0.364 & 0.576 & 0.594 & 0.663 & 0.673 & 0.364 \\
& CasSeqGCN & 0.576 & 0.594 & 0.663 & 0.672 & 0.365 & 0.491 & 0.504 & 0.566 & 0.583 & 0.351 & 0.355 & 0.368 & 0.424 & 0.444 & 0.300 & 0.300 & 0.317 & 0.373 & 0.391 & 0.261 \\
& Graph-LSTM & 0.554 & 0.565 & 0.622 & 0.627 & 0.364 & 0.470 & 0.478 & 0.532 & 0.545 & \underline{0.341} & \underline{0.337} & \underline{0.350} & \underline{0.400} & \underline{0.418} & \textbf{0.279} & \underline{0.287} & \underline{0.306} & \underline{0.359} & \underline{0.375} & \underline{0.246} \\
& \textbf{MMG-PopNet} & \textbf{0.485} & \textbf{0.493} & \textbf{0.541} & \textbf{0.542} & 0.379 & \textbf{0.421} & \textbf{0.421} & \textbf{0.460} & \textbf{0.468} & \textbf{0.329} & \textbf{0.328} & \textbf{0.334} & \textbf{0.374} & \textbf{0.386} & \textbf{0.279} & \textbf{0.278} & \textbf{0.291} & \textbf{0.336} & \textbf{0.346} & \textbf{0.240} \\
\addlinespace[1mm]
\midrule

\multirow{6}{*}{\makecell[l]{\textcolor{googleblue}{\bf \scshape Structural}\\\textcolor{googleblue}{\bf \scshape Virality}}}
& MLP & \underline{0.259} & \underline{0.257} & \underline{0.282} & \underline{0.284} & \textbf{0.147} & \underline{0.232} & \underline{0.232} & \underline{0.256} & \underline{0.264} & \underline{0.143} & \underline{0.181} & \underline{0.184} & \underline{0.210} & \underline{0.219} & 0.123 & 0.157 & 0.166 & 0.193 & 0.206 & 0.113 \\
& DeepHawkes & 0.324 & 0.324 & 0.355 & 0.357 & 0.174 & 0.302 & 0.296 & 0.326 & 0.330 & 0.170 & 0.224 & 0.244 & 0.257 & 0.291 & 0.141 & 0.210 & 0.211 & 0.238 & 0.247 & 0.135 \\
& DeepCas & 0.307 & 0.308 & 0.337 & 0.338 & \underline{0.155} & 0.317 & 0.320 & 0.349 & 0.349 & 0.159 & 0.318 & 0.319 & 0.349 & 0.350 & 0.157 & 0.318 & 0.319 & 0.349 & 0.350 & 0.157 \\
& CasSeqGCN & 0.317 & 0.319 & 0.349 & 0.349 & 0.157 & 0.265 & 0.266 & 0.295 & 0.303 & 0.150 & 0.189 & 0.195 & 0.223 & 0.236 & 0.126 & 0.160 & 0.168 & 0.195 & 0.208 & 0.111 \\
& Graph-LSTM & 0.306 & 0.304 & 0.328 & 0.327 & 0.158 & 0.256 & 0.256 & 0.280 & 0.287 & 0.146 & 0.182 & 0.187 & 0.212 & 0.222 & \underline{0.118} & \underline{0.152} & \underline{0.163} & \underline{0.189} & \underline{0.199} & \underline{0.105} \\
& \textbf{MMG-PopNet} & \textbf{0.257} & \textbf{0.255} & \textbf{0.276} & \textbf{0.274} & 0.161 & \textbf{0.220} & \textbf{0.216} & \textbf{0.234} & \textbf{0.240} & \textbf{0.137} & \textbf{0.172} & \textbf{0.174} & \textbf{0.193} & \textbf{0.201} & \textbf{0.116} & \textbf{0.147} & \textbf{0.153} & \textbf{0.175} & \textbf{0.183} & \textbf{0.102} \\
\addlinespace[1mm]
\midrule

\multirow{6}{*}{\textcolor{googleblue}{\bf \scshape Size}}
& MLP & \underline{0.842} & \underline{0.957} & \underline{1.123} & \underline{1.238} & \textbf{0.694} & \underline{0.726} & \underline{0.833} & \underline{0.992} & \underline{1.113} & \underline{0.655} & 0.542 & 0.639 & \underline{0.778} & \underline{0.885} & 0.557 & 0.475 & 0.577 & 0.711 & 0.817 & 0.512 \\
& DeepHawkes & 1.262 & 1.490 & 1.816 & 2.007 & 0.967 & 1.084 & 1.261 & 1.545 & 1.740 & 0.852 & 0.630 & 0.802 & 0.991 & 1.169 & 0.626 & 0.520 & 0.619 & 0.782 & 0.894 & 0.518 \\
& DeepCas & 1.129 & 1.315 & 1.561 & 1.718 & 0.830 & 1.232 & 1.447 & 1.717 & 1.870 & 0.878 & 1.236 & 1.448 & 1.717 & 1.876 & 0.881 & 1.236 & 1.448 & 1.717 & 1.874 & 0.881 \\
& CasSeqGCN & 1.237 & 1.448 & 1.717 & 1.874 & 0.881 & 0.890 & 1.041 & 1.253 & 1.420 & 0.752 & 0.529 & 0.652 & 0.819 & 0.958 & 0.552 & 0.408 & 0.519 & 0.668 & 0.800 & 0.463 \\
& Graph-LSTM & 1.173 & 1.348 & 1.568 & 1.705 & 0.865 & 0.857 & 0.990 & 1.175 & 1.302 & 0.731 & \underline{0.509} & \underline{0.629} & 0.784 & 0.910 & \underline{0.532} & \underline{0.404} & \underline{0.512} & \underline{0.648} & \underline{0.761} & \underline{0.454} \\
& \textbf{MMG-PopNet} & \textbf{0.829} & \textbf{0.936} & \textbf{1.075} & \textbf{1.169} & \underline{0.704} & \textbf{0.690} & \textbf{0.762} & \textbf{0.865} & \textbf{0.957} & \textbf{0.605} & \textbf{0.482} & \textbf{0.558} & \textbf{0.669} & \textbf{0.760} & \textbf{0.483} & \textbf{0.387} & \textbf{0.464} & \textbf{0.567} & \textbf{0.657} & \textbf{0.408} \\
\addlinespace[1mm]
\midrule

\multirow{6}{*}{\makecell[l]{\textcolor{googlegreen}{\bf \scshape Unique}\\\textcolor{googlegreen}{\bf \scshape Users}}}
& MLP & \underline{0.511} & \underline{0.606} & \underline{0.735} & \underline{0.843} & \underline{0.464} & \underline{0.445} & \underline{0.531} & \underline{0.660} & \underline{0.762} & \underline{0.435} & 0.345 & \underline{0.425} & \underline{0.528} & \underline{0.621} & 0.376 & 0.310 & 0.384 & 0.477 & 0.562 & 0.348 \\
& DeepHawkes & 0.907 & 1.119 & 1.417 & 1.620 & 0.782 & 0.761 & 0.944 & 1.203 & 1.400 & 0.678 & 0.424 & 0.560 & 0.746 & 0.887 & 0.467 & 0.312 & 0.405 & 0.537 & 0.637 & 0.361 \\
& DeepCas & 0.780 & 0.948 & 1.156 & 1.320 & 0.660 & 0.886 & 1.082 & 1.319 & 1.484 & 0.721 & 0.890 & 1.082 & 1.319 & 1.489 & 0.723 & 0.892 & 1.084 & 1.318 & 1.486 & 0.724 \\
& CasSeqGCN & 0.890 & 1.085 & 1.319 & 1.486 & 0.723 & 0.615 & 0.754 & 0.941 & 1.097 & 0.591 & 0.339 & 0.446 & 0.579 & 0.699 & 0.397 & 0.252 & 0.341 & 0.455 & 0.564 & 0.324 \\
& Graph-LSTM & 0.845 & 1.008 & 1.208 & 1.355 & 0.704 & 0.586 & 0.709 & 0.874 & 0.991 & 0.558 & \underline{0.326} & 0.430 & 0.557 & 0.668 & \underline{0.375} & \underline{0.245} & \underline{0.330} & \underline{0.434} & \underline{0.532} & \underline{0.305} \\
& \textbf{MMG-PopNet} & \textbf{0.510} & \textbf{0.594} & \textbf{0.703} & \textbf{0.798} & \textbf{0.454} & \textbf{0.432} & \textbf{0.495} & \textbf{0.583} & \textbf{0.665} & \textbf{0.397} & \textbf{0.301} & \textbf{0.366} & \textbf{0.455} & \textbf{0.537} & \textbf{0.313} & \textbf{0.235} & \textbf{0.293} & \textbf{0.367} & \textbf{0.444} & \textbf{0.258} \\
\addlinespace[1mm]
\midrule

\multirow{6}{*}{\makecell[l]{\textcolor{googlepurple}{\bf \scshape Like}\\\textcolor{googlepurple}{\bf \scshape Score}}}
& MLP & -- & -- & -- & -- & \underline{1.312} & -- & -- & -- & -- & \underline{1.299} & -- & -- & -- & -- & \underline{1.260} & -- & -- & -- & -- & \underline{1.225} \\
& DeepHawkes & -- & -- & -- & -- & 2.533 & -- & -- & -- & -- & 2.398 & -- & -- & -- & -- & 2.114 & -- & -- & -- & -- & 1.898 \\
& DeepCas & -- & -- & -- & -- & 2.347 & -- & -- & -- & -- & 2.502 & -- & -- & -- & -- & 2.510 & -- & -- & -- & -- & 2.514 \\
& CasSeqGCN & -- & -- & -- & -- & 2.509 & -- & -- & -- & -- & 2.280 & -- & -- & -- & -- & 1.904 & -- & -- & -- & -- & 1.756 \\
& Graph-LSTM & -- & -- & -- & -- & 2.469 & -- & -- & -- & -- & 2.205 & -- & -- & -- & -- & 1.839 & -- & -- & -- & -- & 1.649 \\
& \textbf{MMG-PopNet} & -- & -- & -- & -- & \textbf{1.294} & -- & -- & -- & -- & \textbf{1.140} & -- & -- & -- & -- & \textbf{1.069} & -- & -- & -- & -- & \textbf{1.006} \\

\bottomrule
\end{tabular}
}
\vspace{1ex}
\begin{flushleft} 

\normalsize

\textbf{Description: } The results demonstrate that MMG-PopNet consistently achieves the lowest prediction error across all targets. The data reveals a steep drop in MSE as the observation window expands from 0 minutes ("Root Only") to 20 minutes, highlighting the value of early engagement data. Notably, while baseline models like DeepHawkes and DeepCas struggle significantly where they frequently exceed an MSE of 1.0 on taregts \textsc{Size} and \textsc{Unique Users}. Here, MMG-PopNet maintains a substantial advantage over these targets.

\end{flushleft}
\end{table}

\begin{table}[t]
\centering
\caption{r/AMA future-horizon prediction MSE across observation windows and popularity targets. \textbf{Lower is better}. Best values are \textbf{bolded} and second-best values are \underline{underlined}. Column groups correspond to Root Only, 15 min, 30 min, and 60 min observation windows, and columns within each group report prediction error at $\{4\text{h}, 8\text{h}, 16\text{h}, 24\text{h}\}$ and at the final cascade state. \textsc{Like Score} is reported only for the final state because its intermediate-horizon labels are unavailable.}
\label{tab:ama_future_horizon}

\vspace{1mm}
\small
\setlength{\tabcolsep}{3.2pt}
\renewcommand{\arraystretch}{1.18}

\resizebox{\textwidth}{!}{%
\begin{tabular}{l l ccccc ccccc ccccc ccccc}
\toprule
\multirow{2}{*}{\textbf{Task}} &
\multirow{2}{*}{\textbf{Model}} &
\multicolumn{5}{c}{\textbf{Root Only}} &
\multicolumn{5}{c}{\textbf{15 min}} &
\multicolumn{5}{c}{\textbf{30 min}} &
\multicolumn{5}{c}{\textbf{60 min}} \\
\cmidrule(lr){3-7}
\cmidrule(lr){8-12}
\cmidrule(lr){13-17}
\cmidrule(lr){18-22}
& &
\textbf{4h} & \textbf{8h} & \textbf{16h} & \textbf{24h} & \textbf{Final} &
\textbf{4h} & \textbf{8h} & \textbf{16h} & \textbf{24h} & \textbf{Final} &
\textbf{4h} & \textbf{8h} & \textbf{16h} & \textbf{24h} & \textbf{Final} &
\textbf{4h} & \textbf{8h} & \textbf{16h} & \textbf{24h} & \textbf{Final} \\
\midrule

\multirow{6}{*}{\textcolor{googleblue}{\bf \scshape Max Width}}
& MLP & \underline{0.545} & \underline{0.647} & \underline{0.807} & \underline{0.960} & \underline{0.670} & 0.383 & 0.478 & \underline{0.622} & \underline{0.743} & 0.533 & 0.324 & 0.425 & 0.565 & 0.687 & 0.493 & 0.264 & 0.343 & 0.463 & 0.552 & 0.407 \\
& DeepHawkes & 0.609 & 0.719 & 0.908 & 1.079 & 0.717 & 0.419 & 0.511 & 0.666 & 0.803 & 0.546 & 0.366 & 0.464 & 0.631 & 0.777 & 0.514 & 0.207 & 0.295 & 0.428 & 0.553 & 0.361 \\
& DeepCas & 0.790 & 0.900 & 1.066 & 1.228 & 0.898 & 0.627 & 0.724 & 0.887 & 1.068 & 0.744 & 0.660 & 0.784 & 1.007 & 1.139 & 0.796 & 0.581 & 0.692 & 0.862 & 1.001 & 0.707 \\
& CasSeqGCN & 0.606 & 0.714 & 0.891 & 1.057 & 0.714 & 0.393 & 0.492 & 0.655 & 0.798 & 0.528 & 0.273 & 0.386 & 0.533 & \underline{0.652} & 0.447 & \underline{0.155} & \underline{0.236} & \underline{0.346} & \underline{0.436} & \underline{0.307} \\
& Graph-LSTM & 0.574 & 0.671 & 0.836 & 0.984 & 0.681 & \underline{0.359} & \underline{0.467} & 0.626 & 0.756 & \underline{0.507} & \underline{0.260} & \underline{0.375} & \underline{0.524} & 0.657 & \underline{0.438} & 0.173 & 0.259 & 0.373 & 0.470 & 0.316 \\
& \textbf{MMG-PopNet} & \textbf{0.536} & \textbf{0.630} & \textbf{0.787} & \textbf{0.947} & \textbf{0.663} & \textbf{0.348} & \textbf{0.440} & \textbf{0.579} & \textbf{0.694} & \textbf{0.491} & \textbf{0.259} & \textbf{0.359} & \textbf{0.489} & \textbf{0.586} & \textbf{0.426} & \textbf{0.153} & \textbf{0.223} & \textbf{0.318} & \textbf{0.392} & \textbf{0.285} \\
\addlinespace[1mm]
\midrule

\multirow{6}{*}{\textcolor{googleblue}{\bf \scshape Max Depth}}
& MLP & \textbf{0.371} & \textbf{0.364} & \underline{0.364} & \underline{0.361} & \textbf{0.321} & 0.280 & 0.286 & \underline{0.289} & \underline{0.294} & 0.268 & 0.223 & \underline{0.224} & \underline{0.225} & \underline{0.234} & 0.228 & 0.161 & 0.179 & 0.187 & 0.198 & 0.199 \\
& DeepHawkes & 0.394 & 0.388 & 0.388 & 0.390 & 0.334 & 0.326 & 0.321 & 0.316 & 0.321 & 0.287 & 0.304 & 0.282 & 0.275 & 0.293 & 0.274 & 0.212 & 0.223 & 0.227 & 0.242 & 0.231 \\
& DeepCas & 0.556 & 0.535 & 0.523 & 0.511 & 0.514 & 0.420 & 0.401 & 0.389 & 0.392 & 0.365 & 0.436 & 0.410 & 0.409 & 0.400 & 0.372 & 0.378 & 0.375 & 0.361 & 0.366 & 0.339 \\
& CasSeqGCN & 0.394 & 0.388 & 0.384 & 0.384 & 0.328 & 0.299 & 0.298 & 0.299 & 0.309 & 0.273 & 0.224 & 0.226 & 0.230 & 0.243 & 0.224 & 0.151 & 0.168 & \underline{0.175} & \underline{0.188} & 0.187 \\
& Graph-LSTM & 0.381 & 0.376 & 0.368 & 0.366 & \underline{0.323} & \underline{0.278} & \underline{0.285} & \underline{0.289} & 0.295 & \underline{0.263} & \underline{0.221} & 0.225 & 0.228 & 0.250 & \underline{0.223} & \underline{0.148} & \underline{0.166} & \underline{0.175} & 0.192 & \underline{0.186} \\
& \textbf{MMG-PopNet} & \underline{0.378} & \underline{0.369} & \textbf{0.355} & \textbf{0.351} & 0.330 & \textbf{0.268} & \textbf{0.275} & \textbf{0.275} & \textbf{0.284} & \textbf{0.262} & \textbf{0.207} & \textbf{0.207} & \textbf{0.210} & \textbf{0.218} & \textbf{0.215} & \textbf{0.133} & \textbf{0.149} & \textbf{0.156} & \textbf{0.170} & \textbf{0.177} \\
\addlinespace[1mm]
\midrule

\multirow{6}{*}{\makecell[l]{\textcolor{googleblue}{\bf \scshape Structural}\\\textcolor{googleblue}{\bf \scshape Virality}}}
& MLP & \textbf{0.171} & \textbf{0.158} & \underline{0.151} & \underline{0.141} & \textbf{0.130} & \underline{0.125} & \underline{0.121} & \underline{0.118} & \underline{0.114} & \underline{0.106} & \underline{0.101} & 0.095 & \underline{0.090} & \underline{0.093} & 0.090 & 0.071 & 0.075 & 0.073 & 0.074 & 0.075 \\
& DeepHawkes & 0.184 & 0.170 & 0.162 & 0.154 & 0.137 & 0.150 & 0.139 & 0.130 & 0.125 & 0.114 & 0.142 & 0.121 & 0.114 & 0.120 & 0.111 & 0.091 & 0.092 & 0.089 & 0.094 & 0.089 \\
& DeepCas & 0.291 & 0.269 & 0.248 & 0.229 & 0.247 & 0.215 & 0.190 & 0.170 & 0.162 & 0.164 & 0.237 & 0.208 & 0.198 & 0.180 & 0.181 & 0.191 & 0.174 & 0.156 & 0.150 & 0.146 \\
& CasSeqGCN & 0.184 & 0.170 & 0.160 & 0.150 & 0.134 & 0.137 & 0.128 & 0.122 & 0.119 & 0.108 & \underline{0.101} & \underline{0.094} & 0.092 & 0.098 & \underline{0.088} & 0.065 & 0.069 & 0.067 & \underline{0.072} & 0.069 \\
& Graph-LSTM & \underline{0.177} & \underline{0.164} & 0.153 & 0.144 & \underline{0.132} & 0.126 & 0.122 & 0.120 & 0.115 & \textbf{0.105} & 0.104 & 0.098 & 0.094 & 0.104 & 0.091 & \underline{0.063} & \underline{0.068} & \underline{0.066} & \underline{0.072} & \underline{0.068} \\
& \textbf{MMG-PopNet} & 0.179 & 0.165 & \textbf{0.147} & \textbf{0.138} & 0.139 & \textbf{0.121} & \textbf{0.117} & \textbf{0.112} & \textbf{0.110} & \textbf{0.105} & \textbf{0.094} & \textbf{0.087} & \textbf{0.085} & \textbf{0.088} & \textbf{0.087} & \textbf{0.060} & \textbf{0.063} & \textbf{0.061} & \textbf{0.065} & \textbf{0.066} \\
\addlinespace[1mm]
\midrule

\multirow{6}{*}{\textcolor{googleblue}{\bf \scshape Size}}
& MLP & \underline{0.838} & \underline{0.934} & \underline{1.110} & \underline{1.265} & \textbf{0.923} & 0.553 & \underline{0.661} & \underline{0.827} & \underline{0.960} & 0.714 & 0.443 & 0.551 & 0.687 & \underline{0.828} & 0.631 & 0.331 & 0.433 & 0.559 & 0.661 & 0.511 \\
& DeepHawkes & 0.940 & 1.051 & 1.254 & 1.434 & 0.987 & 0.639 & 0.732 & 0.896 & 1.049 & 0.739 & 0.553 & 0.627 & 0.785 & 0.963 & 0.670 & 0.273 & 0.384 & 0.534 & 0.696 & 0.456 \\
& DeepCas & 1.398 & 1.472 & 1.634 & 1.792 & 1.460 & 0.996 & 1.073 & 1.236 & 1.435 & 1.057 & 1.089 & 1.189 & 1.411 & 1.538 & 1.139 & 0.934 & 1.051 & 1.218 & 1.374 & 1.011 \\
& CasSeqGCN & 0.937 & 1.045 & 1.235 & 1.407 & 0.980 & 0.607 & 0.711 & 0.888 & 1.049 & 0.726 & 0.418 & 0.540 & 0.691 & 0.835 & 0.596 & \underline{0.226} & \underline{0.331} & \underline{0.447} & \underline{0.560} & \underline{0.413} \\
& Graph-LSTM & 0.887 & 0.985 & 1.149 & 1.300 & 0.942 & \underline{0.544} & 0.666 & 0.842 & 0.981 & \underline{0.693} & \underline{0.409} & \underline{0.539} & \underline{0.686} & 0.860 & \underline{0.595} & 0.261 & 0.367 & 0.487 & 0.605 & 0.436 \\
& \textbf{MMG-PopNet} & \textbf{0.832} & \textbf{0.919} & \textbf{1.075} & \textbf{1.242} & \underline{0.925} & \textbf{0.517} & \textbf{0.617} & \textbf{0.768} & \textbf{0.897} & \textbf{0.668} & \textbf{0.383} & \textbf{0.487} & \textbf{0.618} & \textbf{0.732} & \textbf{0.565} & \textbf{0.209} & \textbf{0.298} & \textbf{0.398} & \textbf{0.490} & \textbf{0.378} \\
\addlinespace[1mm]
\midrule

\multirow{6}{*}{\makecell[l]{\textcolor{googlegreen}{\bf \scshape Unique}\\\textcolor{googlegreen}{\bf \scshape Users}}}
& MLP & \underline{0.529} & \underline{0.661} & \underline{0.857} & \underline{1.037} & \textbf{0.663} & 0.372 & 0.495 & \underline{0.669} & \underline{0.817} & 0.532 & 0.317 & 0.440 & 0.612 & 0.767 & 0.502 & 0.251 & 0.351 & 0.497 & 0.600 & 0.410 \\
& DeepHawkes & 0.590 & 0.737 & 0.966 & 1.165 & 0.712 & 0.411 & 0.531 & 0.718 & 0.887 & 0.546 & 0.358 & 0.484 & 0.687 & 0.870 & 0.529 & 0.198 & 0.303 & 0.462 & 0.603 & 0.370 \\
& DeepCas & 0.800 & 0.937 & 1.129 & 1.306 & 0.937 & 0.612 & 0.739 & 0.938 & 1.140 & 0.747 & 0.658 & 0.811 & 1.072 & 1.238 & 0.818 & 0.554 & 0.694 & 0.896 & 1.057 & 0.714 \\
& CasSeqGCN & 0.586 & 0.731 & 0.951 & 1.145 & 0.706 & 0.396 & 0.527 & 0.723 & 0.897 & 0.537 & 0.290 & 0.425 & 0.606 & 0.765 & 0.470 & \underline{0.161} & \underline{0.257} & \underline{0.389} & \underline{0.499} & \underline{0.321} \\
& Graph-LSTM & 0.556 & 0.687 & 0.886 & 1.059 & 0.677 & \underline{0.352} & \underline{0.488} & 0.677 & 0.834 & \underline{0.510} & \underline{0.273} & \underline{0.410} & \underline{0.589} & \underline{0.759} & \underline{0.460} & 0.184 & 0.286 & 0.426 & 0.539 & 0.338 \\
& \textbf{MMG-PopNet} & \textbf{0.523} & \textbf{0.650} & \textbf{0.839} & \textbf{1.027} & \underline{0.665} & \textbf{0.344} & \textbf{0.459} & \textbf{0.624} & \textbf{0.765} & \textbf{0.494} & \textbf{0.267} & \textbf{0.384} & \textbf{0.541} & \textbf{0.671} & \textbf{0.441} & \textbf{0.160} & \textbf{0.241} & \textbf{0.356} & \textbf{0.445} & \textbf{0.298} \\
\addlinespace[1mm]
\midrule

\multirow{6}{*}{\makecell[l]{\textcolor{googlepurple}{\bf \scshape Like}\\\textcolor{googlepurple}{\bf \scshape Score}}}
& MLP & -- & -- & -- & -- & \underline{1.367} & -- & -- & -- & -- & \underline{1.316} & -- & -- & -- & -- & \underline{1.283} & -- & -- & -- & -- & \underline{1.204} \\
& DeepHawkes & -- & -- & -- & -- & 1.441 & -- & -- & -- & -- & 1.406 & -- & -- & -- & -- & 1.430 & -- & -- & -- & -- & 1.338 \\
& DeepCas & -- & -- & -- & -- & 1.468 & -- & -- & -- & -- & 1.422 & -- & -- & -- & -- & 1.497 & -- & -- & -- & -- & 1.419 \\
& CasSeqGCN & -- & -- & -- & -- & 1.436 & -- & -- & -- & -- & 1.405 & -- & -- & -- & -- & 1.403 & -- & -- & -- & -- & 1.305 \\
& Graph-LSTM & -- & -- & -- & -- & \textbf{1.359} & -- & -- & -- & -- & 1.350 & -- & -- & -- & -- & 1.389 & -- & -- & -- & -- & 1.281 \\
& \textbf{MMG-PopNet} & -- & -- & -- & -- & 1.403 & -- & -- & -- & -- & \textbf{1.281} & -- & -- & -- & -- & \textbf{1.239} & -- & -- & -- & -- & \textbf{1.086} \\

\bottomrule
\end{tabular}
}
\vspace{1ex}
\begin{flushleft} 

\normalsize

\textbf{Description: } While MMG-PopNet establishes dominance in long horizon predictions and larger observation windows, the data reveals a unique trend in the "Root Only" setting where the simpler MLP model frequently matches or outperforms complex graph models in predicting \textsc{Max Depth} and \textsc{Structural Virality}.

\end{flushleft}
\end{table}

\begin{table}[t]
\centering
\caption{r/Gaming future-horizon prediction MSE across observation windows and popularity targets. \textbf{Lower is better}. Best values are \textbf{bolded} and second-best values are \underline{underlined}. Column groups correspond to Root Only, 20 min, 50 min, and 90 min observation windows, and columns within each group report prediction error at $\{4\text{h}, 8\text{h}, 16\text{h}, 24\text{h}\}$ and at the final cascade state. \textsc{Like Score} is reported only for the final state because its intermediate-horizon labels are unavailable.}
\label{tab:gaming_future_horizon}

\vspace{1mm}
\small
\setlength{\tabcolsep}{3.2pt}
\renewcommand{\arraystretch}{1.18}

\resizebox{\textwidth}{!}{%
\begin{tabular}{l l ccccc ccccc ccccc ccccc}
\toprule
\multirow{2}{*}{\textbf{Task}} &
\multirow{2}{*}{\textbf{Model}} &
\multicolumn{5}{c}{\textbf{Root Only}} &
\multicolumn{5}{c}{\textbf{20 min}} &
\multicolumn{5}{c}{\textbf{50 min}} &
\multicolumn{5}{c}{\textbf{90 min}} \\
\cmidrule(lr){3-7}
\cmidrule(lr){8-12}
\cmidrule(lr){13-17}
\cmidrule(lr){18-22}
& &
\textbf{4h} & \textbf{8h} & \textbf{16h} & \textbf{24h} & \textbf{Final} &
\textbf{4h} & \textbf{8h} & \textbf{16h} & \textbf{24h} & \textbf{Final} &
\textbf{4h} & \textbf{8h} & \textbf{16h} & \textbf{24h} & \textbf{Final} &
\textbf{4h} & \textbf{8h} & \textbf{16h} & \textbf{24h} & \textbf{Final} \\
\midrule

\multirow{6}{*}{\textcolor{googleblue}{\bf \scshape Max Width}}
& MLP & \underline{1.301} & \underline{1.623} & \underline{1.882} & \underline{2.067} & \underline{1.778} & 0.701 & 0.952 & 1.176 & 1.300 & 1.174 & 0.506 & 0.715 & 0.868 & 0.970 & 0.879 & 0.406 & 0.567 & 0.693 & 0.783 & 0.715 \\
& DeepHawkes & 1.401 & 1.757 & 2.117 & 2.393 & 1.919 & 0.933 & 1.111 & 1.368 & 1.552 & 1.295 & 0.594 & 0.974 & 1.279 & 1.580 & 1.256 & 0.354 & 0.573 & 0.841 & 1.073 & 0.866 \\
& DeepCas & 1.996 & 2.317 & 2.722 & 2.898 & 2.353 & 1.367 & 1.669 & 1.975 & 2.147 & 1.815 & 1.336 & 1.652 & 1.900 & 1.940 & 1.786 & 1.250 & 1.538 & 1.757 & 1.913 & 1.700 \\
& CasSeqGCN & 1.393 & 1.713 & 1.992 & 2.164 & 1.868 & \underline{0.664} & \underline{0.931} & 1.186 & 1.345 & 1.135 & 0.388 & 0.649 & 0.870 & 0.971 & 0.855 & 0.225 & \underline{0.393} & \underline{0.572} & \underline{0.665} & 0.590 \\
& Graph-LSTM & 1.345 & 1.651 & 1.941 & 2.101 & 1.815 & \textbf{0.622} & \textbf{0.893} & \underline{1.137} & \underline{1.285} & \underline{1.104} & \textbf{0.334} & \underline{0.589} & \underline{0.785} & \underline{0.897} & \underline{0.760} & \textbf{0.211} & 0.411 & 0.619 & 0.738 & \underline{0.587} \\
& \textbf{MMG-PopNet} & \textbf{1.160} & \textbf{1.442} & \textbf{1.687} & \textbf{1.818} & \textbf{1.554} & 0.718 & 0.940 & \textbf{1.116} & \textbf{1.213} & \textbf{1.103} & \underline{0.343} & \textbf{0.511} & \textbf{0.659} & \textbf{0.733} & \textbf{0.662} & \underline{0.214} & \textbf{0.348} & \textbf{0.494} & \textbf{0.575} & \textbf{0.490} \\
\addlinespace[1mm]
\midrule

\multirow{6}{*}{\textcolor{googleblue}{\bf \scshape Max Depth}}
& MLP & \textbf{0.295} & \underline{0.320} & \underline{0.342} & \underline{0.366} & \underline{0.342} & \textbf{0.215} & \textbf{0.238} & \underline{0.273} & \underline{0.296} & \underline{0.280} & 0.157 & \underline{0.175} & \underline{0.188} & \underline{0.192} & \underline{0.208} & 0.126 & 0.150 & 0.168 & 0.173 & 0.176 \\
& DeepHawkes & 0.317 & 0.332 & 0.366 & 0.409 & 0.354 & 0.275 & 0.282 & 0.315 & 0.346 & 0.300 & 0.198 & 0.232 & 0.264 & 0.305 & 0.279 & 0.199 & 0.210 & 0.228 & 0.243 & 0.234 \\
& DeepCas & 0.577 & 0.564 & 0.605 & 0.600 & 0.589 & 0.337 & 0.349 & 0.380 & 0.397 & 0.371 & 0.338 & 0.346 & 0.361 & 0.353 & 0.375 & 0.313 & 0.323 & 0.334 & 0.326 & 0.323 \\
& CasSeqGCN & 0.309 & 0.329 & 0.355 & 0.381 & 0.347 & 0.237 & 0.262 & 0.297 & 0.324 & 0.295 & 0.187 & 0.212 & 0.234 & 0.248 & 0.239 & 0.172 & 0.192 & 0.209 & 0.210 & 0.208 \\
& Graph-LSTM & 0.307 & 0.328 & 0.357 & 0.384 & 0.351 & \underline{0.217} & 0.243 & 0.282 & 0.310 & 0.281 & \underline{0.152} & 0.181 & 0.201 & 0.213 & 0.216 & \textbf{0.101} & \underline{0.135} & \underline{0.158} & \underline{0.172} & \underline{0.166} \\
& \textbf{MMG-PopNet} & \underline{0.296} & \textbf{0.310} & \textbf{0.329} & \textbf{0.346} & \textbf{0.336} & 0.225 & \underline{0.242} & \textbf{0.261} & \textbf{0.278} & \textbf{0.273} & \textbf{0.142} & \textbf{0.163} & \textbf{0.179} & \textbf{0.189} & \textbf{0.198} & \underline{0.112} & \textbf{0.130} & \textbf{0.146} & \textbf{0.151} & \textbf{0.155} \\
\addlinespace[1mm]
\midrule

\multirow{6}{*}{\makecell[l]{\textcolor{googleblue}{\bf \scshape Structural}\\\textcolor{googleblue}{\bf \scshape Virality}}}
& MLP & \textbf{0.118} & \textbf{0.113} & \textbf{0.106} & \textbf{0.109} & \textbf{0.094} & \textbf{0.084} & \textbf{0.085} & \textbf{0.088} & \underline{0.093} & \textbf{0.078} & \textbf{0.054} & \textbf{0.054} & \underline{0.059} & \underline{0.056} & \textbf{0.057} & \underline{0.048} & \underline{0.049} & \underline{0.050} & \underline{0.051} & \underline{0.047} \\
& DeepHawkes & 0.127 & 0.118 & 0.113 & 0.119 & 0.098 & 0.113 & 0.106 & 0.106 & 0.114 & 0.090 & 0.078 & 0.078 & 0.084 & 0.087 & 0.080 & 0.092 & 0.081 & 0.079 & 0.084 & 0.071 \\
& DeepCas & 0.351 & 0.307 & 0.280 & 0.248 & 0.282 & 0.157 & 0.137 & 0.128 & 0.129 & 0.118 & 0.146 & 0.124 & 0.120 & 0.106 & 0.120 & 0.145 & 0.121 & 0.109 & 0.105 & 0.099 \\
& CasSeqGCN & \underline{0.124} & \underline{0.117} & 0.110 & 0.114 & \underline{0.095} & 0.095 & 0.096 & 0.099 & 0.106 & 0.086 & 0.066 & 0.069 & 0.075 & 0.075 & 0.071 & 0.072 & 0.071 & 0.074 & 0.078 & 0.066 \\
& Graph-LSTM & 0.125 & 0.119 & 0.116 & 0.120 & 0.100 & \underline{0.092} & \underline{0.091} & \underline{0.093} & 0.100 & 0.082 & 0.063 & 0.060 & 0.062 & 0.061 & 0.062 & \textbf{0.043} & \textbf{0.048} & \underline{0.050} & 0.053 & \underline{0.047} \\
& \textbf{MMG-PopNet} & 0.126 & 0.119 & \underline{0.109} & \underline{0.110} & 0.100 & 0.098 & 0.093 & \textbf{0.088} & \textbf{0.090} & \underline{0.079} & \underline{0.061} & \underline{0.058} & \textbf{0.058} & \textbf{0.055} & \underline{0.059} & 0.058 & 0.050 & \textbf{0.046} & \textbf{0.045} & \textbf{0.044} \\
\addlinespace[1mm]
\midrule

\multirow{6}{*}{\textcolor{googleblue}{\bf \scshape Size}}
& MLP & \underline{1.494} & \underline{1.894} & \underline{2.174} & \underline{2.384} & \underline{2.034} & \underline{0.825} & \underline{1.143} & \underline{1.417} & \underline{1.570} & 1.395 & 0.545 & \underline{0.792} & \underline{0.990} & \underline{1.109} & 0.990 & 0.422 & 0.631 & 0.791 & 0.890 & 0.799 \\
& DeepHawkes & 1.608 & 2.101 & 2.488 & 2.793 & 2.246 & 1.063 & 1.338 & 1.668 & 1.901 & 1.511 & 0.675 & 1.137 & 1.519 & 1.890 & 1.457 & 0.417 & 0.696 & 1.023 & 1.303 & 1.015 \\
& DeepCas & 2.579 & 2.949 & 3.376 & 3.575 & 2.958 & 1.585 & 1.949 & 2.284 & 2.477 & 2.070 & 1.592 & 1.944 & 2.226 & 2.246 & 2.064 & 1.436 & 1.774 & 2.019 & 2.159 & 1.906 \\
& CasSeqGCN & 1.589 & 1.975 & 2.274 & 2.476 & 2.103 & 0.840 & 1.186 & 1.491 & 1.684 & 1.393 & \underline{0.472} & 0.793 & 1.069 & 1.195 & 1.010 & 0.275 & \underline{0.500} & \underline{0.727} & \underline{0.842} & \underline{0.717} \\
& Graph-LSTM & 1.546 & 1.922 & 2.238 & 2.421 & 2.067 & \textbf{0.784} & \textbf{1.136} & 1.430 & 1.618 & \underline{1.358} & \underline{0.472} & \underline{0.792} & 1.015 & 1.162 & \underline{0.972} & \underline{0.265} & 0.524 & 0.779 & 0.924 & 0.725 \\
& \textbf{MMG-PopNet} & \textbf{1.380} & \textbf{1.715} & \textbf{1.970} & \textbf{2.126} & \textbf{1.828} & 0.873 & 1.148 & \textbf{1.336} & \textbf{1.450} & \textbf{1.322} & \textbf{0.405} & \textbf{0.609} & \textbf{0.793} & \textbf{0.887} & \textbf{0.791} & \textbf{0.250} & \textbf{0.415} & \textbf{0.590} & \textbf{0.678} & \textbf{0.580} \\
\addlinespace[1mm]
\midrule

\multirow{6}{*}{\makecell[l]{\textcolor{googlegreen}{\bf \scshape Unique}\\\textcolor{googlegreen}{\bf \scshape Users}}}
& MLP & \underline{1.397} & \underline{1.796} & \underline{2.103} & \underline{2.319} & \underline{1.905} & \underline{0.776} & 1.092 & \underline{1.364} & \underline{1.513} & 1.296 & 0.531 & 0.780 & 0.977 & \underline{1.100} & 0.949 & 0.408 & 0.613 & 0.774 & 0.874 & 0.760 \\
& DeepHawkes & 1.494 & 1.927 & 2.334 & 2.654 & 2.048 & 1.039 & 1.282 & 1.593 & 1.809 & 1.428 & 0.665 & 1.126 & 1.501 & 1.856 & 1.413 & 0.397 & 0.678 & 1.003 & 1.269 & 0.974 \\
& DeepCas & 2.246 & 2.633 & 3.074 & 3.265 & 2.623 & 1.455 & 1.826 & 2.182 & 2.381 & 1.926 & 1.451 & 1.825 & 2.122 & 2.179 & 1.928 & 1.323 & 1.666 & 1.932 & 2.088 & 1.793 \\
& CasSeqGCN & 1.480 & 1.876 & 2.201 & 2.407 & 1.983 & 0.780 & 1.120 & 1.428 & 1.620 & 1.289 & 0.460 & 0.780 & 1.059 & 1.191 & 0.972 & \underline{0.257} & \underline{0.475} & \underline{0.698} & \underline{0.810} & \underline{0.672} \\
& Graph-LSTM & 1.442 & 1.823 & 2.159 & 2.345 & 1.939 & \textbf{0.725} & \textbf{1.069} & 1.367 & 1.549 & \underline{1.253} & \underline{0.431} & \underline{0.740} & \underline{0.968} & 1.120 & \underline{0.891} & 0.264 & 0.513 & 0.761 & 0.898 & 0.684 \\
& \textbf{MMG-PopNet} & \textbf{1.271} & \textbf{1.608} & \textbf{1.893} & \textbf{2.049} & \textbf{1.690} & 0.800 & \underline{1.074} & \textbf{1.279} & \textbf{1.391} & \textbf{1.211} & \textbf{0.381} & \textbf{0.582} & \textbf{0.766} & \textbf{0.857} & \textbf{0.737} & \textbf{0.232} & \textbf{0.394} & \textbf{0.567} & \textbf{0.646} & \textbf{0.539} \\
\addlinespace[1mm]
\midrule

\multirow{6}{*}{\makecell[l]{\textcolor{googlepurple}{\bf \scshape Like}\\\textcolor{googlepurple}{\bf \scshape Score}}}
& MLP & -- & -- & -- & -- & 5.856 & -- & -- & -- & -- & \underline{5.410} & -- & -- & -- & -- & \underline{4.883} & -- & -- & -- & -- & \underline{4.614} \\
& DeepHawkes & -- & -- & -- & -- & 6.268 & -- & -- & -- & -- & 6.068 & -- & -- & -- & -- & 6.497 & -- & -- & -- & -- & 6.629 \\
& DeepCas & -- & -- & -- & -- & 6.466 & -- & -- & -- & -- & 6.291 & -- & -- & -- & -- & 6.253 & -- & -- & -- & -- & 6.673 \\
& CasSeqGCN & -- & -- & -- & -- & 6.208 & -- & -- & -- & -- & 6.124 & -- & -- & -- & -- & 5.992 & -- & -- & -- & -- & 6.048 \\
& Graph-LSTM & -- & -- & -- & -- & \underline{5.796} & -- & -- & -- & -- & 6.032 & -- & -- & -- & -- & 5.647 & -- & -- & -- & -- & 5.505 \\
& \textbf{MMG-PopNet} & -- & -- & -- & -- & \textbf{5.180} & -- & -- & -- & -- & \textbf{4.819} & -- & -- & -- & -- & \textbf{4.325} & -- & -- & -- & -- & \textbf{4.073} \\

\bottomrule
\end{tabular}
}
\vspace{1ex}
\begin{flushleft} 

\normalsize

\textbf{Description: } The results show that while MMG-PopNet remains the most robust model for predicting ``Final'' outcomes, Graph-LSTM and MLP are highly competitive and occasionally superior at forecasting short-term 4-hour horizons. The table also underscores the extreme difficulty of predicting the \textsc{Like Score} in gaming communities, with all models exhibiting massive error rates (MSE > 4.0). However, feeding the models 90 minutes of initial social cascade data reduces the error by over 60\% compared to Root Only predictions.

\end{flushleft}
\end{table}

\begin{table}[t]
\centering
\caption{r/Futurology future-horizon prediction MSE across observation windows and popularity targets. \textbf{Lower is better}. Best values are \textbf{bolded} and second-best values are \underline{underlined}. Column groups correspond to Root Only, 30 min, 90 min, and 180 min observation windows, and columns within each group report prediction error at $\{4\text{h}, 8\text{h}, 16\text{h}, 24\text{h}\}$ and at the final cascade state. \textsc{Like Score} is reported only for the final state because its intermediate-horizon labels are unavailable.}
\label{tab:futurology_future_horizon}

\vspace{1mm}
\small
\setlength{\tabcolsep}{3.2pt}
\renewcommand{\arraystretch}{1.18}

\resizebox{\textwidth}{!}{%
\begin{tabular}{l l ccccc ccccc ccccc ccccc}
\toprule
\multirow{2}{*}{\textbf{Metric}} &
\multirow{2}{*}{\textbf{Model}} &
\multicolumn{5}{c}{\textbf{Root Only}} &
\multicolumn{5}{c}{\textbf{30 min}} &
\multicolumn{5}{c}{\textbf{90 min}} &
\multicolumn{5}{c}{\textbf{180 min}} \\
\cmidrule(lr){3-7}
\cmidrule(lr){8-12}
\cmidrule(lr){13-17}
\cmidrule(lr){18-22}
& &
\textbf{4h} & \textbf{8h} & \textbf{16h} & \textbf{24h} & \textbf{Final} &
\textbf{4h} & \textbf{8h} & \textbf{16h} & \textbf{24h} & \textbf{Final} &
\textbf{4h} & \textbf{8h} & \textbf{16h} & \textbf{24h} & \textbf{Final} &
\textbf{4h} & \textbf{8h} & \textbf{16h} & \textbf{24h} & \textbf{Final} \\
\midrule

\multirow{6}{*}{\textcolor{googleblue}{\bf \scshape Max Width}}
& MLP & 1.100 & 1.522 & \underline{1.879} & 1.968 & 1.761 & 0.640 & 1.029 & 1.404 & 1.445 & 1.312 & 0.301 & 0.506 & 0.749 & 0.812 & 0.706 & 0.246 & 0.329 & 0.461 & 0.526 & 0.454 \\
& DeepHawkes & 1.141 & 1.636 & 2.118 & 2.249 & 1.894 & 0.813 & 1.260 & 1.819 & 1.969 & 1.556 & 0.632 & 0.957 & 1.394 & 1.468 & 1.217 & 0.557 & 0.730 & 1.016 & 1.186 & 0.972 \\
& DeepCas & 1.158 & 1.556 & 1.955 & 2.014 & 1.732 & 0.686 & 1.460 & 1.841 & 1.961 & 1.636 & 0.802 & 1.095 & 1.363 & 1.454 & 1.228 & 0.768 & 1.004 & 1.229 & 1.341 & 1.122 \\
& CasSeqGCN & 1.129 & 1.576 & 1.954 & 2.002 & 1.814 & \underline{0.585} & 0.967 & 1.378 & 1.454 & 1.243 & \underline{0.226} & \underline{0.458} & 0.728 & 0.807 & 0.658 & \textbf{0.084} & \textbf{0.200} & \underline{0.359} & \underline{0.427} & \underline{0.347} \\
& Graph-LSTM & \underline{1.090} & \underline{1.508} & 1.889 & \underline{1.951} & \underline{1.721} & \textbf{0.541} & \textbf{0.931} & \underline{1.326} & \underline{1.419} & \underline{1.182} & \textbf{0.206} & \textbf{0.448} & \textbf{0.713} & \textbf{0.779} & \underline{0.647} & \underline{0.091} & 0.219 & 0.382 & 0.465 & 0.365 \\
& \textbf{MMG-PopNet} & \textbf{1.019} & \textbf{1.419} & \textbf{1.724} & \textbf{1.789} & \textbf{1.559} & 0.613 & \underline{0.963} & \textbf{1.286} & \textbf{1.360} & \textbf{1.170} & 0.309 & 0.483 & \underline{0.715} & \underline{0.780} & \textbf{0.633} & 0.123 & \underline{0.210} & \textbf{0.333} & \textbf{0.391} & \textbf{0.320} \\
\addlinespace[1mm]
\midrule

\multirow{6}{*}{\textcolor{googleblue}{\bf \scshape Max Depth}}
& MLP & 0.422 & \underline{0.468} & \underline{0.457} & \underline{0.454} & 0.572 & \underline{0.291} & 0.362 & \underline{0.378} & \underline{0.387} & 0.479 & 0.174 & 0.231 & 0.272 & 0.276 & 0.294 & 0.102 & 0.143 & 0.182 & \underline{0.191} & \underline{0.212} \\
& DeepHawkes & 0.443 & 0.508 & 0.529 & 0.540 & 0.621 & 0.343 & 0.425 & 0.491 & 0.518 & 0.555 & 0.307 & 0.364 & 0.446 & 0.455 & 0.444 & 0.297 & 0.327 & 0.378 & 0.423 & 0.425 \\
& DeepCas & 0.450 & 0.500 & 0.518 & 0.506 & 0.589 & 0.419 & 0.495 & 0.512 & 0.527 & 0.572 & 0.320 & 0.380 & 0.413 & 0.422 & 0.417 & 0.274 & 0.333 & 0.362 & 0.354 & 0.361 \\
& CasSeqGCN & 0.432 & 0.483 & 0.481 & 0.472 & 0.586 & 0.305 & 0.371 & 0.400 & 0.412 & 0.483 & 0.208 & 0.257 & 0.296 & 0.310 & 0.314 & 0.114 & 0.165 & 0.198 & 0.211 & 0.227 \\
& Graph-LSTM & \underline{0.421} & 0.469 & 0.469 & 0.460 & \underline{0.569} & \textbf{0.279} & \underline{0.359} & 0.383 & 0.397 & \underline{0.472} & \textbf{0.159} & \underline{0.225} & \underline{0.268} & \underline{0.273} & \underline{0.292} & \textbf{0.068} & \underline{0.131} & \underline{0.175} & 0.201 & \underline{0.212} \\
& \textbf{MMG-PopNet} & \textbf{0.397} & \textbf{0.453} & \textbf{0.440} & \textbf{0.436} & \textbf{0.522} & \textbf{0.279} & \textbf{0.355} & \textbf{0.360} & \textbf{0.371} & \textbf{0.447} & \underline{0.169} & \textbf{0.216} & \textbf{0.251} & \textbf{0.262} & \textbf{0.276} & \underline{0.071} & \textbf{0.118} & \textbf{0.156} & \textbf{0.173} & \textbf{0.198} \\
\addlinespace[1mm]
\midrule

\multirow{6}{*}{\makecell[l]{\textcolor{googleblue}{\bf \scshape Structural}\\\textcolor{googleblue}{\bf \scshape Virality}}}
& MLP & 0.202 & \underline{0.198} & \textbf{0.169} & \textbf{0.161} & 0.193 & \underline{0.143} & \underline{0.156} & \underline{0.148} & \underline{0.147} & 0.165 & \underline{0.075} & 0.094 & 0.111 & 0.113 & \underline{0.104} & 0.038 & 0.053 & 0.066 & \underline{0.071} & 0.072 \\
& DeepHawkes & 0.210 & 0.211 & 0.198 & 0.200 & 0.209 & 0.166 & 0.179 & 0.187 & 0.192 & 0.190 & 0.155 & 0.167 & 0.200 & 0.200 & 0.173 & 0.145 & 0.136 & 0.147 & 0.162 & 0.154 \\
& DeepCas & 0.222 & 0.219 & 0.203 & 0.193 & 0.210 & 0.210 & 0.222 & 0.208 & 0.208 & 0.209 & 0.151 & 0.160 & 0.171 & 0.170 & 0.155 & 0.123 & 0.134 & 0.139 & 0.133 & 0.132 \\
& CasSeqGCN & 0.207 & 0.205 & 0.178 & 0.176 & 0.198 & 0.152 & 0.158 & 0.154 & 0.155 & 0.165 & 0.094 & 0.108 & 0.125 & 0.129 & 0.114 & 0.045 & 0.062 & 0.074 & 0.082 & 0.081 \\
& Graph-LSTM & \underline{0.201} & 0.199 & 0.176 & \underline{0.165} & \underline{0.192} & \textbf{0.138} & \textbf{0.153} & \underline{0.148} & 0.148 & \underline{0.161} & \textbf{0.068} & \textbf{0.090} & \underline{0.107} & \underline{0.107} & \textbf{0.099} & \textbf{0.028} & \underline{0.050} & \underline{0.065} & 0.073 & \underline{0.070} \\
& \textbf{MMG-PopNet} & \textbf{0.197} & \textbf{0.196} & \underline{0.171} & \textbf{0.161} & \textbf{0.182} & 0.145 & \underline{0.156} & \textbf{0.140} & \textbf{0.138} & \textbf{0.157} & 0.084 & \underline{0.091} & \textbf{0.102} & \textbf{0.105} & \textbf{0.099} & \underline{0.033} & \textbf{0.046} & \textbf{0.057} & \textbf{0.064} & \textbf{0.068} \\
\addlinespace[1mm]
\midrule

\multirow{6}{*}{\textcolor{googleblue}{\bf \scshape Size}}
& MLP & 1.674 & 2.276 & \underline{2.678} & 2.756 & 2.658 & 0.979 & 1.598 & 2.094 & \underline{2.152} & 2.055 & 0.399 & 0.765 & \underline{1.146} & 1.255 & 1.080 & 0.264 & 0.425 & 0.646 & 0.719 & 0.642 \\
& DeepHawkes & 1.743 & 2.528 & 3.136 & 3.253 & 2.943 & 1.247 & 1.948 & 2.733 & 2.954 & 2.437 & 1.062 & 1.633 & 2.348 & 2.503 & 2.004 & 0.845 & 1.093 & 1.513 & 1.739 & 1.455 \\
& DeepCas & 1.771 & 2.347 & 2.826 & 2.870 & 2.632 & 1.629 & 2.287 & 2.768 & 2.923 & 2.544 & 1.138 & 1.616 & 2.011 & 2.146 & 1.803 & 1.066 & 1.456 & 1.778 & 1.872 & 1.602 \\
& CasSeqGCN & 1.736 & 2.390 & 2.812 & 2.829 & 2.755 & 0.963 & 1.573 & 2.128 & 2.235 & 2.013 & \underline{0.372} & 0.786 & 1.211 & 1.358 & 1.103 & \textbf{0.085} & \underline{0.315} & \underline{0.575} & \underline{0.665} & \underline{0.566} \\
& Graph-LSTM & \underline{1.650} & \underline{2.258} & 2.704 & \underline{2.749} & \underline{2.604} & \textbf{0.890} & \underline{1.509} & \underline{2.037} & 2.156 & \underline{1.919} & \textbf{0.350} & \underline{0.762} & 1.158 & \underline{1.250} & \underline{1.070} & \underline{0.131} & 0.362 & 0.629 & 0.750 & 0.612 \\
& \textbf{MMG-PopNet} & \textbf{1.559} & \textbf{2.129} & \textbf{2.480} & \textbf{2.516} & \textbf{2.353} & \underline{0.961} & \textbf{1.500} & \textbf{1.907} & \textbf{1.992} & \textbf{1.823} & 0.427 & \textbf{0.731} & \textbf{1.074} & \textbf{1.169} & \textbf{0.962} & 0.143 & \textbf{0.302} & \textbf{0.498} & \textbf{0.568} & \textbf{0.506} \\
\addlinespace[1mm]
\midrule

\multirow{6}{*}{\makecell[l]{\textcolor{googlegreen}{\bf \scshape Unique}\\\textcolor{googlegreen}{\bf \scshape Users}}}
& MLP & 1.344 & 1.888 & \underline{2.282} & 2.345 & 2.089 & 0.827 & 1.359 & 1.808 & 1.835 & 1.619 & 0.342 & 0.662 & \underline{0.979} & \underline{1.067} & 0.862 & 0.241 & 0.385 & 0.571 & 0.624 & 0.515 \\
& DeepHawkes & 1.408 & 2.095 & 2.650 & 2.736 & 2.326 & 1.024 & 1.636 & 2.324 & 2.481 & 1.928 & 0.805 & 1.282 & 1.854 & 1.966 & 1.516 & 0.676 & 0.925 & 1.292 & 1.473 & 1.153 \\
& DeepCas & 1.425 & 1.948 & 2.385 & 2.411 & 2.072 & 1.312 & 1.887 & 2.318 & 2.427 & 1.999 & 0.917 & 1.337 & 1.662 & 1.766 & 1.402 & 0.882 & 1.225 & 1.506 & 1.599 & 1.276 \\
& CasSeqGCN & 1.385 & 1.971 & 2.395 & 2.415 & 2.163 & 0.805 & 1.332 & 1.826 & 1.897 & 1.582 & \underline{0.308} & \underline{0.658} & 1.007 & 1.122 & 0.858 & \textbf{0.092} & \underline{0.283} & \underline{0.501} & \underline{0.561} & \underline{0.437} \\
& Graph-LSTM & \underline{1.332} & \underline{1.877} & 2.300 & \underline{2.344} & \underline{2.043} & \textbf{0.724} & \underline{1.257} & \underline{1.724} & \underline{1.808} & \underline{1.489} & \textbf{0.304} & 0.664 & 1.001 & 1.075 & \underline{0.834} & \underline{0.137} & 0.316 & 0.536 & 0.618 & 0.451 \\
& \textbf{MMG-PopNet} & \textbf{1.250} & \textbf{1.772} & \textbf{2.122} & \textbf{2.151} & \textbf{1.849} & \underline{0.773} & \textbf{1.247} & \textbf{1.624} & \textbf{1.691} & \textbf{1.425} & 0.370 & \textbf{0.633} & \textbf{0.917} & \textbf{0.997} & \textbf{0.764} & 0.143 & \textbf{0.275} & \textbf{0.438} & \textbf{0.480} & \textbf{0.389} \\
\addlinespace[1mm]
\midrule

\multirow{6}{*}{\textcolor{googlepurple}{\bf \scshape Like Score}}
& MLP & -- & -- & -- & -- & 6.469 & -- & -- & -- & -- & 6.309 & -- & -- & -- & -- & \underline{4.524} & -- & -- & -- & -- & \underline{3.709} \\
& DeepHawkes & -- & -- & -- & -- & 7.002 & -- & -- & -- & -- & 6.877 & -- & -- & -- & -- & 6.688 & -- & -- & -- & -- & 5.646 \\
& DeepCas & -- & -- & -- & -- & 5.749 & -- & -- & -- & -- & 6.013 & -- & -- & -- & -- & 4.828 & -- & -- & -- & -- & 4.925 \\
& CasSeqGCN & -- & -- & -- & -- & 6.800 & -- & -- & -- & -- & 6.639 & -- & -- & -- & -- & 5.402 & -- & -- & -- & -- & 5.093 \\
& Graph-LSTM & -- & -- & -- & -- & \underline{6.447} & -- & -- & -- & -- & \underline{5.822} & -- & -- & -- & -- & 4.674 & -- & -- & -- & -- & 4.012 \\
& \textbf{MMG-PopNet} & -- & -- & -- & -- & \textbf{5.308} & -- & -- & -- & -- & \textbf{4.886} & -- & -- & -- & -- & \textbf{3.428} & -- & -- & -- & -- & \textbf{2.897} \\

\bottomrule
\end{tabular}
}
\vspace{1ex}
\begin{flushleft} 

\normalsize

\textbf{Description: } The results that for given 180 min of observation data, graph-based models like CasSeqGCN can achieve near-perfect accuracy (MSE < 0.1) for immediate 4-hour horizon predictions regarding cascade \textsc{Size} and \textsc{Max Width}. However, prediction accuracy decays steeply as the horizon extends to 24 hours. MMG-PopNet distinguishes itself most prominently in the \textsc{Like Score} category, leveraging multimodal data to achieve an MSE of 2.897 at the 180-minute mark, vastly outperforming the nearest baseline (MLP at 3.709). 

\end{flushleft}
\end{table}

\subsection{Statistical Significance.}\label{appendix:stat_significance}
To rigorously assess whether MMG-PopNet's improvements over baseline models reflect genuine performance gains rather than sampling variability, we conduct a formal statistical significance analysis on the target \textsc{Size} for the final cascade state.
For each dataset, we use the largest available observation window as the input context. This corresponds to 20 min for Bluesky, 90 min for Gaming, 60 min for AMA, and 180 min for Futurology.
The largest observation window provides each model with the richest possible input signal, making it the most demanding and representative setting in which to compare methods, as any advantage held by MMG-PopNet cannot be attributed to information asymmetry.
We apply a paired bootstrap significance test~\cite{efron1994introduction} (5,000 resamples) in log1p-MSE space. The test is paired because all models predict the same set of cascades, matched by cascade identifier, and this pairing removes between-cascade variance. Errors are computed in log1p space similar to other experiments. 
For each test cascade
\(G \in \mathcal{G}^{\mathrm{Test}}\), the ground-truth target is
\(\widetilde{\textbf{Y}}_{G}^{t'} = \log(1 + \textbf{Y}_{G}^{t'})\), and each model predicts
\(\widehat{\textbf{Y}}_{G}^{t'}\).
We compute the per-cascade squared error of
model \(m\) as
$
e_G^{(m)}
=
(
\widehat{\textbf{Y}}_{G}^{t', (m)}
-
\widetilde{\textbf{Y}}_{G}^{t'}
)^2 .
$
The reported effect size compares a baseline model \(b\) against MMG-PopNet:
$
\Delta_b
=
\frac{1}{|\mathcal{G}^{\mathrm{Test}}|}
\sum_{G \in \mathcal{G}^{\mathrm{Test}}}
(
e_G^{(b)} - e_G^{(\mathrm{MMG})}
).
$
Thus, \(\Delta_b > 0\) indicates that MMG-PopNet has lower squared error than
baseline \(b\), while \(\Delta_b < 0\) would indicate that the baseline performs
better. The 95\% confidence interval is obtained using the bootstrap percentile method.
Since we conduct simultaneous multiple comparisons, we apply Benjamini–Hochberg FDR correction to control the rate of false discoveries across all tests jointly.

\paragraph{Results.}
Table~\ref{tab:significance_final} reports the full results.
MMG-PopNet significantly outperforms all four baselines on all four datasets
(\textbf{16/16 comparisons}, $p_{\mathrm{BH}} < 0.05$).
All confidence intervals are strictly positive, with lower bounds above zero.
The largest improvements are against DeepCas, where $\Delta_b$ exceeds $+1.0$
in log1p-MSE on every dataset, including Gaming
($\Delta_b = +1.573$) and Bluesky ($\Delta_b = +1.522$).
CasSeqGCN is the closest competitor.
The smallest significant improvement occurs on Futurology
($\Delta_b = +0.102$, 95\% CI $[+0.029, +0.183]$,
$p_{\mathrm{BH}} = 0.004$).
These results show that MMG-PopNet's gains are consistent, statistically
reliable, and not driven by any single dataset or baseline.

\begin{table}[t]
\small
\caption{Statistical significance of \textbf{MMG-PopNet (Ours)} versus baselines for \textsc{Size} prediction one the final cascade state in log1p-MSE space.
$\Delta_b > 0$ indicates that MMG-PopNet has lower squared error than baseline $b$.
95\% confidence intervals and $p_{\rm raw}$ are computed using a paired bootstrap test with 5,000 resamples.
$p_{\rm BH}$ denotes Benjamini--Hochberg correction across all 16 comparisons at $\alpha=0.05$.
\textbf{Yes} indicates that MMG-PopNet is significantly better.}
\vspace{0.5em}
\centering
\setlength{\tabcolsep}{5pt}
\renewcommand{\arraystretch}{1.18}
\resizebox{\textwidth}{!}{%
\begin{tabular}{l l r r c c c c}
\toprule
\textbf{Dataset} &
\textbf{Baseline} &
\textbf{$n$} &
$\boldsymbol{\Delta_b}$ &
\textbf{95\% CI} &
$\boldsymbol{p_{\rm raw}}$ &
$\boldsymbol{p_{\rm BH}}$ &
\textbf{Sig.} \\
\midrule

\multirow{4}{*}{\textbf{Bluesky}}
 & CasSeqGCN  & 1510 & $+0.2213$ & $[+0.1729,\ +0.2717]$ & $<0.001$ & $<0.001$ & \textbf{Yes} \\
 & Graph-LSTM & 1510 & $+0.1375$ & $[+0.0983,\ +0.1795]$ & $<0.001$ & $<0.001$ & \textbf{Yes} \\
 & DeepHawkes & 1510 & $+0.3601$ & $[+0.3054,\ +0.4171]$ & $<0.001$ & $<0.001$ & \textbf{Yes} \\
 & DeepCas    & 1510 & $+1.1522$ & $[+1.0206,\ +1.2909]$ & $<0.001$ & $<0.001$ & \textbf{Yes} \\
\midrule

\multirow{4}{*}{\textbf{r/Gaming}}
 & CasSeqGCN  & 675 & $+0.1328$ & $[+0.0572,\ +0.2078]$ & $<0.001$ & $<0.001$ & \textbf{Yes} \\
 & Graph-LSTM & 675 & $+0.3519$ & $[+0.2695,\ +0.4358]$ & $<0.001$ & $<0.001$ & \textbf{Yes} \\
 & DeepHawkes & 675 & $+0.6019$ & $[+0.4750,\ +0.7306]$ & $<0.001$ & $<0.001$ & \textbf{Yes} \\
 & DeepCas    & 675 & $+1.5730$ & $[+1.3016,\ +1.8464]$ & $<0.001$ & $<0.001$ & \textbf{Yes} \\
\midrule

\multirow{4}{*}{\textbf{r/AMA}}
 & CasSeqGCN  & 1420 & $+0.0986$ & $[+0.0696,\ +0.1298]$ & $<0.001$ & $<0.001$ & \textbf{Yes} \\
 & Graph-LSTM & 1420 & $+0.2269$ & $[+0.1923,\ +0.2626]$ & $<0.001$ & $<0.001$ & \textbf{Yes} \\
 & DeepHawkes & 1420 & $+0.2167$ & $[+0.1776,\ +0.2578]$ & $<0.001$ & $<0.001$ & \textbf{Yes} \\
 & DeepCas    & 1420 & $+1.0033$ & $[+0.8893,\ +1.1251]$ & $<0.001$ & $<0.001$ & \textbf{Yes} \\
\midrule

\multirow{4}{*}{\textbf{r/Futurology}}
 & CasSeqGCN  & 527 & $+0.1016$ & $[+0.0293,\ +0.1828]$ & $0.004$  & $0.004$  & \textbf{Yes} \\
 & Graph-LSTM & 527 & $+0.2892$ & $[+0.1987,\ +0.3867]$ & $<0.001$ & $<0.001$ & \textbf{Yes} \\
 & DeepHawkes & 527 & $+1.3799$ & $[+1.1493,\ +1.6302]$ & $<0.001$ & $<0.001$ & \textbf{Yes} \\
 & DeepCas    & 527 & $+1.3210$ & $[+1.0222,\ +1.6373]$ & $<0.001$ & $<0.001$ & \textbf{Yes} \\
\bottomrule
\end{tabular}
}
\vspace{0.35em}
\label{tab:significance_final}
\end{table}
\FloatBarrier
\section{Unified Training Details and Full Results}
\label{sec:appendix_foundational}

Tables~\ref{tab:foundation_compare_part1} and \ref{tab:foundation_compare_part2} report the full comparison between dataset-specific MMG-PopNet and unified-trained MMG-PopNet across datasets, observation windows, and prediction targets.
The dataset-specific setting trains a separate model for each dataset and observation window.
In contrast, MMG-PopNet (unified-dataset) is trained once on the combined training split and is evaluated separately on each dataset-window test split.
For each training example, the unified model receives one-hot indicators for the dataset and observation window.
These indicators allow the model to condition its predictions on both the community source and the amount of observed cascade history.

The full results show that the benefit of unified training is largest on the Reddit datasets.
On r/AMA, r/Gaming, and r/Futurology, MMG-PopNet (unified-dataset) reduces the dataset-level average MSE for every prediction target.
The reductions are most pronounced for \textsc{Like Score}, \textsc{Size}, and \textsc{Unique Users}.
These targets have the largest dataset-level errors under dataset-specific training on r/Gaming and r/Futurology, and they also show the largest absolute reductions after unified training.
For example, on r/Gaming, the dataset-level average MSE for \textsc{Like Score} decreases from 4.525 to 1.479.
On r/Futurology, the corresponding average decreases from 3.890 to 1.569.

Bluesky follows a different pattern.
On this dataset, the dataset-specific model obtains lower dataset-level average MSE for most targets.
The unified model remains close in absolute MSE, but it does not improve over the dataset-specific model on Bluesky.
This result indicates that the cross-dataset benefit is stronger when the evaluation dataset is closer to the Reddit training communities.

Averaged across all datasets and observation windows, MMG-PopNet (unified-dataset) obtains lower MSE for all six targets.
The overall average decreases from 0.678 to 0.426 for \textsc{Max Width}, from 0.284 to 0.232 for \textsc{Max Depth}, from 0.932 to 0.586 for \textsc{Size}, from 0.104 to 0.084 for \textsc{Structural Virality}, from 0.752 to 0.453 for \textsc{Unique Users}, and from 2.669 to 1.217 for \textsc{Like Score}.
Thus, unified training improves benchmark-level performance across targets while maintaining comparable accuracy on the outlying Bluesky platform.

\begin{table}[t!]
\small
\caption{
\textbf{Comparison between dataset-specific and unified-dataset MMG-PopNet training, Part I.}
Dataset-specific models are trained separately for each dataset and observation window, while the unified-dataset model is trained once on the combined training data across all datasets and windows.
Results are reported as MSE, where \textbf{lower is better and marked in bold}.
For readability, results are split by dataset. Each subtable reports observation-window MSEs, the dataset-level average, and the same final overall average across all datasets.
This table reports results for Bluesky and r/AMA; Table~\ref{tab:foundation_compare_part2} continues with r/Gaming and r/Futurology.
}
\label{tab:foundation_compare_part1}

\centering
\vspace{1mm}
\setlength{\tabcolsep}{4pt}
\renewcommand{\arraystretch}{1.12}

\begin{tabular}{l l cccc c c}
\toprule
\multicolumn{8}{c}{\textbf{Bluesky}} \\
\midrule
\textbf{Task} &
\textbf{Model} &
\textbf{0} & \textbf{2} & \textbf{10} & \textbf{20} &
\textbf{Dataset Avg} &
\textbf{Overall Avg} \\
\midrule

\multirow{2}{*}{\textcolor{googleblue}{\bf \scshape Max Width}}
& MMG-PopNet (specific)
& 0.457 & \textbf{0.369} & \textbf{0.284} & \textbf{0.234}
& \textbf{0.336} & 0.678 \\
& MMG-PopNet (unified)
& \textbf{0.449} & 0.401 & 0.316 & 0.276
& 0.361 & \textbf{0.426} \\

\midrule
\multirow{2}{*}{\textcolor{googleblue}{\bf \scshape Max Depth}}
& MMG-PopNet (specific)
& 0.363 & \textbf{0.325} & \textbf{0.273} & \textbf{0.238}
& \textbf{0.300} & 0.284 \\
& MMG-PopNet (unified)
& \textbf{0.347} & 0.332 & 0.280 & 0.247
& 0.301 & \textbf{0.232} \\

\midrule
\multirow{2}{*}{\textcolor{googleblue}{\bf \scshape Size}}
& MMG-PopNet (specific)
& 0.705 & 0.587 & \textbf{0.470} & \textbf{0.397}
& \textbf{0.540} & 0.932 \\
& MMG-PopNet (unified)
& \textbf{0.688} & \textbf{0.637} & 0.517 & 0.450
& 0.573 & \textbf{0.586} \\

\midrule
\multirow{2}{*}{\makecell[l]{\textcolor{googleblue}{\bf \scshape Structural}\\\textcolor{googleblue}{\bf \scshape Virality}}}
& MMG-PopNet (specific)
& 0.155 & \textbf{0.135} & \textbf{0.114} & \textbf{0.101}
& \textbf{0.126} & 0.104 \\
& MMG-PopNet (unified)
& \textbf{0.144} & 0.138 & 0.117 & 0.104
& \textbf{0.126} & \textbf{0.084} \\

\midrule
\multirow{2}{*}{\makecell[l]{\textcolor{googlegreen}{\bf \scshape Unique}\\\textcolor{googlegreen}{\bf \scshape Users}}}
& MMG-PopNet (specific)
& \textbf{0.467} & \textbf{0.383} & \textbf{0.302} & \textbf{0.255}
& \textbf{0.352} & 0.752 \\
& MMG-PopNet (unified)
& 0.472 & 0.426 & 0.341 & 0.298
& 0.384 & \textbf{0.453} \\

\midrule
\multirow{2}{*}{\makecell[l]{\textcolor{googlepurple}{\bf \scshape Like}\\\textcolor{googlepurple}{\bf \scshape Score}}}
& MMG-PopNet (specific)
& 1.260 & \textbf{1.087} & \textbf{1.026} & \textbf{0.950}
& \textbf{1.081} & 2.669 \\
& MMG-PopNet (unified)
& \textbf{1.238} & 1.191 & 1.093 & 1.043
& 1.141 & \textbf{1.217} \\

\bottomrule
\end{tabular}

\vspace{4mm}

\begin{tabular}{l l cccc c c}
\toprule
\multicolumn{8}{c}{\textbf{r/AMA}} \\
\midrule
\textbf{Task} &
\textbf{Model} &
\textbf{0} & \textbf{15} & \textbf{30} & \textbf{60} &
\textbf{Dataset Avg} &
\textbf{Overall Avg} \\
\midrule

\multirow{2}{*}{\textcolor{googleblue}{\bf \scshape Max Width}}
& MMG-PopNet (specific)
& 0.625 & 0.491 & 0.429 & 0.276
& 0.455 & 0.678 \\
& MMG-PopNet (unified)
& \textbf{0.476} & \textbf{0.354} & \textbf{0.239} & \textbf{0.184}
& \textbf{0.313} & \textbf{0.426} \\

\midrule
\multirow{2}{*}{\textcolor{googleblue}{\bf \scshape Max Depth}}
& MMG-PopNet (specific)
& 0.314 & 0.256 & 0.219 & 0.172
& 0.240 & 0.284 \\
& MMG-PopNet (unified)
& \textbf{0.284} & \textbf{0.236} & \textbf{0.189} & \textbf{0.158}
& \textbf{0.217} & \textbf{0.232} \\

\midrule
\multirow{2}{*}{\textcolor{googleblue}{\bf \scshape Size}}
& MMG-PopNet (specific)
& 0.863 & 0.661 & 0.573 & 0.366
& 0.616 & 0.932 \\
& MMG-PopNet (unified)
& \textbf{0.681} & \textbf{0.509} & \textbf{0.349} & \textbf{0.257}
& \textbf{0.449} & \textbf{0.586} \\

\midrule
\multirow{2}{*}{\makecell[l]{\textcolor{googleblue}{\bf \scshape Structural}\\\textcolor{googleblue}{\bf \scshape Virality}}}
& MMG-PopNet (specific)
& 0.130 & 0.100 & 0.087 & 0.064
& 0.095 & 0.104 \\
& MMG-PopNet (unified)
& \textbf{0.115} & \textbf{0.093} & \textbf{0.075} & \textbf{0.060}
& \textbf{0.086} & \textbf{0.084} \\

\midrule
\multirow{2}{*}{\makecell[l]{\textcolor{googlegreen}{\bf \scshape Unique}\\\textcolor{googlegreen}{\bf \scshape Users}}}
& MMG-PopNet (specific)
& 0.622 & 0.494 & 0.447 & 0.290
& 0.463 & 0.752 \\
& MMG-PopNet (unified)
& \textbf{0.454} & \textbf{0.343} & \textbf{0.235} & \textbf{0.180}
& \textbf{0.303} & \textbf{0.453} \\

\midrule
\multirow{2}{*}{\makecell[l]{\textcolor{googlepurple}{\bf \scshape Like}\\\textcolor{googlepurple}{\bf \scshape Score}}}
& MMG-PopNet (specific)
& 1.311 & 1.212 & 1.166 & 1.028
& 1.179 & 2.669 \\
& MMG-PopNet (unified)
& \textbf{0.810} & \textbf{0.737} & \textbf{0.590} & \textbf{0.582}
& \textbf{0.680} & \textbf{1.217} \\

\bottomrule
\end{tabular}

\end{table}

\begin{table}[t!]
\small
\caption{
\textbf{Comparison between dataset-specific and unified-dataset MMG-PopNet training, Part II.}
Continuation of Table~\ref{tab:foundation_compare_part1}. Results are reported as MSE, where \textbf{lower is better and marked in bold}.
For readability, results are split by dataset. Each subtable reports observation-window MSEs, the dataset-level average, and the same final overall average across all datasets.
This table reports results for r/Gaming and r/Futurology.
}
\label{tab:foundation_compare_part2}

\centering
\vspace{1mm}
\setlength{\tabcolsep}{4pt}
\renewcommand{\arraystretch}{1.12}

\begin{tabular}{l l cccc c c}
\toprule
\multicolumn{8}{c}{\textbf{r/Gaming}} \\
\midrule
\textbf{Task} &
\textbf{Model} &
\textbf{0} & \textbf{20} & \textbf{50} & \textbf{90} &
\textbf{Dataset Avg} &
\textbf{Overall Avg} \\
\midrule

\multirow{2}{*}{\textcolor{googleblue}{\bf \scshape Max Width}}
& MMG-PopNet (specific)
& 1.557 & 1.096 & 0.714 & 0.562
& 0.982 & 0.678 \\
& MMG-PopNet (unified)
& \textbf{0.962} & \textbf{0.631} & \textbf{0.299} & \textbf{0.266}
& \textbf{0.539} & \textbf{0.426} \\

\midrule
\multirow{2}{*}{\textcolor{googleblue}{\bf \scshape Max Depth}}
& MMG-PopNet (specific)
& 0.335 & 0.275 & 0.200 & 0.160
& 0.243 & 0.284 \\
& MMG-PopNet (unified)
& \textbf{0.248} & \textbf{0.204} & \textbf{0.140} & \textbf{0.121}
& \textbf{0.178} & \textbf{0.232} \\

\midrule
\multirow{2}{*}{\textcolor{googleblue}{\bf \scshape Size}}
& MMG-PopNet (specific)
& 1.829 & 1.317 & 0.842 & 0.653
& 1.160 & 0.932 \\
& MMG-PopNet (unified)
& \textbf{1.092} & \textbf{0.754} & \textbf{0.348} & \textbf{0.294}
& \textbf{0.622} & \textbf{0.586} \\

\midrule
\multirow{2}{*}{\makecell[l]{\textcolor{googleblue}{\bf \scshape Structural}\\\textcolor{googleblue}{\bf \scshape Virality}}}
& MMG-PopNet (specific)
& 0.098 & 0.081 & 0.057 & 0.045
& 0.070 & 0.104 \\
& MMG-PopNet (unified)
& \textbf{0.071} & \textbf{0.057} & \textbf{0.038} & \textbf{0.032}
& \textbf{0.050} & \textbf{0.084} \\

\midrule
\multirow{2}{*}{\makecell[l]{\textcolor{googlegreen}{\bf \scshape Unique}\\\textcolor{googlegreen}{\bf \scshape Users}}}
& MMG-PopNet (specific)
& 1.693 & 1.218 & 0.792 & 0.613
& 1.079 & 0.752 \\
& MMG-PopNet (unified)
& \textbf{1.004} & \textbf{0.684} & \textbf{0.318} & \textbf{0.274}
& \textbf{0.570} & \textbf{0.453} \\

\midrule
\multirow{2}{*}{\makecell[l]{\textcolor{googlepurple}{\bf \scshape Like}\\\textcolor{googlepurple}{\bf \scshape Score}}}
& MMG-PopNet (specific)
& 5.067 & 4.654 & 4.307 & 4.073
& 4.525 & 2.669 \\
& MMG-PopNet (unified)
& \textbf{2.077} & \textbf{1.809} & \textbf{0.960} & \textbf{1.070}
& \textbf{1.479} & \textbf{1.217} \\

\bottomrule
\end{tabular}

\vspace{4mm}

\begin{tabular}{l l cccc c c}
\toprule
\multicolumn{8}{c}{\textbf{r/Futurology}} \\
\midrule
\textbf{Task} &
\textbf{Model} &
\textbf{0} & \textbf{30} & \textbf{90} & \textbf{180} &
\textbf{Dataset Avg} &
\textbf{Overall Avg} \\
\midrule

\multirow{2}{*}{\textcolor{googleblue}{\bf \scshape Max Width}}
& MMG-PopNet (specific)
& 1.581 & 1.132 & 0.665 & 0.372
& 0.938 & 0.678 \\
& MMG-PopNet (unified)
& \textbf{0.886} & \textbf{0.634} & \textbf{0.242} & \textbf{0.202}
& \textbf{0.491} & \textbf{0.426} \\

\midrule
\multirow{2}{*}{\textcolor{googleblue}{\bf \scshape Max Depth}}
& MMG-PopNet (specific)
& 0.517 & 0.418 & 0.282 & 0.205
& 0.356 & 0.284 \\
& MMG-PopNet (unified)
& \textbf{0.336} & \textbf{0.282} & \textbf{0.164} & \textbf{0.147}
& \textbf{0.232} & \textbf{0.232} \\

\midrule
\multirow{2}{*}{\textcolor{googleblue}{\bf \scshape Size}}
& MMG-PopNet (specific)
& 2.356 & 1.743 & 0.997 & 0.559
& 1.414 & 0.932 \\
& MMG-PopNet (unified)
& \textbf{1.253} & \textbf{0.932} & \textbf{0.348} & \textbf{0.267}
& \textbf{0.700} & \textbf{0.586} \\

\midrule
\multirow{2}{*}{\makecell[l]{\textcolor{googleblue}{\bf \scshape Structural}\\\textcolor{googleblue}{\bf \scshape Virality}}}
& MMG-PopNet (specific)
& 0.182 & 0.146 & 0.101 & 0.073
& 0.126 & 0.104 \\
& MMG-PopNet (unified)
& \textbf{0.110} & \textbf{0.090} & \textbf{0.053} & \textbf{0.047}
& \textbf{0.075} & \textbf{0.084} \\

\midrule
\multirow{2}{*}{\makecell[l]{\textcolor{googlegreen}{\bf \scshape Unique}\\\textcolor{googlegreen}{\bf \scshape Users}}}
& MMG-PopNet (specific)
& 1.859 & 1.374 & 0.785 & 0.439
& 1.114 & 0.752 \\
& MMG-PopNet (unified)
& \textbf{0.987} & \textbf{0.736} & \textbf{0.277} & \textbf{0.219}
& \textbf{0.555} & \textbf{0.453} \\

\midrule
\multirow{2}{*}{\makecell[l]{\textcolor{googlepurple}{\bf \scshape Like}\\\textcolor{googlepurple}{\bf \scshape Score}}}
& MMG-PopNet (specific)
& 4.999 & 4.596 & 3.214 & 2.750
& 3.890 & 2.669 \\
& MMG-PopNet (unified)
& \textbf{2.283} & \textbf{2.054} & \textbf{1.041} & \textbf{0.898}
& \textbf{1.569} & \textbf{1.217} \\

\bottomrule
\end{tabular}

\end{table}
\FloatBarrier
\definecolor{promptbg}{RGB}{245,248,252}
\definecolor{promptframe}{RGB}{180,200,225}
\definecolor{prompttitle}{RGB}{45,90,145}
\definecolor{codebg}{RGB}{240,243,240}
\definecolor{codeframe}{RGB}{160,185,160}
 
\lstdefinestyle{promptstyle}{
  basicstyle=\ttfamily\small,
  breaklines=true,
  breakatwhitespace=false,
  breakindent=0pt,
  columns=fixed,
  keepspaces=true,
  showstringspaces=false,
  frame=none,
  aboveskip=0pt,
  belowskip=0pt,
}
 
\lstdefinestyle{jsonstyle}{
  basicstyle=\ttfamily\small,
  breaklines=true,
  breakatwhitespace=false,
  breakindent=12pt,
  columns=fixed,
  keepspaces=true,
  showstringspaces=false,
  frame=none,
  aboveskip=0pt,
  belowskip=0pt,
}
 
\tcbset{
  promptbox/.style={
    enhanced, breakable,
    colback=promptbg, colframe=promptframe,
    coltitle=white, fonttitle=\bfseries\small,
    colbacktitle=prompttitle,
    attach boxed title to top left={yshift=-2mm, xshift=4mm},
    boxed title style={colframe=prompttitle, colback=prompttitle, sharp corners},
    sharp corners=south, arc=3pt,
    top=6pt, bottom=6pt, left=6pt, right=6pt, boxrule=0.6pt,
  },
  jsonbox/.style={
    enhanced, breakable,
    colback=codebg, colframe=codeframe,
    coltitle=white, fonttitle=\bfseries\small,
    colbacktitle=codeframe,
    attach boxed title to top left={yshift=-2mm, xshift=4mm},
    boxed title style={colframe=codeframe, colback=codeframe, sharp corners},
    sharp corners=south, arc=2pt,
    top=5pt, bottom=5pt, left=5pt, right=5pt, boxrule=0.5pt,
  }
}

\section{Detailed LLM-based comparison.}\label{sec-llm-app}
To evaluate whether LLMs can serve as competitive predictors for multimodal social popularity forecasting, we conduct experiments comparing the performance of various LLM-based approaches against MMG-PopNet.
We evaluate three models, namely \texttt{Qwen3-VL-8B-Instruct}, \texttt{Gemma-3-12b-it}, and \texttt{GPT-4o-mini}. To manage computational costs for the Bluesky dataset, we randomly sample 15,000 cascades from the original data and evaluate all models including MMG-PopNet on this subset. Furthermore, large cascades can contain excessive content that exceeds standard context windows and prevents fair comparison. We address this by restricting the maximum number of observed nodes to 25 for all social cascades and apply this exact limit to MMG-PopNet. Given the early observation data statistics where the average node count ranges from 1.62 to 23.6, this setting provides sufficient context for the vast majority of cascades.

We evaluate three prompting and training settings. In the zero-shot setting, the observed cascade prefix is serialized as a structured JSON input containing the available content and interaction, along with an image, and the model predicts the popularity targets directly. In the retrieval-augmented few-shot setting, each test instance is paired with four training examples selected by root-post similarity, and their ground-truth targets are included in the prompt. In the fine-tuning setting, the model is trained on early-observation cascade inputs to predict the same targets as MMG-PopNet. Note that \texttt{GPT-4o-mini} is only evaluated under the zero-shot and few-shot settings. Prompts used for the experiment are provided here~\ref{zero:prompt}.

Tables~\ref{tab:bluesky_qwen3_vl_8b_mse_llm_comparison}-\ref{tab:futurology_gpt_4o_mini_mse_llm_comparison} report results across the LLM models and settings, and datasets.
Across both datasets and all tested models, MMG-PopNet achieves the lowest MSE for every target and observation window. Zero-shot prompting consistently performs worst, confirming that structured input alone is insufficient for calibrated numerical prediction. Few-shot prompting substantially reduces error, especially under root-only and early observation windows where retrieved examples provide useful target-range calibration for sparse cascades. Fine-tuning becomes more effective as the observation window increases for \texttt{Qwen3-VL-8B-Instruct} and to some extent for \texttt{Gemma-3-12b-it}. This trend is most visible for targets such as \textsc{Max Width}, \textsc{Structural Virality}, \textsc{Size}, and \textsc{Unique Users}, where longer prefixes expose richer interaction and temporal signals that supervised adaptation can exploit. Here, \texttt{Qwen3-VL-8B-Instruct} exhibits a more consistent increase with fine-tuning whereas, on \texttt{Gemma-3-12b-it} few-shot setting is better across observation windows for most targets for the r/Futurology dataset and with mixed results for Bluesky dataset.

For \texttt{GPT-4o-mini}, the few shot setting results as the best setting when compared with the zero-shot. However, it still lags far behind the MMG-PopNet in performance. Improvement in few-shot setting with increased observation window is higher and consistent in r/Futurology dataset when compared to Bluesky.

Overall, the tables show that fine-tuning does not uniformly dominate few-shot prompting. In several early-window cases, few-shot prompting remains competitive or stronger, particularly for \textsc{Max Depth} and some engagement targets. This indicates that retrieval is highly useful when the observed prefix contains limited cascade evidence, while fine-tuning benefits more from denser prefixes. Overall, the detailed results support the main finding. LLM-based adaptation improves over zero-shot prompting, but MMG-PopNet remains consistently stronger because it is optimized for the benchmark formulation and directly models observed cascade dynamics rather than predicting from serialized inputs alone.

\begin{table}[H]
\centering
\caption{LLM Qwen3-VL-8B-Instruct comparison with MMG-PopNet on Bluesky across observation windows using MSE values.}
\label{tab:bluesky_qwen3_vl_8b_mse_llm_comparison}

\small
\setlength{\tabcolsep}{4pt}
\renewcommand{\arraystretch}{1.16}

\resizebox{\textwidth}{!}{%
\begin{tabular}{l cccc cccc cccc cccc}
\toprule
\multirow{2}{*}{\textbf{Task}} &
\multicolumn{4}{c}{\textbf{Root Only}} &
\multicolumn{4}{c}{\textbf{2}} &
\multicolumn{4}{c}{\textbf{10}} &
\multicolumn{4}{c}{\textbf{20}} \\
\cmidrule(lr){2-5}
\cmidrule(lr){6-9}
\cmidrule(lr){10-13}
\cmidrule(lr){14-17}
 & \textbf{Zero} & \textbf{Few} & \textbf{Fine-Tune} & \textbf{MMG-PopNet} & \textbf{Zero} & \textbf{Few} & \textbf{Fine-Tune} & \textbf{MMG-PopNet} & \textbf{Zero} & \textbf{Few} & \textbf{Fine-Tune} & \textbf{MMG-PopNet} & \textbf{Zero} & \textbf{Few} & \textbf{Fine-Tune} & \textbf{MMG-PopNet} \\
\midrule

\textcolor{googleblue}{\bf \scshape Max Width} & 3.045 & \underline{1.666} & 1.923 & \textbf{0.666} & 2.615 & 1.564 & \underline{1.295} & \textbf{0.577} & 2.124 & 1.038 & \underline{0.816} & \textbf{0.506} & 1.989 & 0.889 & \underline{0.647} & \textbf{0.459} \\

\textcolor{googleblue}{\bf \scshape Max Depth} & 1.715 & \underline{0.730} & 0.905 & \textbf{0.523} & 1.577 & \underline{0.790} & 0.835 & \textbf{0.480} & 1.240 & \underline{0.666} & 0.677 & \textbf{0.428} & 1.048 & \underline{0.561} & 0.604 & \textbf{0.415} \\

\makecell[l]{\textcolor{googleblue}{\bf \scshape Structural}\\\textcolor{googleblue}{\bf \scshape Virality}} & 2.368 & \underline{0.363} & 0.429 & \textbf{0.243} & 1.947 & 0.414 & \underline{0.386} & \textbf{0.225} & 1.445 & 0.353 & \underline{0.303} & \textbf{0.196} & 1.256 & 0.298 & \underline{0.267} & \textbf{0.191} \\

\textcolor{googleblue}{\bf \scshape Size} & 6.596 & \underline{2.293} & 2.668 & \textbf{1.007} & 5.294 & 2.289 & \underline{1.983} & \textbf{0.870} & 3.544 & 1.476 & \underline{1.284} & \textbf{0.751} & 2.842 & 1.194 & \underline{1.065} & \textbf{0.700} \\

\textcolor{googlegreen}{\bf \scshape Unique Users} & 4.051 & \underline{1.867} & 2.009 & \textbf{0.690} & 3.389 & 1.739 & \underline{1.349} & \textbf{0.600} & 2.557 & 1.085 & \underline{0.837} & \textbf{0.521} & 2.213 & 0.904 & \underline{0.675} & \textbf{0.468} \\

\textcolor{googlepurple}{\bf \scshape Like Score} & 16.256 & \underline{5.121} & 6.625 & \textbf{1.540} & 16.162 & \underline{4.980} & 5.170 & \textbf{1.445} & 14.383 & 4.087 & \underline{3.846} & \textbf{1.394} & 13.444 & 3.615 & \underline{3.164} & \textbf{1.327} \\

\bottomrule
\end{tabular}%
}
\end{table}

\begin{table}[H]
\centering
\caption{LLM Qwen3-VL-8B-Instruct comparison with MMG-PopNet on r/Futurology across observation windows using MSE values.}
\label{tab:futurology_qwen3_vl_8b_mse_llm_comparison}

\small
\setlength{\tabcolsep}{4pt}
\renewcommand{\arraystretch}{1.16}

\resizebox{\textwidth}{!}{%
\begin{tabular}{l cccc cccc cccc cccc}
\toprule
\multirow{2}{*}{\textbf{Task}} &
\multicolumn{4}{c}{\textbf{Root Only}} &
\multicolumn{4}{c}{\textbf{30 min}} &
\multicolumn{4}{c}{\textbf{90 min}} &
\multicolumn{4}{c}{\textbf{180 min}} \\
\cmidrule(lr){2-5}
\cmidrule(lr){6-9}
\cmidrule(lr){10-13}
\cmidrule(lr){14-17}
 & \textbf{Zero} & \textbf{Few} & \textbf{Fine-Tune} & \textbf{MMG-PopNet} & \textbf{Zero} & \textbf{Few} & \textbf{Fine-Tune} & \textbf{MMG-PopNet} & \textbf{Zero} & \textbf{Few} & \textbf{Fine-Tune} & \textbf{MMG-PopNet} & \textbf{Zero} & \textbf{Few} & \textbf{Fine-Tune} & \textbf{MMG-PopNet} \\
\midrule

\textcolor{googleblue}{\bf \scshape Max Width} & 4.912 & \underline{2.274} & 2.466 & \textbf{1.581} & 3.661 & 2.005 & \underline{1.774} & \textbf{1.321} & 2.952 & 1.417 & \underline{0.971} & \textbf{0.794} & 2.758 & 1.155 & \underline{0.782} & \textbf{0.457} \\

\textcolor{googleblue}{\bf \scshape Max Depth} & 1.605 & \underline{0.785} & 1.012 & \textbf{0.517} & 1.151 & \underline{0.685} & 0.782 & \textbf{0.496} & 0.769 & \underline{0.439} & 0.517 & \textbf{0.296} & 0.551 & \underline{0.357} & 0.425 & \textbf{0.242} \\

\makecell[l]{\textcolor{googleblue}{\bf \scshape Structural}\\\textcolor{googleblue}{\bf \scshape Virality}} & 2.164 & \underline{0.264} & 0.343 & \textbf{0.182} & 1.198 & 0.265 & \underline{0.262} & \textbf{0.178} & 0.667 & 0.188 & \underline{0.179} & \textbf{0.104} & 0.434 & 0.160 & \underline{0.141} & \textbf{0.091} \\

\textcolor{googleblue}{\bf \scshape Size} & 9.238 & \underline{3.895} & 4.080 & \textbf{2.356} & 5.706 & 3.096 & \underline{2.953} & \textbf{1.921} & 3.572 & 1.986 & \underline{1.721} & \textbf{1.141} & 2.630 & 1.466 & \underline{1.358} & \textbf{0.700} \\

\textcolor{googlegreen}{\bf \scshape Unique Users} & 6.693 & \underline{3.005} & 3.028 & \textbf{1.859} & 4.365 & 2.440 & \underline{2.181} & \textbf{1.551} & 2.798 & 1.616 & \underline{1.273} & \textbf{0.917} & 2.067 & 1.217 & \underline{1.038} & \textbf{0.552} \\

\textcolor{googlepurple}{\bf \scshape Like Score} & 21.135 & \underline{8.859} & 8.863 & \textbf{4.999} & 20.018 & \underline{8.178} & 8.572 & \textbf{5.030} & 17.077 & \underline{6.933} & 6.944 & \textbf{3.400} & 13.981 & 7.052 & \underline{6.016} & \textbf{3.039} \\

\bottomrule
\end{tabular}%
}
\end{table}

\begin{table}[H]
\centering
\caption{LLM Gemma-3-12b-it comparison with MMG-PopNet on Bluesky across observation windows using MSE values.}
\label{tab:bluesky_gemma_3_12b_mse_llm_comparison}

\small
\setlength{\tabcolsep}{4pt}
\renewcommand{\arraystretch}{1.16}

\resizebox{\textwidth}{!}{%
\begin{tabular}{l cccc cccc cccc cccc}
\toprule
\multirow{2}{*}{\textbf{Task}} &
\multicolumn{4}{c}{\textbf{Root Only}} &
\multicolumn{4}{c}{\textbf{2}} &
\multicolumn{4}{c}{\textbf{10}} &
\multicolumn{4}{c}{\textbf{20}} \\
\cmidrule(lr){2-5}
\cmidrule(lr){6-9}
\cmidrule(lr){10-13}
\cmidrule(lr){14-17}
 & \textbf{Zero} & \textbf{Few} & \textbf{Fine-Tune} & \textbf{MMG-PopNet} & \textbf{Zero} & \textbf{Few} & \textbf{Fine-Tune} & \textbf{MMG-PopNet} & \textbf{Zero} & \textbf{Few} & \textbf{Fine-Tune} & \textbf{MMG-PopNet} & \textbf{Zero} & \textbf{Few} & \textbf{Fine-Tune} & \textbf{MMG-PopNet} \\
\midrule

\textcolor{googleblue}{\bf \scshape Max Width} & 5.366 & \underline{1.312} & 1.993 & \textbf{0.666} & 3.796 & \underline{1.320} & 1.391 & \textbf{0.577} & 2.545 & 0.969 & \underline{0.812} & \textbf{0.506} & 2.179 & 0.860 & \underline{0.657} & \textbf{0.459} \\

\textcolor{googleblue}{\bf \scshape Max Depth} & 2.285 & \underline{0.628} & 0.961 & \textbf{0.523} & 1.655 & \underline{0.659} & 0.800 & \textbf{0.480} & 1.269 & \underline{0.577} & 0.721 & \textbf{0.428} & 1.105 & \underline{0.533} & 0.603 & \textbf{0.415} \\

\makecell[l]{\textcolor{googleblue}{\bf \scshape Structural}\\\textcolor{googleblue}{\bf \scshape Virality}} & 2.362 & \underline{0.286} & 0.490 & \textbf{0.243} & 1.828 & \underline{0.314} & 0.405 & \textbf{0.225} & 1.276 & \underline{0.280} & 0.320 & \textbf{0.196} & 1.067 & \underline{0.264} & 0.277 & \textbf{0.191} \\

\textcolor{googleblue}{\bf \scshape Size} & 6.577 & \underline{1.625} & 2.819 & \textbf{1.007} & 5.186 & \underline{1.656} & 2.038 & \textbf{0.870} & 3.406 & \underline{1.200} & 1.355 & \textbf{0.751} & 2.675 & \underline{1.061} & 1.102 & \textbf{0.700} \\

\textcolor{googlegreen}{\bf \scshape Unique Users} & 4.041 & \underline{1.454} & 1.968 & \textbf{0.690} & 3.132 & \underline{1.348} & 1.395 & \textbf{0.600} & 1.956 & 0.960 & \underline{0.833} & \textbf{0.521} & 1.491 & 0.825 & \underline{0.672} & \textbf{0.468} \\

\textcolor{googlepurple}{\bf \scshape Like Score} & 16.254 & \underline{4.149} & 7.586 & \textbf{1.540} & 15.135 & \underline{3.998} & 5.782 & \textbf{1.445} & 14.355 & \underline{3.547} & 3.614 & \textbf{1.394} & 14.018 & 3.462 & \underline{3.221} & \textbf{1.327} \\

\bottomrule
\end{tabular}%
}
\end{table}

\begin{table}[H]
\centering
\caption{LLM Gemma-3-12b-it comparison with MMG-PopNet on r/Futurology across observation windows using MSE values.}
\label{tab:futurology_gemma_3_12b_mse_llm_comparison}

\small
\setlength{\tabcolsep}{4pt}
\renewcommand{\arraystretch}{1.16}

\resizebox{\textwidth}{!}{%
\begin{tabular}{l cccc cccc cccc cccc}
\toprule
\multirow{2}{*}{\textbf{Task}} &
\multicolumn{4}{c}{\textbf{Root Only}} &
\multicolumn{4}{c}{\textbf{30 min}} &
\multicolumn{4}{c}{\textbf{90 min}} &
\multicolumn{4}{c}{\textbf{180 min}} \\
\cmidrule(lr){2-5}
\cmidrule(lr){6-9}
\cmidrule(lr){10-13}
\cmidrule(lr){14-17}
 & \textbf{Zero} & \textbf{Few} & \textbf{Fine-Tune} & \textbf{MMG-PopNet} & \textbf{Zero} & \textbf{Few} & \textbf{Fine-Tune} & \textbf{MMG-PopNet} & \textbf{Zero} & \textbf{Few} & \textbf{Fine-Tune} & \textbf{MMG-PopNet} & \textbf{Zero} & \textbf{Few} & \textbf{Fine-Tune} & \textbf{MMG-PopNet} \\
\midrule

\textcolor{googleblue}{\bf \scshape Max Width} & 7.837 & \underline{1.984} & 2.831 & \textbf{1.581} & 4.027 & 1.948 & \underline{1.900} & \textbf{1.321} & 3.058 & 1.350 & \underline{1.274} & \textbf{0.794} & 2.680 & 1.026 & \underline{1.017} & \textbf{0.457} \\

\textcolor{googleblue}{\bf \scshape Max Depth} & 2.439 & \underline{0.687} & 1.299 & \textbf{0.517} & 1.311 & \underline{0.650} & 0.891 & \textbf{0.496} & 0.974 & \underline{0.457} & 0.589 & \textbf{0.296} & 0.766 & \underline{0.321} & 0.509 & \textbf{0.242} \\

\makecell[l]{\textcolor{googleblue}{\bf \scshape Structural}\\\textcolor{googleblue}{\bf \scshape Virality}} & 2.204 & \underline{0.241} & 0.429 & \textbf{0.182} & 0.969 & \underline{0.237} & 0.296 & \textbf{0.178} & 0.637 & \underline{0.177} & 0.196 & \textbf{0.104} & 0.449 & \underline{0.132} & 0.170 & \textbf{0.091} \\

\textcolor{googleblue}{\bf \scshape Size} & 9.238 & \underline{3.226} & 4.869 & \textbf{2.356} & 5.575 & \underline{2.948} & 3.305 & \textbf{1.921} & 3.415 & \underline{1.930} & 2.202 & \textbf{1.141} & 2.510 & \underline{1.361} & 1.816 & \textbf{0.700} \\

\textcolor{googlegreen}{\bf \scshape Unique Users} & 6.693 & \underline{2.497} & 3.470 & \textbf{1.859} & 4.319 & \underline{2.283} & 2.576 & \textbf{1.551} & 2.676 & \underline{1.507} & 1.731 & \textbf{0.917} & 1.951 & \underline{1.074} & 1.406 & \textbf{0.552} \\

\textcolor{googlepurple}{\bf \scshape Like Score} & 21.116 & \underline{7.747} & 12.986 & \textbf{4.999} & 19.077 & \underline{8.055} & 13.202 & \textbf{5.030} & 18.979 & \underline{6.803} & 12.123 & \textbf{3.400} & 18.766 & \underline{6.938} & 10.649 & \textbf{3.039} \\

\bottomrule
\end{tabular}%
}
\end{table}

\begin{table}[H]
\centering
\caption{LLM GPT-4o-mini comparison with MMG-PopNet on Bluesky across observation windows using MSE values.}
\label{tab:bluesky_gpt_4o_mini_mse_llm_comparison}

\small
\setlength{\tabcolsep}{4pt}
\renewcommand{\arraystretch}{1.16}

\resizebox{\textwidth}{!}{%
\begin{tabular}{l ccc ccc ccc ccc}
\toprule
\multirow{2}{*}{\textbf{Task}} &
\multicolumn{3}{c}{\textbf{Root Only}} &
\multicolumn{3}{c}{\textbf{2}} &
\multicolumn{3}{c}{\textbf{10}} &
\multicolumn{3}{c}{\textbf{20}} \\
\cmidrule(lr){2-4}
\cmidrule(lr){5-7}
\cmidrule(lr){8-10}
\cmidrule(lr){11-13}
 & \textbf{Zero} & \textbf{Few} & \textbf{MMG-PopNet} & \textbf{Zero} & \textbf{Few} & \textbf{MMG-PopNet} & \textbf{Zero} & \textbf{Few} & \textbf{MMG-PopNet} & \textbf{Zero} & \textbf{Few} & \textbf{MMG-PopNet} \\
\midrule

\textcolor{googleblue}{\bf \scshape Max Width} & 2.742 & \underline{1.666} & \textbf{0.666} & 2.268 & \underline{1.562} & \textbf{0.577} & 1.632 & \underline{1.486} & \textbf{0.506} & \underline{1.386} & 1.395 & \textbf{0.459} \\

\textcolor{googleblue}{\bf \scshape Max Depth} & 1.577 & \underline{0.590} & \textbf{0.523} & 1.398 & \underline{0.598} & \textbf{0.480} & 1.055 & \underline{0.539} & \textbf{0.428} & 0.906 & \underline{0.484} & \textbf{0.415} \\

\makecell[l]{\textcolor{googleblue}{\bf \scshape Structural}\\\textcolor{googleblue}{\bf \scshape Virality}} & 0.799 & \underline{0.275} & \textbf{0.243} & 0.746 & \underline{0.281} & \textbf{0.225} & 0.614 & \underline{0.248} & \textbf{0.196} & 0.532 & \underline{0.223} & \textbf{0.191} \\

\textcolor{googleblue}{\bf \scshape Size} & 5.531 & \underline{1.880} & \textbf{1.007} & 4.545 & \underline{1.789} & \textbf{0.870} & 3.096 & \underline{1.464} & \textbf{0.751} & 2.446 & \underline{1.307} & \textbf{0.700} \\

\textcolor{googlegreen}{\bf \scshape Unique Users} & 3.472 & \underline{1.690} & \textbf{0.690} & 2.767 & \underline{1.530} & \textbf{0.600} & 1.796 & \underline{1.266} & \textbf{0.521} & 1.392 & \underline{1.115} & \textbf{0.468} \\

\textcolor{googlepurple}{\bf \scshape Like Score} & 12.566 & \underline{4.002} & \textbf{1.540} & 14.130 & \underline{3.852} & \textbf{1.445} & 13.828 & \underline{3.478} & \textbf{1.394} & 13.415 & \underline{3.212} & \textbf{1.327} \\

\bottomrule
\end{tabular}%
}
\end{table}

\begin{table}[H]
\centering
\caption{LLM GPT-4o-mini comparison with MMG-PopNet on r/Futurology across observation windows using MSE values.}
\label{tab:futurology_gpt_4o_mini_mse_llm_comparison}

\small
\setlength{\tabcolsep}{4pt}
\renewcommand{\arraystretch}{1.16}

\resizebox{\textwidth}{!}{%
\begin{tabular}{l ccc ccc ccc ccc}
\toprule
\multirow{2}{*}{\textbf{Task}} &
\multicolumn{3}{c}{\textbf{Root Only}} &
\multicolumn{3}{c}{\textbf{30 min}} &
\multicolumn{3}{c}{\textbf{90 min}} &
\multicolumn{3}{c}{\textbf{180 min}} \\
\cmidrule(lr){2-4}
\cmidrule(lr){5-7}
\cmidrule(lr){8-10}
\cmidrule(lr){11-13}
 & \textbf{Zero} & \textbf{Few} & \textbf{MMG-PopNet} & \textbf{Zero} & \textbf{Few} & \textbf{MMG-PopNet} & \textbf{Zero} & \textbf{Few} & \textbf{MMG-PopNet} & \textbf{Zero} & \textbf{Few} & \textbf{MMG-PopNet} \\
\midrule

\textcolor{googleblue}{\bf \scshape Max Width} & 4.853 & \underline{2.179} & \textbf{1.581} & 3.388 & \underline{2.173} & \textbf{1.321} & 2.449 & \underline{1.813} & \textbf{0.794} & 2.118 & \underline{1.522} & \textbf{0.457} \\

\textcolor{googleblue}{\bf \scshape Max Depth} & 1.476 & \underline{0.697} & \textbf{0.517} & 1.126 & \underline{0.623} & \textbf{0.496} & 0.749 & \underline{0.431} & \textbf{0.296} & 0.551 & \underline{0.360} & \textbf{0.242} \\

\makecell[l]{\textcolor{googleblue}{\bf \scshape Structural}\\\textcolor{googleblue}{\bf \scshape Virality}} & 0.847 & \underline{0.237} & \textbf{0.182} & 0.692 & \underline{0.221} & \textbf{0.178} & 0.526 & \underline{0.180} & \textbf{0.104} & 0.405 & \underline{0.158} & \textbf{0.091} \\

\textcolor{googleblue}{\bf \scshape Size} & 8.418 & \underline{3.451} & \textbf{2.356} & 5.444 & \underline{3.163} & \textbf{1.921} & 3.226 & \underline{2.208} & \textbf{1.141} & 2.238 & \underline{1.612} & \textbf{0.700} \\

\textcolor{googlegreen}{\bf \scshape Unique Users} & 6.123 & \underline{2.648} & \textbf{1.859} & 4.056 & \underline{2.511} & \textbf{1.551} & 2.460 & \underline{1.758} & \textbf{0.917} & 1.689 & \underline{1.318} & \textbf{0.552} \\

\textcolor{googlepurple}{\bf \scshape Like Score} & 20.291 & \underline{7.341} & \textbf{4.999} & 20.546 & \underline{7.930} & \textbf{5.030} & 20.385 & \underline{6.932} & \textbf{3.400} & 19.958 & \underline{7.070} & \textbf{3.039} \\

\bottomrule
\end{tabular}%
}
\end{table}

 \newpage
\paragraph*{Zero-Shot Prompt}
 
The zero-shot prompt is constructed once per query with no retrieved examples. The placeholder \texttt{<WINDOW>} is replaced by the observation window size, and \texttt{<EARLY\_CONVERSATION\_TREE\_JSON>} is replaced by the serialised conversation tree.
 
\vspace{4pt}
 
\begin{tcolorbox}[promptbox, title={Prompt Template --- Zero-Shot}]\label{zero:prompt}
\begin{lstlisting}[style=promptstyle]
You are analyzing a social media conversation tree to predict its final growth metrics. You will be provided with the first <WINDOW> of a social media conversation thread. This conversation thread continued after this.
 
Based on the content, timing, and structural patterns of these early interactions, predict the FINAL state of this conversation tree when it reaches saturation (that is, stops growing).
 
METRIC DEFINITIONS:
- max_width: Maximum number of replies at any single depth level.
- max_depth: Maximum depth of the conversation tree.
- structural_virality: Average distance between all pairs of nodes (Low = Broadcast, High = Viral).
- num_posts: Total number of posts in the final conversation (root + replies).
- num_unique_users: Total number of unique users in the final conversation.
- root_score: Engagement score of the intial root post.
 
You will be given a NESTED JSON TREE. Each reply node has text, time_since_root, time_since_parent, and children. The root node has text, timestamp, and children.
 
OUTPUT FORMAT:
Return a single JSON object. Do not include markdown formatting.
{"max_width": <number>, "max_depth": <number>, "structural_virality": <number>, "num_posts": <number>, "num_unique_users": <number>, "root_score": <number>}
 
EARLY CONVERSATION TREE:
<EARLY_CONVERSATION_TREE_JSON>
 
INSTRUCTIONS:
Using only the conversation tree above as evidence, predict the FINAL values.
Do NOT include any explanation or extra fields.
Output exactly one JSON object with these keys only: max_width, max_depth, structural_virality, num_posts, num_unique_users, root_score.
 
Your JSON Prediction:
\end{lstlisting}
\end{tcolorbox}
 
\vspace{6pt}
 
Below is a minimal example of the input that populates \texttt{<EARLY\_CONVERSATION\_TREE\_JSON>} when only the root post has been observed.
 
\vspace{4pt}
 
\begin{tcolorbox}[jsonbox, title={Example Input --- Root-Only Tree}]
\begin{lstlisting}[style=jsonstyle]
{
  "text": "Nvidia partner says it can cut data center energy use by 50% as AI boom strains power grid",
  "timestamp": "2024-08-27 11:17:18",
  "children": []
}
\end{lstlisting}
\end{tcolorbox}

 
\paragraph{Few-Shot RAG Prompt}
 
The few-shot prompt augments the zero-shot template with $k$ retrieved examples (default $k=4$), selected via embedding-based nearest-neighbour search over the training set. Each placeholder \texttt{<RETRIEVED\_EXAMPLE\_TREE\_$i$\_JSON>} and \texttt{<RETRIEVED\_EXAMPLE\_$i$\_TARGETS\_JSON>} is replaced by the corresponding retrieved tree and its ground-truth final metrics; \texttt{<TARGET\_TREE\_JSON>} is the query conversation.
 
\vspace{4pt}
 
\begin{tcolorbox}[promptbox, title={Prompt Template --- Few-Shot RAG}]\label{few:prompt}
\begin{lstlisting}[style=promptstyle]
You are analyzing a social media conversation tree to predict its final growth metrics. You will be provided with the first <WINDOW> of a social media conversation thread. This conversation thread continued after this.
 
Based on the content, timing, and structural patterns of these early interactions, predict the FINAL state of this conversation tree when it reaches saturation (that is, stops growing).
 
METRIC DEFINITIONS:
- max_width: Maximum number of replies at any single depth level.
- max_depth: Maximum depth of the conversation tree.
- structural_virality: Average distance between all pairs of nodes (Low = Broadcast, High = Viral).
- num_posts: Total number of posts in the final conversation (root + replies).
- num_unique_users: Total number of unique users in the final conversation.
- root_score: Engagement score of the intial root post.
 
Each conversation is represented as a NESTED JSON TREE.
 
I will show you several examples of early trees and their final metrics, then ask you to predict for a new tree.
 
====================
FEW-SHOT EXAMPLES
====================
 
--- Example 1 ---
Early conversation tree:
<RETRIEVED_EXAMPLE_TREE_1_JSON>
Final metrics:
<RETRIEVED_EXAMPLE_1_TARGETS_JSON>
 
--- Example 2 ---
Early conversation tree:
<RETRIEVED_EXAMPLE_TREE_2_JSON>
Final metrics:
<RETRIEVED_EXAMPLE_2_TARGETS_JSON>
 
[... k examples total ...]
 
====================
NEW CONVERSATION (PREDICT THIS)
====================
Early conversation tree:
<TARGET_TREE_JSON>
Final metrics:
\end{lstlisting}
\end{tcolorbox}

\FloatBarrier
\newpage
\section{Modality Ablation Details}
\label{sec:appendix_modality}
In Section~\ref{sec:modality_analysis}, to evaluate the specific contributions of each modality in the MMG-PopNet model, we conducted comprehensive ablation studies. By removing one input source at a time, we measured the relative Mean Squared Error (MSE) increase over the full model. The full MMG-PopNet architecture combines four distinct information sources: text (node-level post text), image (root-post visual content), topology (reply-tree structure), and temporal (node-level timing features).

\textbf{Implementation Details:} To isolate each modality, we modified the model architecture and inputs as follows (w/o means ``without''):

\begin{itemize}[leftmargin=*, itemsep=0pt]
    \item \textbf{w/o Text (Semantic Ablation):} and its corresponding projection layer are completely disabled. To maintain architectural compatibility with the downstream Graph Neural Network (GNN), the text embedding for every node is replaced with a zero-initialized vector of the exact same dimensionality. Transformer parameters are frozen and excluded from optimization. This isolates the contribution of actual semantic content. By passing zero-vectors into an otherwise unchanged text feature slot, any performance drop directly reflects the loss of node-level language cues.

    \item \textbf{w/o Image (Visual Ablation):} The CLIP vision encoder is bypassed and all raw image pixels are ignored. However, the graph-level image fusion slot is preserved. Every cascade is instead assigned an identical, learnable ``dummy'' image embedding that is optimized alongside the rest of the network parameters.

    \item \textbf{w/o Topology (Structural Ablation):} We completely disable the neighbor aggregation mechanism of the bidirectional GraphSAGE network by ignoring the reply edges entirely. Instead of message passing, the network treats the cascade as a disconnected set of independent nodes. To ensure any performance drop is strictly due to the loss of structural connectivity and not a reduction in parameter capacity, the GraphSAGE layers are substituted with standard MLPs. These MLPs are applied independently to each node and strictly match the layer count and hidden feature dimensions of the original GNN. This isolates the value of explicit cascade interactions. Because the replacement MLPs retain the exact same per-node transformation capacity as the original model, this ablation cleanly measures the impact of structural message passing.

    \item \textbf{w/o Temporal (Timing Ablation):} The explicit node-level cascade timing features, specifically time since root and time since parent, are stripped from the input feature matrix before entering the GNN. The nodes are processed using only their text features and topological connections. This specifically targets localized reaction speeds and the pace of the cascade. Global posting metadata may remain, but the precise timing of individual user interactions is entirely removed.
\end{itemize}

Tables~\ref{tab:ablation_mse},\ref{tab:ablation_mse_gaming},\ref{tab:ablation_mse_futurology} report the detailed modality-to-target sensitivity analysis across the r/Gaming and r/Futurology datasets. Table~\ref{tab:ablation_mse} aggregates the MSE performance for both datasets to highlight the trade-offs between semantic and structural signals. Table~\ref{tab:ablation_mse_gaming} and \ref{tab:ablation_mse_futurology}provide the detailed mean and standard deviation across three random seed runs (42, 1042, and 2042) for r/Gaming and r/Futurology, respectively.

The detailed results confirm the main trend in Section~\ref{sec:modality_analysis}.
The full MMG-PopNet model consistently achieves the best average MSE across every target, confirming that each modality contributes valuable predictive information. However, the sensitivity to specific modalities varies distinctly between the two communities

On \textbf{r/Gaming}, temporal information is the most influential signal for most structural and participation targets. 
Temporal information is the most influential signal for structural and participation metrics. Removing temporal features causes the sharpest performance degradation for \textsc{Max Width} (21.2\%), \textsc{Unique Users} (21.0\%), \textsc{Size} (18.7\%), \textsc{Max Depth}, and \textsc{Structural Virality}. This indicates that the pace and timing of early responses are critical for predicting how gaming discussions grow and branch. Conversely, removing image features has the smallest overall effect, indicating that root visual content plays a more modest, complementary role in this specific community.

On \textbf{r/Futurology}, the ablation effects are more distributed across modalities. 
Temporal and topology features remain important for structural targets, but text and image removals also produce visible degradation across several targets. 
This suggests that r/Futurology popularity prediction depends on a broader mixture of semantic, temporal, and structural signals, rather than being dominated by a single modality. 

Across both datasets, text is consistently the most important modality for \textsc{Like Score}. 
Removing textual semantics increases the average \textsc{Like Score} MSE from 4.300 to 4.876 on r/Gaming and from 3.397 to 4.088 on r/Futurology.
These results are consistent with the text-centered nature of the dataset platforms of the benchmark., where discussion content is primarily text based. This suggests a broader hypothesis that the dominant modality may shift with platform design. On platforms where images or videos are the primary medium of interaction, visual features may play a larger role in predicting engagement and cascade growth.

\begin{table}[t]
\small
\caption{\textbf{Modality-to-Target Sensitivity Analysis.} MSE performance of the full MMG-PopNet model versus single-modality ablations across the r/Gaming and r/Futurology datasets. This table isolates the contribution of each modality across the six distinct predictive targets, highlighting the trade-offs between semantic and structural signals. Lower is better. The best value is \textbf{bolded}; the second-best is \underline{underlined}.}
\label{tab:ablation_mse}

\centering
\vspace{1mm}
\setlength{\tabcolsep}{4pt}
\renewcommand{\arraystretch}{1.18}

\resizebox{\textwidth}{!}{%
\begin{tabular}{l l cccc cccc c}
\toprule
\multirow{2}{*}{\textbf{Task}} &
\multirow{2}{*}{\textbf{Model}} &
\multicolumn{4}{c}{\textbf{r/Gaming}} &
\multicolumn{4}{c}{\textbf{r/Futurology}} &
\multirow{2}{*}{\textbf{Avg}} \\
\cmidrule(lr){3-6}
\cmidrule(lr){7-10}
&  &
\textbf{20} & \textbf{50} & \textbf{90} & \textbf{Avg} &
\textbf{30} & \textbf{90} & \textbf{180} & \textbf{Avg} & \\
\midrule

\textcolor{googleblue}{\bf \scshape Max Width}
 & MMG-PopNet         & \textbf{1.054} & \textbf{0.652} & \underline{0.509} & \textbf{0.739} & \textbf{1.093} & \underline{0.620} & 0.348 & \textbf{0.687} & \textbf{0.713} \\
 & w/o Text      & 1.079 & 0.674 & \textbf{0.487} & \underline{0.747} & 1.202 & \textbf{0.617} & \textbf{0.319} & 0.713 & 0.730 \\
 & w/o Image     & \underline{1.056} & \underline{0.653} & \underline{0.509} & \textbf{0.739} & 1.145 & 0.650 & \underline{0.340} & \underline{0.711} & \underline{0.725} \\
 & w/o Temporal  & 1.083 & 1.184 & 0.641 & 0.969 & \underline{1.127} & 0.711 & 0.441 & 0.759 & 0.864 \\
 & w/o Topology  & 1.061 & 0.706 & 0.551 & 0.773 & 1.202 & 0.693 & 0.377 & 0.757 & 0.765 \\
\midrule
\addlinespace[1mm]

\textcolor{googleblue}{\bf \scshape Max Depth}
 & MMG-PopNet         & \textbf{0.266} & \underline{0.196} & \underline{0.152} & \textbf{0.205} & \textbf{0.413} & \textbf{0.271} & \textbf{0.202} & \textbf{0.295} & \textbf{0.250} \\
 & w/o Text      & 0.269 & \underline{0.196} & \textbf{0.150} & \textbf{0.205} & 0.457 & \textbf{0.271} & 0.205 & 0.311 & 0.258 \\
 & w/o Image     & \underline{0.267} & \textbf{0.194} & 0.155 & \textbf{0.205} & 0.422 & \underline{0.280} & \underline{0.204} & \underline{0.302} & \underline{0.254} \\
 & w/o Temporal  & 0.269 & 0.296 & 0.163 & 0.243 & \underline{0.418} & 0.285 & 0.211 & 0.305 & 0.274 \\
 & w/o Topology  & 0.277 & 0.214 & 0.192 & \underline{0.228} & 0.440 & 0.295 & 0.218 & 0.318 & 0.273 \\
\midrule
\addlinespace[1mm]

\textcolor{googleblue}{\bf \scshape Size}
 & MMG-PopNet         & \underline{1.272} & \textbf{0.784} & \underline{0.598} & \textbf{0.885} & \textbf{1.703} & \textbf{0.949} & 0.524 & \textbf{1.059} & \textbf{0.972} \\
 & w/o Text      & 1.293 & 0.813 & \textbf{0.562} & 0.889 & 1.899 & \underline{0.962} & \textbf{0.509} & 1.123 & 1.006 \\
 & w/o Image     & \underline{1.272} & \textbf{0.784} & 0.603 & \underline{0.886} & 1.783 & 0.995 & \underline{0.520} & \underline{1.099} & \underline{0.993} \\
 & w/o Temporal  & 1.302 & 1.391 & 0.745 & 1.146 & \underline{1.749} & 1.086 & 0.649 & 1.161 & 1.154 \\
 & w/o Topology  & \textbf{1.267} & \underline{0.807} & 0.644 & 0.906 & 1.854 & 1.050 & 0.548 & 1.151 & 1.028 \\
\midrule
\addlinespace[1mm]

\makecell[l]{\textcolor{googleblue}{\bf \scshape Structural}\\\textcolor{googleblue}{\bf \scshape Virality}}
 & MMG-PopNet         & \textbf{0.076} & \underline{0.057} & \underline{0.045} & \textbf{0.059} & \textbf{0.143} & \textbf{0.097} & \textbf{0.072} & \textbf{0.104} & \textbf{0.082} \\
 & w/o Text      & 0.078 & 0.058 & \underline{0.045} & \underline{0.060} & 0.158 & \underline{0.098} & \textbf{0.072} & 0.109 & 0.085 \\
 & w/o Image     & \underline{0.077} & \textbf{0.056} & \textbf{0.044} & \textbf{0.059} & 0.147 & 0.100 & \textbf{0.072} & \underline{0.106} & \underline{0.083} \\
 & w/o Temporal  & \textbf{0.076} & 0.117 & \underline{0.045} & 0.079 & \underline{0.146} & 0.102 & \underline{0.073} & 0.107 & 0.093 \\
 & w/o Topology  & 0.082 & 0.068 & 0.061 & 0.070 & 0.152 & 0.107 & 0.078 & 0.112 & 0.091 \\
\midrule
\addlinespace[1mm]

\makecell[l]{\textcolor{googlegreen}{\bf \scshape Unique}\\\textcolor{googlegreen}{\bf \scshape Users}}
 & MMG-PopNet         & \underline{1.170} & \textbf{0.730} & \underline{0.556} & \textbf{0.819} & \textbf{1.334} & \textbf{0.747} & 0.407 & \textbf{0.829} & \textbf{0.824} \\
 & w/o Text      & 1.197 & 0.764 & \textbf{0.525} & 0.829 & 1.485 & \underline{0.756} & \textbf{0.392} & 0.878 & 0.853 \\
 & w/o Image     & 1.172 & \underline{0.735} & 0.561 & \underline{0.822} & 1.402 & 0.789 & \underline{0.406} & \underline{0.866} & \underline{0.844} \\
 & w/o Temporal  & 1.198 & 1.331 & 0.697 & 1.075 & \underline{1.380} & 0.863 & 0.516 & 0.920 & 0.997 \\
 & w/o Topology  & \textbf{1.166} & 0.771 & 0.597 & 0.845 & 1.467 & 0.830 & 0.436 & 0.911 & 0.878 \\
\midrule
\addlinespace[1mm]

\makecell[l]{\textcolor{googlepurple}{\bf \scshape Like}\\\textcolor{googlepurple}{\bf \scshape Score}}
 & MMG-PopNet         & \textbf{4.659} & \textbf{4.237} & \textbf{4.003} & \textbf{4.300} & \textbf{4.527} & \textbf{3.032} & \textbf{2.632} & \textbf{3.397} & \textbf{3.848} \\
 & w/o Text      & 5.215 & 5.016 & 4.397 & 4.876 & 5.347 & 3.680 & 3.237 & 4.088 & 4.482 \\
 & w/o Image     & 4.714 & \underline{4.327} & \underline{4.023} & \underline{4.355} & 4.783 & \underline{3.271} & \underline{2.674} & \underline{3.576} & \underline{3.965} \\
 & w/o Temporal  & 4.726 & 5.041 & 4.284 & 4.683 & \underline{4.597} & 3.454 & 3.166 & 3.739 & 4.211 \\
 & w/o Topology  & \underline{4.702} & 4.348 & 4.113 & 4.387 & 4.812 & 3.292 & 2.728 & 3.611 & 3.999 \\
\bottomrule
\end{tabular}
}
\end{table}

\begin{table}[t]
\small
\caption{\textbf{Modality-to-Target Sensitivity Analysis on r/Gaming.} MSE performance of full MMG-PopNet model versus single-modality ablations on the r/Gaming dataset. Results are averaged across three random seeds: 42, 1042, and 2042, and are reported as mean $\pm$ standard deviation. Lower is better. The best value is \textbf{bolded}; the second-best is \underline{underlined}.}
\label{tab:ablation_mse_gaming}

\centering
\vspace{1mm}
\setlength{\tabcolsep}{6pt}
\renewcommand{\arraystretch}{1.18}

\resizebox{\textwidth}{!}{%
\begin{tabular}{l l cccc}
\toprule
\textbf{Task} & \textbf{Model} &
\textbf{20} & \textbf{50} & \textbf{90} & \textbf{Avg} \\
\midrule

\textcolor{googleblue}{\bf \scshape Max Width}
 & MMG-PopNet         & \textbf{1.054 $\pm$ 0.008} & \textbf{0.652 $\pm$ 0.013} & \underline{0.509 $\pm$ 0.014} & \textbf{0.739 $\pm$ 0.009} \\
 & w/o Text      & 1.079 $\pm$ 0.015 & 0.674 $\pm$ 0.003 & \textbf{0.487 $\pm$ 0.030} & \underline{0.747 $\pm$ 0.013} \\
 & w/o Image     & \underline{1.056 $\pm$ 0.012} & \underline{0.653 $\pm$ 0.006} & \underline{0.509 $\pm$ 0.022} & \textbf{0.739 $\pm$ 0.006} \\
 & w/o Temporal  & 1.083 $\pm$ 0.008 & 1.184 $\pm$ 0.708 & 0.641 $\pm$ 0.007 & 0.969 $\pm$ 0.234 \\
 & w/o Topology  & 1.061 $\pm$ 0.008 & 0.706 $\pm$ 0.004 & 0.551 $\pm$ 0.012 & 0.773 $\pm$ 0.007 \\
\midrule

\textcolor{googleblue}{\bf \scshape Max Depth}
 & MMG-PopNet         & \textbf{0.266 $\pm$ 0.003} & \underline{0.196 $\pm$ 0.003} & \underline{0.152 $\pm$ 0.006} & \textbf{0.205 $\pm$ 0.001} \\
 & w/o Text      & 0.269 $\pm$ 0.002 & \underline{0.196 $\pm$ 0.003} & \textbf{0.150 $\pm$ 0.001} & \textbf{0.205 $\pm$ 0.000} \\
 & w/o Image     & \underline{0.267 $\pm$ 0.004} & \textbf{0.194 $\pm$ 0.003} & 0.155 $\pm$ 0.005 & \textbf{0.205 $\pm$ 0.002} \\
 & w/o Temporal  & 0.269 $\pm$ 0.002 & 0.296 $\pm$ 0.153 & 0.163 $\pm$ 0.002 & 0.243 $\pm$ 0.052 \\
 & w/o Topology  & 0.277 $\pm$ 0.003 & 0.214 $\pm$ 0.007 & 0.192 $\pm$ 0.005 & \underline{0.228 $\pm$ 0.001} \\
\midrule

\textcolor{googleblue}{\bf \scshape Size}
 & MMG-PopNet         & \underline{1.272 $\pm$ 0.012} & \textbf{0.784 $\pm$ 0.004} & \underline{0.598 $\pm$ 0.023} & \textbf{0.885 $\pm$ 0.010} \\
 & w/o Text      & 1.293 $\pm$ 0.016 & 0.813 $\pm$ 0.008 & \textbf{0.562 $\pm$ 0.028} & 0.889 $\pm$ 0.012 \\
 & w/o Image     & \underline{1.272 $\pm$ 0.014} & \textbf{0.784 $\pm$ 0.005} & 0.603 $\pm$ 0.020 & \underline{0.886 $\pm$ 0.004} \\
 & w/o Temporal  & 1.302 $\pm$ 0.014 & 1.391 $\pm$ 0.820 & 0.745 $\pm$ 0.004 & 1.146 $\pm$ 0.277 \\
 & w/o Topology  & \textbf{1.267 $\pm$ 0.013} & \underline{0.807 $\pm$ 0.011} & 0.644 $\pm$ 0.022 & 0.906 $\pm$ 0.011 \\
\midrule

\makecell[l]{\textcolor{googleblue}{\bf \scshape Structural}\\\textcolor{googleblue}{\bf \scshape Virality}}
 & MMG-PopNet         & \textbf{0.076 $\pm$ 0.003} & \underline{0.057 $\pm$ 0.000} & \underline{0.045 $\pm$ 0.003} & \textbf{0.059 $\pm$ 0.001} \\
 & w/o Text      & 0.078 $\pm$ 0.001 & 0.058 $\pm$ 0.002 & \underline{0.045 $\pm$ 0.003} & \underline{0.060 $\pm$ 0.001} \\
 & w/o Image     & \underline{0.077 $\pm$ 0.003} & \textbf{0.056 $\pm$ 0.002} & \textbf{0.044 $\pm$ 0.003} & \textbf{0.059 $\pm$ 0.002} \\
 & w/o Temporal  & \textbf{0.076 $\pm$ 0.001} & 0.117 $\pm$ 0.096 & \underline{0.045 $\pm$ 0.001} & 0.079 $\pm$ 0.032 \\
 & w/o Topology  & 0.082 $\pm$ 0.004 & 0.068 $\pm$ 0.004 & 0.061 $\pm$ 0.003 & 0.070 $\pm$ 0.001 \\
\midrule

\makecell[l]{\textcolor{googlegreen}{\bf \scshape Unique}\\\textcolor{googlegreen}{\bf \scshape Users}}
 & MMG-PopNet         & \underline{1.170 $\pm$ 0.010} & \textbf{0.730 $\pm$ 0.011} & \underline{0.556 $\pm$ 0.019} & \textbf{0.819 $\pm$ 0.010} \\
 & w/o Text      & 1.197 $\pm$ 0.016 & 0.764 $\pm$ 0.005 & \textbf{0.525 $\pm$ 0.024} & 0.829 $\pm$ 0.011 \\
 & w/o Image     & 1.172 $\pm$ 0.016 & \underline{0.735 $\pm$ 0.006} & 0.561 $\pm$ 0.017 & \underline{0.822 $\pm$ 0.003} \\
 & w/o Temporal  & 1.198 $\pm$ 0.007 & 1.331 $\pm$ 0.820 & 0.697 $\pm$ 0.007 & 1.075 $\pm$ 0.273 \\
 & w/o Topology  & \textbf{1.166 $\pm$ 0.012} & 0.771 $\pm$ 0.006 & 0.597 $\pm$ 0.016 & 0.845 $\pm$ 0.008 \\
\midrule

\makecell[l]{\textcolor{googlepurple}{\bf \scshape Like}\\\textcolor{googlepurple}{\bf \scshape Score}}
 & MMG-PopNet         & \textbf{4.659 $\pm$ 0.023} & \textbf{4.237 $\pm$ 0.138} & \textbf{4.003 $\pm$ 0.048} & \textbf{4.300 $\pm$ 0.064} \\
 & w/o Text      & 5.215 $\pm$ 0.017 & 5.016 $\pm$ 0.038 & 4.397 $\pm$ 0.009 & 4.876 $\pm$ 0.013 \\
 & w/o Image     & 4.714 $\pm$ 0.028 & \underline{4.327 $\pm$ 0.136} & \underline{4.023 $\pm$ 0.055} & \underline{4.355 $\pm$ 0.044} \\
 & w/o Temporal  & 4.726 $\pm$ 0.020 & 5.041 $\pm$ 0.988 & 4.284 $\pm$ 0.023 & 4.683 $\pm$ 0.335 \\
 & w/o Topology  & \underline{4.702 $\pm$ 0.050} & 4.348 $\pm$ 0.051 & 4.113 $\pm$ 0.042 & 4.387 $\pm$ 0.047 \\
\bottomrule
\end{tabular}
}
\end{table}

\begin{table}[t]
\small
\caption{\textbf{Modality-to-Target Sensitivity Analysis on r/Futurology.} MSE performance of full MMG-PopNet model versus single-modality ablations on the r/Futurology dataset. Results are averaged across three random seeds: 42, 1042, and 2042, and are reported as mean $\pm$ standard deviation. Lower is better. The best value is \textbf{bolded}; the second-best is \underline{underlined}.}
\label{tab:ablation_mse_futurology}

\centering
\vspace{1mm}
\setlength{\tabcolsep}{6pt}
\renewcommand{\arraystretch}{1.18}

\resizebox{\textwidth}{!}{%
\begin{tabular}{l l cccc}
\toprule
\textbf{Task} & \textbf{Model} &
\textbf{30} & \textbf{90} & \textbf{180} & \textbf{Avg} \\
\midrule

\textcolor{googleblue}{\bf \scshape Max Width}
 & MMG-PopNet         & \textbf{1.093 $\pm$ 0.013} & \underline{0.620 $\pm$ 0.020} & 0.348 $\pm$ 0.008 & \textbf{0.687 $\pm$ 0.009} \\
 & w/o Text      & 1.202 $\pm$ 0.020 & \textbf{0.617 $\pm$ 0.005} & \textbf{0.319 $\pm$ 0.016} & 0.713 $\pm$ 0.011 \\
 & w/o Image     & 1.145 $\pm$ 0.015 & 0.650 $\pm$ 0.010 & \underline{0.340 $\pm$ 0.008} & \underline{0.711 $\pm$ 0.007} \\
 & w/o Temporal  & \underline{1.127 $\pm$ 0.022} & 0.711 $\pm$ 0.012 & 0.441 $\pm$ 0.013 & 0.759 $\pm$ 0.005 \\
 & w/o Topology  & 1.202 $\pm$ 0.009 & 0.693 $\pm$ 0.011 & 0.377 $\pm$ 0.022 & 0.757 $\pm$ 0.005 \\
\midrule

\textcolor{googleblue}{\bf \scshape Max Depth}
 & MMG-PopNet         & \textbf{0.413 $\pm$ 0.001} & \textbf{0.271 $\pm$ 0.003} & \textbf{0.202 $\pm$ 0.005} & \textbf{0.295 $\pm$ 0.003} \\
 & w/o Text      & 0.457 $\pm$ 0.012 & \textbf{0.271 $\pm$ 0.002} & 0.205 $\pm$ 0.001 & 0.311 $\pm$ 0.004 \\
 & w/o Image     & 0.422 $\pm$ 0.005 & \underline{0.280 $\pm$ 0.006} & \underline{0.204 $\pm$ 0.002} & \underline{0.302 $\pm$ 0.004} \\
 & w/o Temporal  & \underline{0.418 $\pm$ 0.009} & 0.285 $\pm$ 0.006 & 0.211 $\pm$ 0.004 & 0.305 $\pm$ 0.004 \\
 & w/o Topology  & 0.440 $\pm$ 0.005 & 0.295 $\pm$ 0.007 & 0.218 $\pm$ 0.003 & 0.318 $\pm$ 0.003 \\
\midrule

\textcolor{googleblue}{\bf \scshape Size}
 & MMG-PopNet         & \textbf{1.703 $\pm$ 0.015} & \textbf{0.949 $\pm$ 0.017} & 0.524 $\pm$ 0.010 & \textbf{1.059 $\pm$ 0.012} \\
 & w/o Text      & 1.899 $\pm$ 0.055 & \underline{0.962 $\pm$ 0.005} & \textbf{0.509 $\pm$ 0.013} & 1.123 $\pm$ 0.021 \\
 & w/o Image     & 1.783 $\pm$ 0.022 & 0.995 $\pm$ 0.019 & \underline{0.520 $\pm$ 0.006} & \underline{1.099 $\pm$ 0.010} \\
 & w/o Temporal  & \underline{1.749 $\pm$ 0.039} & 1.086 $\pm$ 0.022 & 0.649 $\pm$ 0.006 & 1.161 $\pm$ 0.009 \\
 & w/o Topology  & 1.854 $\pm$ 0.025 & 1.050 $\pm$ 0.014 & 0.548 $\pm$ 0.019 & 1.151 $\pm$ 0.007 \\
\midrule

\makecell[l]{\textcolor{googleblue}{\bf \scshape Structural}\\\textcolor{googleblue}{\bf \scshape Virality}}
 & MMG-PopNet         & \textbf{0.143 $\pm$ 0.002} & \textbf{0.097 $\pm$ 0.001} & \textbf{0.072 $\pm$ 0.002} & \textbf{0.104 $\pm$ 0.000} \\
 & w/o Text      & 0.158 $\pm$ 0.003 & \underline{0.098 $\pm$ 0.002} & \textbf{0.072 $\pm$ 0.001} & 0.109 $\pm$ 0.001 \\
 & w/o Image     & 0.147 $\pm$ 0.002 & 0.100 $\pm$ 0.002 & \textbf{0.072 $\pm$ 0.001} & \underline{0.106 $\pm$ 0.000} \\
 & w/o Temporal  & \underline{0.146 $\pm$ 0.005} & 0.102 $\pm$ 0.003 & \underline{0.073 $\pm$ 0.002} & 0.107 $\pm$ 0.002 \\
 & w/o Topology  & 0.152 $\pm$ 0.005 & 0.107 $\pm$ 0.003 & 0.078 $\pm$ 0.001 & 0.112 $\pm$ 0.001 \\
\midrule

\makecell[l]{\textcolor{googlegreen}{\bf \scshape Unique}\\\textcolor{googlegreen}{\bf \scshape Users}}
 & MMG-PopNet         & \textbf{1.334 $\pm$ 0.012} & \textbf{0.747 $\pm$ 0.014} & 0.407 $\pm$ 0.012 & \textbf{0.829 $\pm$ 0.010} \\
 & w/o Text      & 1.485 $\pm$ 0.040 & \underline{0.756 $\pm$ 0.004} & \textbf{0.392 $\pm$ 0.010} & 0.878 $\pm$ 0.015 \\
 & w/o Image     & 1.402 $\pm$ 0.017 & 0.789 $\pm$ 0.017 & \underline{0.406 $\pm$ 0.004} & \underline{0.866 $\pm$ 0.009} \\
 & w/o Temporal  & \underline{1.380 $\pm$ 0.030} & 0.863 $\pm$ 0.009 & 0.516 $\pm$ 0.006 & 0.920 $\pm$ 0.007 \\
 & w/o Topology  & 1.467 $\pm$ 0.021 & 0.830 $\pm$ 0.013 & 0.436 $\pm$ 0.019 & 0.911 $\pm$ 0.001 \\
\midrule

\makecell[l]{\textcolor{googlepurple}{\bf \scshape Like}\\\textcolor{googlepurple}{\bf \scshape Score}}
 & MMG-PopNet         & \textbf{4.527 $\pm$ 0.003} & \textbf{3.032 $\pm$ 0.039} & \textbf{2.632 $\pm$ 0.096} & \textbf{3.397 $\pm$ 0.035} \\
 & w/o Text      & 5.347 $\pm$ 0.050 & 3.680 $\pm$ 0.038 & 3.237 $\pm$ 0.029 & 4.088 $\pm$ 0.035 \\
 & w/o Image     & 4.783 $\pm$ 0.066 & \underline{3.271 $\pm$ 0.007} & \underline{2.674 $\pm$ 0.074} & \underline{3.576 $\pm$ 0.024} \\
 & w/o Temporal  & \underline{4.597 $\pm$ 0.080} & 3.454 $\pm$ 0.075 & 3.166 $\pm$ 0.015 & 3.739 $\pm$ 0.003 \\
 & w/o Topology  & 4.812 $\pm$ 0.055 & 3.292 $\pm$ 0.055 & 2.728 $\pm$ 0.097 & 3.611 $\pm$ 0.004 \\
\bottomrule
\end{tabular}
}
\end{table}

\FloatBarrier 
\section{Graph Classification Tasks Details}\label{sec:classification}

\subsection{Task Formulation and Label Construction}

We extend MMG-Pop to two binary graph-classification tasks using the same cascade representations and observation windows as the regression benchmark. Given an observed cascade prefix $G^t$ and its available node- and thread-level features, the objective is to predict a binary outcome derived from the final popularity targets $Y_G^{t'}$.

\textbf{Future-Shape Classification.}
Following~\cite{10.1145/2566486.2567997}, this task predicts whether the final Structural Virality exceeds the median of the corresponding training distribution. It distinguishes cascades with relatively low versus high structural virality, assessing whether early interactions reveal their eventual diffusion structure.

\textbf{Future-Popularity Classification.}
Following~\cite{info11090453}, this task predicts whether the root post's final Like Score exceeds the 75th percentile of the training distribution. It evaluates whether early observations can identify posts that achieve high engagement relative to other cascades in the same dataset.

For a dataset $D$ and observation window $t$, the binary labels are defined as
\begin{equation}
\begin{aligned}
y_G^{\mathrm{shape}}
&= \mathbb{I}\left[
Y_{G,\mathrm{SV}}^{t'} > m_{D,t}^{\mathrm{SV}}
\right],\\
y_G^{\mathrm{pop}}
&= \mathbb{I}\left[
Y_{G,\mathrm{Like}}^{t'} > q_{D,t}^{0.75}
\right],
\end{aligned}
\end{equation}
where $Y_{G,\mathrm{SV}}^{t'}$ and $Y_{G,\mathrm{Like}}^{t'}$ denote the untransformed final Structural Virality and Like Score, respectively. The thresholds $m_{D,t}^{\mathrm{SV}}$ and $q_{D,t}^{0.75}$ are computed from the eligible training cascades using the median and 75th percentile (with linear interpolation), respectively.

Thresholds are computed separately for each dataset and observation window, since the eligible cascade set varies with $t$. They are then fixed for validation and testing. Target values equal to the threshold are assigned to the negative class.

\subsection{Training and Evaluation Protocol}

We evaluate MLP, Graph-LSTM, and MMG-PopNet using the same window-specific cascade filtering as the regression experiments. The representative observation windows are 10, 30, 50, and 90 minutes for Bluesky, r/AMA, r/Gaming, and r/Futurology, respectively. We additionally evaluate r/Futurology at 30 and 180 minutes to examine the effect of observation duration.

The two tasks are trained independently, without sharing parameters or jointly optimizing regression targets. For each cascade, the model produces a logit
\begin{equation}
z_G = F_\Theta\left(
G^t,\{X_v^t\}_{v\in V^t},X^{\mathrm{Thread},t}
\right),
\qquad
\hat{p}_G = \sigma(z_G),
\end{equation}
where $\hat{p}_G$ denotes the predicted positive-class probability. 
Both tasks use binary cross-entropy with logits. Future-Shape Classification uses an unweighted loss, as the training-set median produces approximately balanced classes. In contrast, the 75th-percentile threshold in Future-Popularity Classification yields roughly 25\% positive samples. To account for this imbalance, we apply a positive-class weight $w_+=n_0/n_1$, where $n_0$ and $n_1$ denote the negative and positive training-class counts, respectively, balancing their contributions to the loss.

For each model, dataset, window, and task, we select a decision threshold $\eta^*$ that maximizes positive-class F1 on the validation split. Candidate thresholds comprise all distinct validation probabilities together with 0 and 1 and ties are resolved by selecting the smallest threshold. Test predictions are positive when $\hat{p}_G\geq\eta^*$. We report Macro-F1, the unweighted mean of both class-wise F1 scores, and ROC-AUC, computed directly from predicted probabilities independently of $\eta^*$.

\textbf{Results Analysis.}
While MMG-PopNet achieves the highest average performance on both tasks, Graph-LSTM slightly outperforms it in shape-classification Macro-F1 on r/Gaming at 50 minutes (0.736 vs. 0.735) and r/Futurology at 30 minutes (0.688 vs. 0.682), with both models tying at 180 minutes (0.816). However, MMG-PopNet consistently achieves higher ROC-AUC, indicating stronger discrimination across classification thresholds. On r/Futurology, its ROC-AUC increases from 0.805 to 0.889 for shape classification and from 0.804 to 0.908 for popularity classification as the observation window expands from 30 to 180 minutes. These results further demonstrate the value of additional early interaction information for predicting both diffusion structure and engagement outcomes.

\FloatBarrier 
\section{Qualitative Case Study}\label{sec-casestudy}
This qualitative case study examines MMG-PopNet predictions for the r/Futurology dataset under a 180-minute observation window, focusing on the \textsc{Size} target, which measures the final number of nodes in a social cascade. The scatter plot compares predicted \textsc{Size} against actual \textsc{Size}. Both axes are shown in log scale and are displayed using powers of ten, such as $10^1$, $10^2$, and $10^3$, to make the heavy-tailed popularity distribution easier to interpret. In this view, points near the red dashed diagonal indicate more accurate predictions, while points above or below the diagonal indicate over-prediction or under-prediction.

For the qualitative analysis, several cascades are selected from different regions of the scatter plot to provide a wide range of examples, including accurate predictions, over-predictions, and under-predictions. These selected examples are shown as colored nodes on the scatter plot. For each selected case, interpretability is applied to the root post content. Specifically, GradSAM~\cite{barkan2021grad} token explanations are used to highlight influential text tokens, and Grad-CAM~\cite{selvaraju2020grad,jacobgilpytorchcam} image explanations are used to highlight influential image regions. This helps provide intuition about how the root post’s text and image content may have contributed to the predicted cascade \textsc{Size}, while also showing that final popularity depends on additional temporal and interaction dynamics captured by the full model.

\newpage

\begin{figure}[H]
\vspace{-2ex}
    \centering
\includegraphics[width=0.6\linewidth]{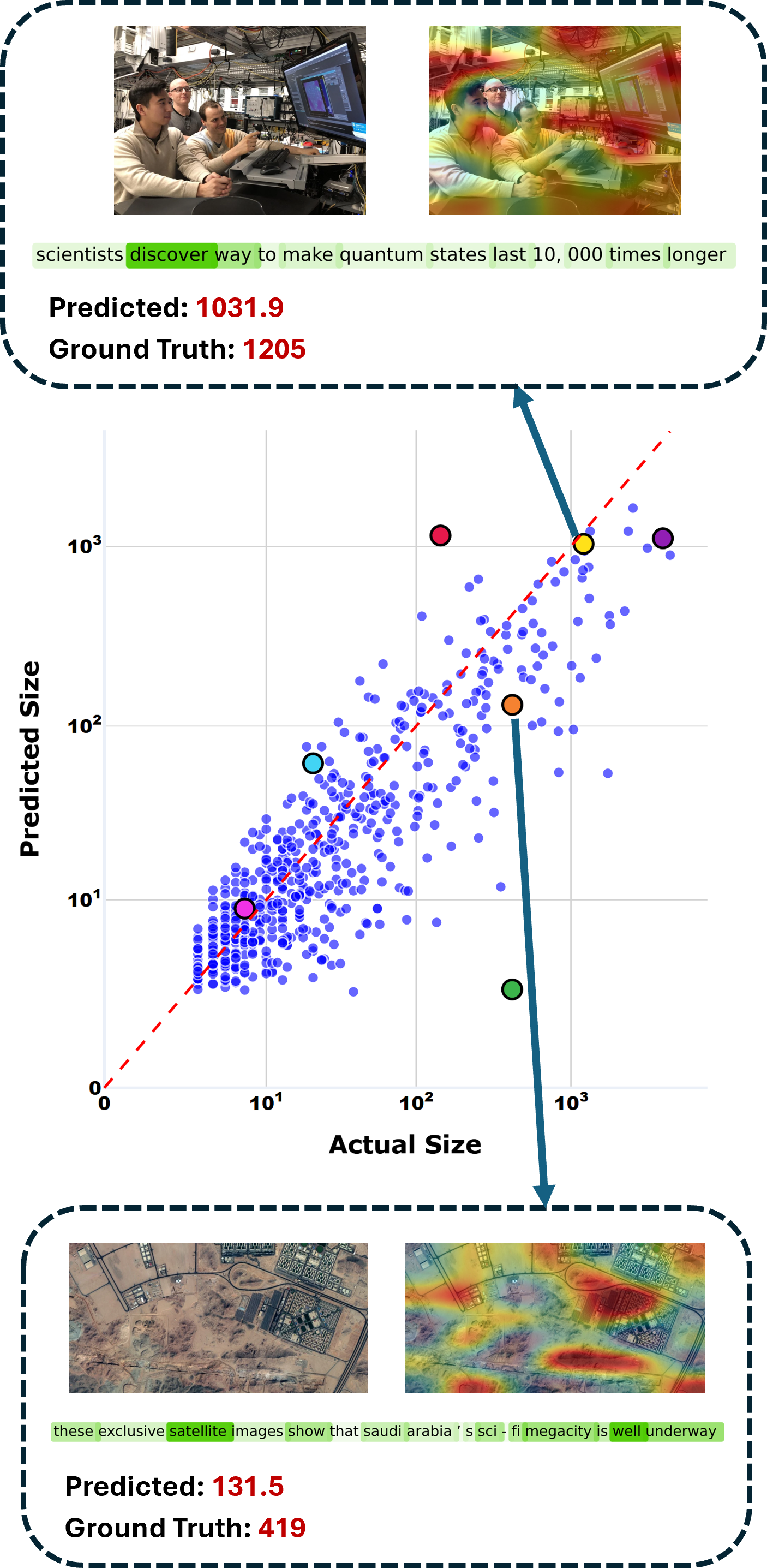}
\caption{The top example shows a relatively accurate high-popularity prediction, where the predicted \textsc{Size} is close to the actual \textsc{Size}. GradSAM highlights root-post tokens such as ``discover'', ``quantum states'', and ``longer'' suggesting that the model attends to scientific novelty and breakthrough-oriented language. Grad-CAM emphasizes parts of the laboratory image, including the people and equipment, which may provide visual cues of scientific credibility. The bottom example shows an under-predicted cascade, where the actual \textsc{Size} is much larger than the predicted \textsc{Size}. GradSAM highlights tokens such as ``exclusive,'' ``satellite,'' ``megacity,'' and ``well underway,'' while Grad-CAM focuses on several regions of the satellite image. This case suggests that although the root post contains visually and textually salient signals, the model may underestimate posts whose later popularity is driven by broader public interest in large-scale infrastructure or geopolitical topics.
}
    \label{fig:qual2}
    \vspace{-1em}
\end{figure}

\begin{figure}[H]
\vspace{-2ex}
    \centering
\includegraphics[width=0.61\linewidth]{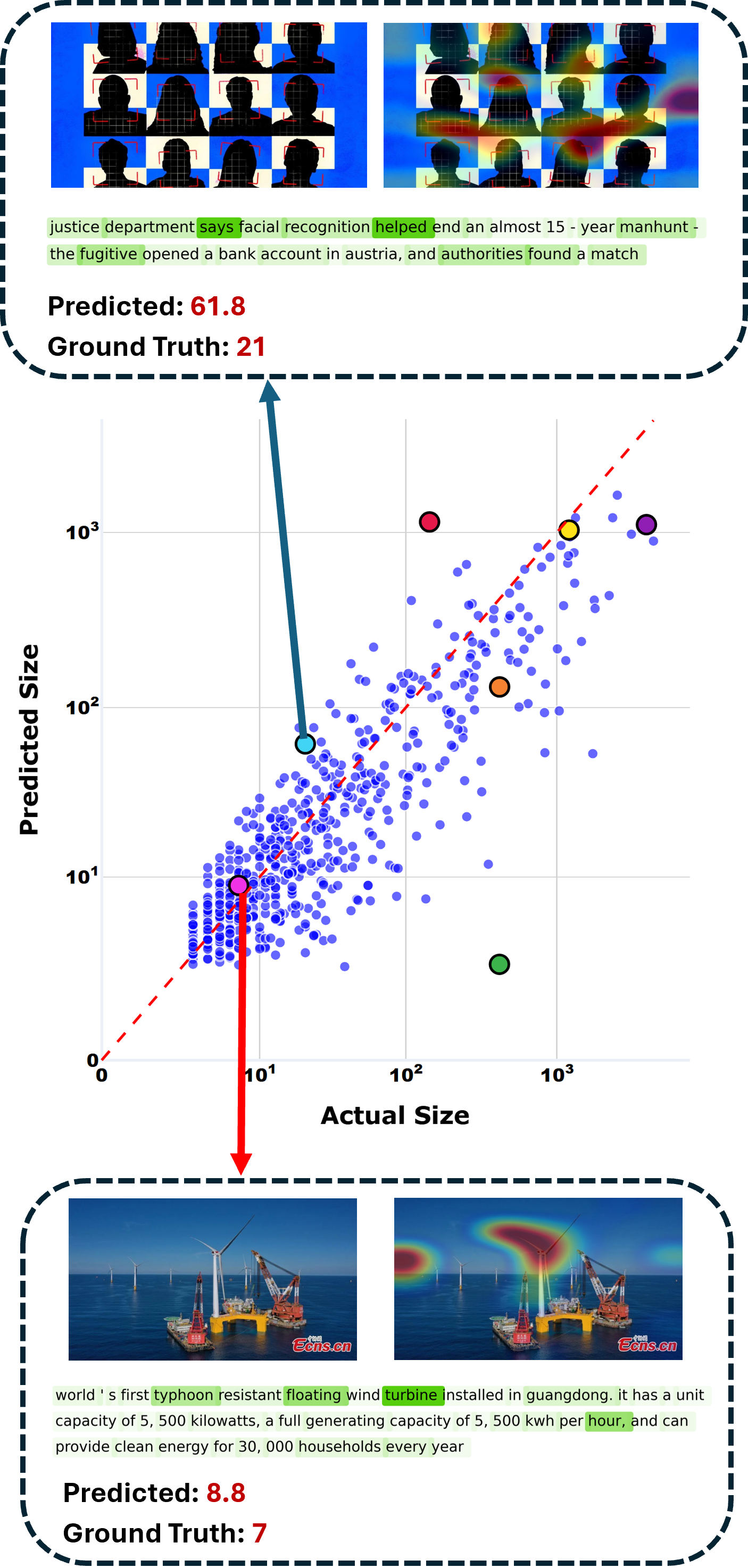}
\caption{The top example shows an over-predicted cascade, where the predicted \textsc{Size} is larger than the actual \textsc{Size}. GradSAM highlights root-post tokens such as ``says,'' ``facial recognition,'' ``helped,'' ``fugitive,'' and ``match,'' suggesting that the model attends to crime, surveillance, and authority-related language. Grad-CAM highlights multiple regions across the facial-recognition image, which may reinforce the post’s technology and public-safety framing. The bottom example shows a relatively accurate low-popularity prediction, where the predicted \textsc{Size} is close to the actual \textsc{Size}. GradSAM highlights tokens such as ``floating,'' ``wind turbine,'' and ``hour,'' while Grad-CAM focuses on parts of the offshore turbine and surrounding scene. This comparison suggests that root-post content can provide meaningful cues for \textsc{Size} prediction, but attention to salient technology-related terms does not always translate into high cascade growth.
}
    \label{fig:qual3}
    \vspace{-1em}
\end{figure}

\begin{figure}[H]
\vspace{-2ex}
    \centering
\includegraphics[width=0.62\linewidth]{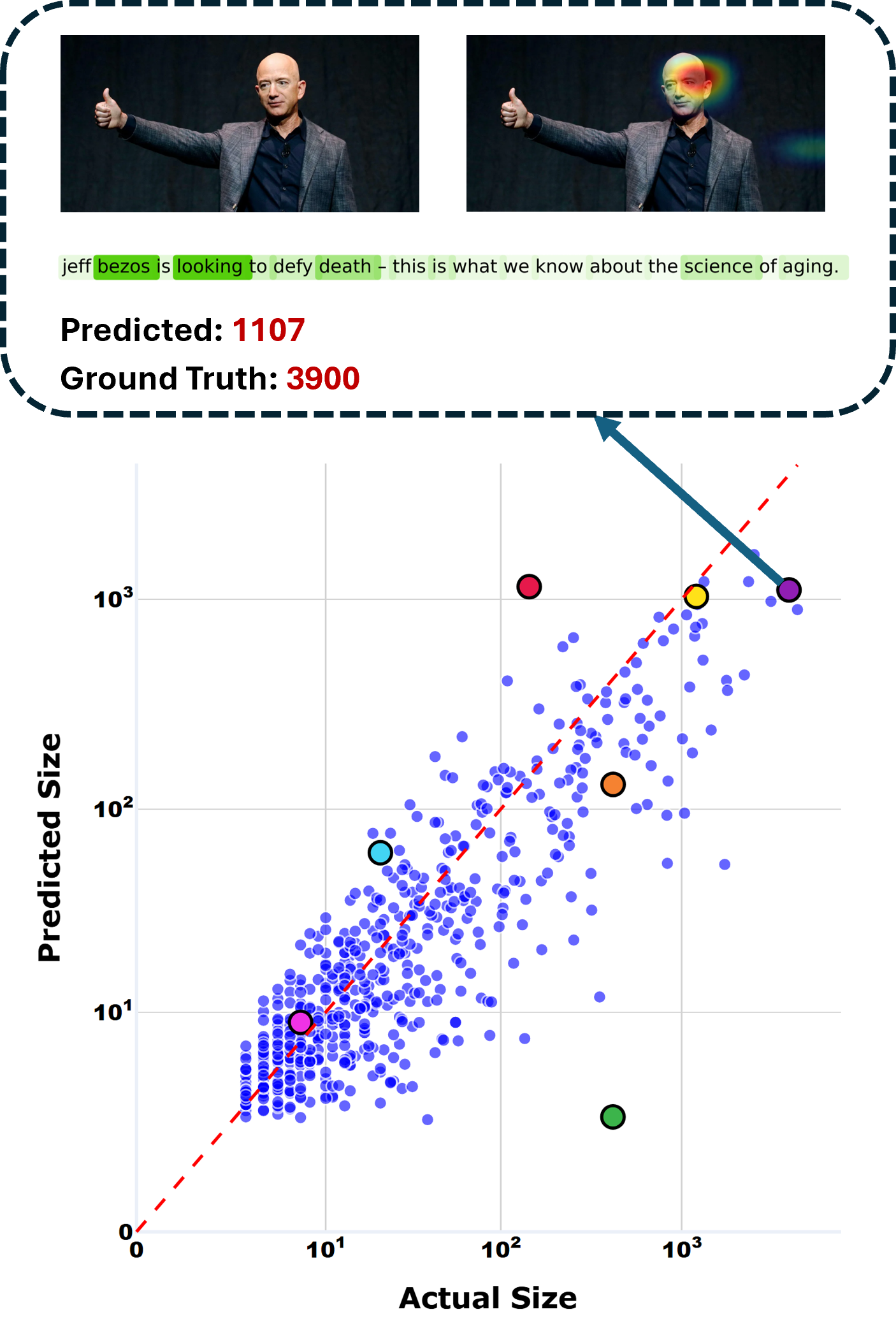}
\caption{The highlighted example shows an under-predicted cascade, where the actual \textsc{Size} is much larger than the predicted \textsc{Size}. GradSAM highlights root-post tokens such as ``bezos,'' looking,'' death,'' and science of aging,'' suggesting that the model attends to the named entity and the longevity-related framing of the post. Grad-CAM focuses strongly on the face of Jeff Bezos, indicating that the visual explanation is concentrated on him in the image. This case suggests that the root post contains salient celebrity and science-related cues, but the model still underestimates the eventual discussion volume, possibly because later cascade growth is driven by broader public debate around wealth, longevity, and aging beyond the root content alone.
}
    \label{fig:qual4}
    \vspace{-1em}
\end{figure}
\FloatBarrier
\section{Limitations.}\label{sec:limit}
Despite the unified design of MMG-Pop and the strong empirical performance of MMG-PopNet, this work has several limitations.

\textbf{First}, the benchmark is constructed from Bluesky and Reddit, which provide diverse but still incomplete coverage of social media ecosystems. Platform-specific moderation policies, recommendation algorithms, user demographics, and interaction norms can substantially affect popularity dynamics. Therefore, conclusions drawn from these datasets may not fully generalize to platforms such as X/Twitter, TikTok, Instagram, YouTube, or private messaging communities.

\textbf{Second}, our formulation represents social cascades primarily as tree-structured reply or interaction graphs. This abstraction captures explicit propagation paths, but it may omit broader network exposure effects, algorithmic ranking effects, cross-platform diffusion (e.g., getting high engagement on one social platform due to a viral event on second social platform), and unobserved impressions. A post may become popular not only because of its visible reply tree, but also because of recommendation systems, external sharing, creator reputation, or coordinated amplification that is not directly observable in the collected cascade.

\textbf{Third}, although MMG-PopNet jointly models text, image, temporal, and structural signals, the available modalities are uneven across platforms and communities. Root visual content contributes modestly in our experiments, but this may partly reflect dataset composition rather than the intrinsic value of visual signals. Similarly, richer video, audio and user-history features are not fully modeled. Future extensions should consider broader media types and more complete user-context features while carefully protecting user privacy.

\textbf{Fourth}, the work of popularity prediction has the potential of being misused by nefarious parties to identify the signals that provide them the highest social engagement to spread hateful or toxic messages or content on the social media. So, one has to be careful and mindful in using such techniques to ensure well being of all. 
\FloatBarrier
\newpage
\section*{NeurIPS Paper Checklist}

The checklist is designed to encourage best practices for responsible machine learning research, addressing issues of reproducibility, transparency, research ethics, and societal impact. Do not remove the checklist: {\bf The papers not including the checklist will be desk rejected.} The checklist should follow the references and follow the (optional) supplemental material.  The checklist does NOT count towards the page
limit. 

Please read the checklist guidelines carefully for information on how to answer these questions. For each question in the checklist:
\begin{itemize}
    \item You should answer \answerYes{}, \answerNo{}, or \answerNA{}.
    \item \answerNA{} means either that the question is Not Applicable for that particular paper or the relevant information is Not Available.
    \item Please provide a short (1--2 sentence) justification right after your answer (even for \answerNA). 
\end{itemize}

{\bf The checklist answers are an integral part of your paper submission.} They are visible to the reviewers, area chairs, senior area chairs, and ethics reviewers. You will also be asked to include it (after eventual revisions) with the final version of your paper, and its final version will be published with the paper.

The reviewers of your paper will be asked to use the checklist as one of the factors in their evaluation. While \answerYes{} is generally preferable to \answerNo{}, it is perfectly acceptable to answer \answerNo{} provided a proper justification is given (e.g., error bars are not reported because it would be too computationally expensive'' or ``we were unable to find the license for the dataset we used''). In general, answering \answerNo{} or \answerNA{} is not grounds for rejection. While the questions are phrased in a binary way, we acknowledge that the true answer is often more nuanced, so please just use your best judgment and write a justification to elaborate. All supporting evidence can appear either in the main paper or the supplemental material, provided in appendix. If you answer \answerYes{} to a question, in the justification please point to the section(s) where related material for the question can be found.

IMPORTANT, please:
\begin{itemize}
    \item {\bf Delete this instruction block, but keep the section heading ``NeurIPS Paper Checklist"},
    \item  {\bf Keep the checklist subsection headings, questions/answers and guidelines below.}
    \item {\bf Do not modify the questions and only use the provided macros for your answers}.
\end{itemize}


\begin{enumerate}

\item {\bf Claims}
    \item[] Question: Do the main claims made in the abstract and introduction accurately reflect the paper's contributions and scope?
    \item[] Answer: \answerYes{}
    \item[] Justification: Our main claims, including that the proposed MMG-PopNet consistently achieves superior social media popularity prediction performance and that current LLM-based approaches exhibit notable limitations in social popularity prediction, constitute the core contributions of this work.

\item {\bf Limitations}
    \item[] Question: Does the paper discuss the limitations of the work performed by the authors?
    \item[] Answer: \answerYes{}
    \item[] Justification: See Appendix~\ref{sec:limit} for the limitations of the work.

\item {\bf Theory assumptions and proofs}
    \item[] Question: For each theoretical result, does the paper provide the full set of assumptions and a complete (and correct) proof?
    \item[] Answer: \answerNA{} 
    \item[] Justification: \answerNA{}

    \item {\bf Experimental result reproducibility}
    \item[] Question: Does the paper fully disclose all the information needed to reproduce the main experimental results of the paper to the extent that it affects the main claims and/or conclusions of the paper (regardless of whether the code and data are provided or not)?
    \item[] Answer: \answerYes{} 
    \item[] Justification: Detailed implementation details, along with experimental and setup information are provided for the proposed method as well as for representative baselines. See Appendix~\ref{sec:appendix_baseline} and \ref{sec:appendix_experiment} for more details.

\item {\bf Open access to data and code}
    \item[] Question: Does the paper provide open access to the data and code, with sufficient instructions to faithfully reproduce the main experimental results, as described in supplemental material?
    \item[] Answer: \answerYes{} 
    \item[] Justification: Yes, link to the code has been provided in the abstract. The link to dataset is at \url{https://huggingface.co/datasets/anonymoususer54829/MMG-Pop}.

\item {\bf Experimental setting/details}
    \item[] Question: Does the paper specify all the training and test details (e.g., data splits, hyperparameters, how they were chosen, type of optimizer) necessary to understand the results?
    \item[] Answer: \answerYes{} 
    \item[] Justification: See Appendix~\ref{sec:appendix_baseline} and \ref{sec:appendix_experiment}.

\item {\bf Experiment statistical significance}
    \item[] Question: Does the paper report error bars suitably and correctly defined or other appropriate information about the statistical significance of the experiments?
    \item[] Answer: \answerYes{}
    \item[] Justification: See Appendix~\ref{appendix:stat_significance} and \ref{sec:appendix_modality}.

\item {\bf Experiments compute resources}
    \item[] Question: For each experiment, does the paper provide sufficient information on the computer resources (type of compute workers, memory, time of execution) needed to reproduce the experiments?
    \item[] Answer: \answerYes{} 
    \item[] Justification: See Appendix~\ref{appendix:split_compute}.
    
\item {\bf Code of ethics}
    \item[] Question: Does the research conducted in the paper conform, in every respect, with the NeurIPS Code of Ethics \url{https://neurips.cc/public/EthicsGuidelines}?
    \item[] Answer: \answerYes{} 
    \item[] Justification: Full efforts have been made to preserve anonymity, along with discussing limitations affecting out work.

\item {\bf Broader impacts}
    \item[] Question: Does the paper discuss both potential positive societal impacts and negative societal impacts of the work performed?
    \item[] Answer: \answerYes{} 
    \item[] Justification: Positive applications of the work are introduced in the introduction, while the potential negative impact is mentioned as fourth point in Appendix~\ref{sec:limit}. 
    
\item {\bf Safeguards}
    \item[] Question: Does the paper describe safeguards that have been put in place for responsible release of data or models that have a high risk for misuse (e.g., pre-trained language models, image generators, or scraped datasets)?
    \item[] Answer: \answerYes{} 
    \item[] Justification: Our work does not release any newly collected or proprietary social media dataset that introduces previously unavailable knowledge or sensitive user information. Instead, the benchmark is curated from well-established publicly available open-source data sources and complies with the corresponding platform terms of service, as discussed in Appendix~\ref{sec:appendix_dataset}. In addition, we do not release personally identifiable information (PII). The curated data only retains anonymized Reddit author identifiers, which are platform-generated random codes and do not reveal personal identity or enable direct tracking of individual users. These efforts help mitigate potential privacy and misuse concerns while supporting reproducible research on social media popularity prediction.

\item {\bf Licenses for existing assets}
    \item[] Question: Are the creators or original owners of assets (e.g., code, data, models), used in the paper, properly credited and are the license and terms of use explicitly mentioned and properly respected?
    \item[] Answer: \answerYes{} 
    \item[] Justification: Yes, creators or original owners are appropriately cited throughout the wor.

\item {\bf New assets}
    \item[] Question: Are new assets introduced in the paper well documented and is the documentation provided alongside the assets?
    \item[] Answer: \answerYes{} 
    \item[] Justification: Steps as to how to use and operate the dataset and the code are mentioned on the anonymous code repository.  

\item {\bf Crowdsourcing and research with human subjects}
    \item[] Question: For crowdsourcing experiments and research with human subjects, does the paper include the full text of instructions given to participants and screenshots, if applicable, as well as details about compensation (if any)? 
    \item[] Answer: \answerNA{} 
    \item[] Justification: \answerNA{}

\item {\bf Institutional review board (IRB) approvals or equivalent for research with human subjects}
    \item[] Question: Does the paper describe potential risks incurred by study participants, whether such risks were disclosed to the subjects, and whether Institutional Review Board (IRB) approvals (or an equivalent approval/review based on the requirements of your country or institution) were obtained?
    \item[] Answer: \answerNA{} 
    \item[] Justification: \answerNA{}

\item {\bf Declaration of LLM usage}
    \item[] Question: Does the paper describe the usage of LLMs if it is an important, original, or non-standard component of the core methods in this research? Note that if the LLM is used only for writing, editing, or formatting purposes and does \emph{not} impact the core methodology, scientific rigor, or originality of the research, declaration is not required.
    \item[] Answer: \answerNA{} 
    \item[] Justification: \answerNA{}

\end{enumerate}

\end{document}